\documentclass[12pt,a4paper]{article}
\usepackage{amsmath,amssymb,mathrsfs, url,cite,ifpdf,slashed,multirow,tabularx,type1cm,booktabs,subfigure}
\usepackage{ascmac}
\usepackage{float}
\usepackage{graphicx, color}

\allowdisplaybreaks

\usepackage{caption}
\definecolor{BlueViolet}{rgb}{0.2, 0.00, 0.7}
\definecolor{Blue}{rgb}{0.15, 0.00, 0.9}
\usepackage[colorlinks=true,linkcolor=Blue,citecolor=Blue,urlcolor=BlueViolet]{hyperref}

\def\thefootnote{\ifnum\c@footnote>\z@\textasteriskcentered\@arabic\c@footnote\fi}
\makeatletter
\renewcommand{\footnoterule}{%
\kern-3\p@
\hrule width 0.4\columnwidth
\kern 2.6\p@}
\def\thefootnote{\ifnum\c@footnote>\z@\@arabic\c@footnote\fi}
\makeatother
\makeatletter
\newcommand{\@authornote}[2]{{\def\thefootnote{\fnsymbol{footnote}}\setcounter{footnote}{#1}#2\setcounter{footnote}{0}}}
\newcommand{\authornotemark}[1]{\@authornote#1{\addtocounter{footnote}{-1}\footnotemark}}
\newcommand{\authornotetext}[2]{\@authornote#1{\footnotetext{#2}}}
\makeatother

\begin{document}


\newcommand{\TeV}{\,{\rm TeV}}
\newcommand{\GeV}{\,{\rm GeV}}
\newcommand{\MeV}{\,{\rm MeV}}
\newcommand{\keV}{\,{\rm keV}}
\newcommand{\invfb}{\,{\rm fb^{-1}}}
\newcommand{\invpb}{\,{\rm pb^{-1}}}
\newcommand{\Slash}[1]{{\ooalign{\hfil \hspace*{-5pt}~#1\hfil\crcr\raise.167ex\hbox{/}}}}
\def\be{\begin{equation}}
\def\ee{\end{equation}}
\def\fr{\frac}
\def\({\left(}
\def\){\right)}
\def\<{\langle}
\def\>{\rangle}
\newcommand{\non}{\nonumber \\ }
\newcommand{\matl}{\left( \begin{array}}
\newcommand{\matr}{\end{array} \right)}
\renewcommand{\sb}{\sin\beta}
\newcommand{\cb}{\cos\beta}
\newcommand{\tb}{\tan\beta}

\def\o{\over}
\def\beq#1\eeq{\begin{align}#1\end{align}}
\newcommand{\gsim}{ \mathop{}_{\textstyle \sim}^{\textstyle >} }
\newcommand{\lsim}{ \mathop{}_{\textstyle \sim}^{\textstyle <} }
\newcommand{\vev}[1]{ \left\langle {#1} \right\rangle }
\newcommand{\bra}[1]{ \langle {#1} | }
\newcommand{\ket}[1]{ | {#1} \rangle }
\newcommand{\EV}{ \text{eV} }
\newcommand{\KEV}{ \text{keV} }
\newcommand{\MEV}{ \text{MeV} }
\newcommand{\GEV}{ \text{GeV} }
\newcommand{\TEV}{ \text{TeV} }
\newcommand{\1}{\mbox{1}\hspace{-0.25em}\mbox{l}}
\newcommand{\headline}[1]{\noindent{\bf #1}}
\def\diag{\mathop\text{diag}\nolimits}
\def\Spin{\mathop\text{Spin}}
\def\SO{\mathop\text{SO}}
\def\O{\mathop\text{O}}
\def\SU{\mathop\text{SU}}
\def\U{\mathop\text{U}}
\def\Sp{\mathop\text{Sp}}
\def\SL{\mathop\text{SL}}
\def\tr{\mathop\text{tr}}
\def\mpl{M_\text{Pl}}

\def\dd{\mathrm{d}}
\def\ff{\mathrm{f}}
\def\BH{\text{BH}}
\def\inf{\text{inf}}
\def\ev{\text{evap}}
\def\eq{\text{eq}}
\def\SM{\text{sm}}
\def\Mpl{M_\text{Pl}}
\def\GeV{\text{GeV}}

\begin{titlepage}
\begin{center}

\hfill TTP16--029\\
\hfill July 2016

\vspace{1.cm}
{\large   \bf 
{Singularity-free Next-to-leading Order}
$\boldsymbol{\Delta S= 1}$ \\ 
\vspace{.15 cm}
Renormalization Group Evolution and \\
\vspace{.25 cm}  $\boldsymbol{\epsilon_{K}^\prime/ \epsilon_{K}}$ 
in the Standard Model and Beyond  
}

\vspace{1.3cm}
{\fontsize{14.pt}{0pt}\selectfont{
{\bf Teppei Kitahara}$^{a,b}$\footnote[0]{${}${\it E-mail:} \textcolor{BlueViolet}{teppei.kitahara@kit.edu}},
{\bf Ulrich Nierste}$^{a}$\footnote[0]{${}${\it E-mail:} \textcolor{BlueViolet}{Ulrich.Nierste@kit.edu}},\\ 
\vspace{.2cm}
and
{\bf Paul Tremper}$^{a}$\footnote[0]{${}${\it E-mail:} \textcolor{BlueViolet}{paul.tremper@kit.edu}}
}}

\vspace{.5cm} {\it
  $^{a}${~Institute for Theoretical Particle Physics (TTP), Karlsruhe Institute of Technology, Engesserstra{\ss}e 7, D-76128 Karlsruhe, Germany}\\
  $^{b}${~Institute for Nuclear Physics (IKP), Karlsruhe Institute of
    Technology, Hermann-von-Helmholtz-Platz 1, D-76344
    Eggenstein-Leopoldshafen, Germany} }

\vspace{1.cm} \abstract{The standard analytic solution of the
  renormalization group (RG) evolution for the $\Delta S = 1$ Wilson
  coefficients involves several singularities, which complicate
    analytic solutions.  In this paper we derive a singularity-free
  solution of the next-to-leading order (NLO) RG equations, which
  greatly facilitates the calculation of $\epsilon_K^\prime$, the
  measure of direct $CP$ violation in $K\to \pi\pi$ decays.  Using our
    new RG evolution and the latest lattice results for the hadronic
    matrix elements, we calculate the ratio
  $\epsilon_{K}'/\epsilon_{K}$ (with $\epsilon_{K}$ quantifying
    indirect $CP$ violation) in the Standard Model (SM) at NLO to
  $\epsilon_{K}'/\epsilon_{K} = \left(1.06 \pm  5.07  \right) \times
  10^{-4} $, which is $2.8\,\sigma$ below the experimental value.  We
  also present the evolution matrix in the high-energy regime for
  calculations of new physics contributions and derive easy-to-use
    approximate formulae.  We find that the RG amplification of
  new-physics contributions to Wilson coefficients of the
  electroweak penguin operators is further enhanced by the NLO
    corrections: If the new contribution is generated at the scale of
  $1$--$10$$\,$TeV, the RG evolution between the new-physics scale
  and the electroweak scale enhances these coefficients by
  $50$--$100$\%. Our solution contains a term of order
    $\alpha_{EM}^2/\alpha_s^2$, which is numerically unimportant for the
    SM case but should be included in studies of high-scale
    new-physics.}
\end{center}
\end{titlepage}
\setcounter{footnote}{0}
\renewcommand{\thefootnote}{\#\arabic{footnote}}
\setcounter{page}{1}

\hrule
\tableofcontents
\vskip .2in
\hrule
\vskip .4in

\section{Introduction}
The parameter $\epsilon_{K}^{\prime}/\epsilon_{K}$ is the ratio of
{the measures} of direct {and} indirect charge-parity ($CP$)
violation in the Kaon system.  While {indirect} $CP$ violation is a
per-mille effect in the Standard Model (SM), {$\epsilon_{K}^{\prime}$ is
  smaller by another three orders of magnitude than $\epsilon_{K}$,
  with} $|\epsilon_{K}^{\prime}|\sim \mathcal{O}(10^{-6})$.  A strong
suppression {by the Glashow-Iliopoulos-Maiani (GIM) mechanism} and an
accidental cancellation of leading contributions in the Standard Model
{makes} $\epsilon_{K}^{\prime}/\epsilon_{K}$ highly sensitive to new
physics. The first element of the SM prediction for
  $\epsilon_{K}^{\prime}$ is the calculation of initial conditions for
  Wilson coefficients and their renormalization group evolution from the
  electroweak scale (of the order of $W$ and top mass) down to the
  hadronic scale of order 1$\,$ GeV, at which hadronic matrix elements
  are calculated. These steps purely involve perturbative methods and
  have been carried out to leading order (LO) in the strong coupling
  constant $\alpha_s$ in {Refs.\cite{Gilman:1978wm, Guberina:1979ix, Hagelin:1983rb, Buras:1987qa}.}
   The next-to-leading order (NLO)
  involves {the electromagnetic coupling $\alpha_{EM}\simeq 1/128$ \cite{Flynn:1989iu,Buchalla:1989we,Paschos:1991as, Lusignoli:1991bm},
  the next higher order in $\alpha_s$ \cite{Buras:1991jm,Buras:1992tc, Ciuchini:1993vr}, and order $\alpha_{EM} \alpha_s $ \cite{Ciuchini:1993vr, Buras:1992zv ,Buras:1993dy}.}  
  In terms
of isospin amplitudes {$\epsilon_{K}^{\prime}$} is given by
({see e.g.\ Ref.}~\cite{Buras:2015yba})
\beq
\frac{\epsilon_{K}^{\prime}}{\epsilon_{K}} = \frac{\omega_{+}}{\sqrt{2}
  \left|\epsilon_{K}\right| \textrm{Re} A_0} \left( \frac{1}{\omega_{+}}
  \textrm{Im} A_2 - (1-\hat{\Omega}_{\textrm{eff}}) \textrm{Im} A_0
\right),
\label{eq:epsilonequation} 
\eeq
where $A_I \equiv \langle (\pi \pi)_I | \mathcal{H}_{\textrm{eff}}^{|
  \Delta S |=1} | K^0 \rangle$ are isospin amplitudes and $\omega_{+} =
(4.53 \pm 0.02) \times 10^{-2}$ (see
  Refs.~\cite{Cirigliano:2003nn,Buras:2015yba} for the precise
  definition), $|\epsilon_{K}|=(2.228 \pm 0.011)\cdot 10^{-3}$,
and $\textrm{Re} A_0 = ( 3.3201\pm 0.0018 )\times 10^{-7}\,\GeV$
are taken from experiment.  $\hat{\Omega}_{\textrm{eff}} = (14.8 \pm
8.0)\times 10^{-2}$ {parameterizes} isospin-violating contributions
\cite{Cirigliano:2003nn, Cirigliano:2003gt}.

The $| \Delta S |=1$ nonleptonic effective Hamiltonian for weak decays
in the Standard Model is given by \cite{Buras:1993dy}
\beq 
\mathcal{H}_{\textrm{eff}}^{ | \Delta S|=1} &= \frac{G_F}{\sqrt{2}}
\lambda_u \sum_{i=1}^{10} Q_i (\mu) \Bigl( \left( 1 - \tau \right) z_i
(\mu) + \tau v_i (\mu) \Bigr) + \textrm{H.c.} \\ &\equiv
\frac{G_F}{\sqrt{2}} \lambda_u \sum_{i=1}^{10} Q_i (\mu) \left( z_i
  (\mu) + \tau y_i (\mu) \right) + \textrm{H.c.},
\label{eq:hamilton}
\eeq 
where $\lambda_u = V_{us}^{\ast} V_{ud} $ and $\tau = - V_{ts}^{\ast}
V_{td} / \left( V_{us}^{\ast} V_{ud} \right)$.  The operator basis $Q_i$
{comprises} ten operators which are defined in Ref.~\cite{Buras:1993dy};
the current-current operators $Q_1$ and $Q_2$ 
\beq 
Q_1 = \left( \bar{s}_{\alpha} u_{\beta}\right)_{V-A} \left( \bar{u}_{
    \beta} d_{\alpha} \right)_{V-A},~~~~~~ Q_2 = \left( \bar{s} u
\right)_{V-A} \left( \bar{u} d\right)_{V-A}, \label{eq:q1} 
\eeq 
the QCD-penguin
operators $Q_3$ to $Q_6$ \beq Q_3 = \left( \bar{s} d \right)_{V-A}
\sum_q \left( \bar{q} q \right)_{V-A},~~~
Q_4 = \left( \bar{s}_{\alpha} d_{\beta} \right)_{V-A} \sum_q \left( \bar{q}_{\beta} q_{\alpha} \right)_{V-A},\\
Q_5 = \left( \bar{s} d \right)_{V-A} \sum_q \left( \bar{q} q
\right)_{V+A},~~~ Q_6 = \left( \bar{s}_{\alpha} d_{\beta} \right)_{V-A}
\sum_q \left( \bar{q}_{\beta} q_{\alpha} \right)_{V+A}, \eeq and the
QED-penguin operators $Q_7$ to $Q_{10}$ \beq Q_7 = \frac{3}{2} \left(
  \bar{s} d \right)_{V-A} \sum_q e_q \left( \bar{q} q \right)_{V+A},~~~
Q_8 = \frac{3}{2} \left( \bar{s}_{\alpha} d_{\beta} \right)_{V-A} \sum_q e_q \left( \bar{q}_{\beta} q_{\alpha} \right)_{V+A},\\
Q_9 = \frac{3}{2} \left( \bar{s} d \right)_{V-A} \sum_q e_q \left(
  \bar{q} q \right)_{V-A},~~~ Q_{10} = \frac{3}{2} \left(
  \bar{s}_{\alpha} d_{\beta} \right)_{V-A} \sum_q e_q \left(
  \bar{q}_{\beta} q_{\alpha} \right)_{V-A},\label{eq:q10}
\eeq
where $V \mp A$ represents $\gamma_{\mu} (1 \mp \gamma_5)$, $\alpha$ and
$\beta$ {denote} color indices, and $e_q$ is the electric charge of the
quark $q$.  The corresponding Wilson coefficients $z_i$ and $v_i$ (or
$y_i$) serve as effective couplings to these effective operators.

By virtue of the framework of effective theories, the parameter $\mu$ splits short distance from  long  distance scales, 
effectively separating the perturbative high energy regime from the non-perturbative realm of low energy QCD.
 Taking up the perturbative part of the calculation, the Wilson
 coefficients have been determined through matching calculations up to
 next-to-leading order  at the {scale $M_W$} \cite{Buras:1993dy}. 
 The calculation of the hadronic matrix elements, being non-perturbative quantities, is a major challenge and has recently been performed on the lattice with unprecedented accuracy \cite{Blum:2011ng,Blum:2012uk, Blum:2015ywa, Bai:2015nea}.

 The combination of these calculations into a prediction for
 $\epsilon_{K}^{\prime}/\epsilon_{K}$ requires a treatment within 
 renormalization group (RG) improved perturbation theory  to
 sum up large logarithms.  However, it is known that the analytic
 determination of the required evolution matrix at the next-to-leading
 order suffers from singularities appearing in intermediate steps of the
 calculation, which make a computational evaluation highly laborious and
 complicated. The standard way to solve the NLO RG equations requires
   the diagonalization of the LO anomalous dimension matrix
   $\hat{\gamma}_s^{(0)}$ and the NLO correction involves fractions
   whose denominators contain the differences of eigenvalues of
   $\hat{\gamma}_s^{(0)}$.  Some of these denominators vanish and are
   usually regulated in the numerical evaluation \cite{Ciuchini:1992tj,
  Ciuchini:1993vr}.
In Ref.~\cite{Huber:2005ig} an analytic solution for the 
RG equations which is free of singularities is presented. 
This solution involves the diagonalization of $\hat{\gamma}_s^{(0)}$
and gives explicit prescriptions to handle the different cases in which the 
formulae of Refs.~\cite{Ciuchini:1992tj,Ciuchini:1993vr} develop 
singularities.
  
In this paper, we present a new singularity-free solution which permits an easy
  and convenient numerical implementation.  Instead of singularities
our analytic formula has undetermined parameters.  However, we will show
that these {spurious} parameters {cancel} and {leave} the evolution
matrix unambiguous. Unlike the solution of Ref.~\cite{Huber:2005ig}
  our new formula requires neither the diagonalization of
  $\hat{\gamma}_s^{(0)}$ nor a distinct treatment of the part of the RG
  evolution which involves the spurious singularities.  Using our
new RG evolution and the latest lattice results
\cite{Blum:2011ng,Blum:2012uk, Blum:2015ywa, Bai:2015nea}, we calculate
the $\epsilon_{K}'/\epsilon_{K}$ in the Standard Model at
next-to-leading order {to find a value which is} below the
experimentally measured quantity {by} 2.8\,$\sigma$.

The second objective of this paper is the derivation of a useful
formula for the calculation of new physics contributions to
$\epsilon_K^\prime / \epsilon_K$, in which we evaluate the evolution
matrices {for scales far above the electroweak scale. To this end 
we identify a contribution of order $\alpha_{EM}^2/\alpha_s^2$ in the
evolution matrix which can become relevant for studies of TeV-scale new
physics, because $\alpha_s$ decreases with increasing scale.}  We observe an
approximately logarithmic behavior of the evolution matrix as a function
of the energy scale above the electroweak scale. 

This paper is organized as follows.  In Sec.~\ref{section2}, we briefly
review the RG evolution of the $|\Delta S| = 1$ effective Hamiltonian at
the next-to-leading order.  We give a detailed analysis of the
evolution matrix and its singularities and provide a new analytic
solution without singularities.  Then we evaluate $\epsilon'_K /
\epsilon_K$ in the Standard Model at the next-to-leading order in
Sec.~\ref{sec:SM}.  In Sec.~\ref{sec:NP}, we work out the evolution
matrices in the high-energy regime explicitly for calculations of new
physics contributions.  The last section is devoted to conclusions and
discussion.  

\section{Renormalization Group Evolution of the 
   $\boldsymbol{\Delta S=1}$ Hamiltonian}
\label{section2}

In this section, we review the singularities in the RG evolution of
the $|\Delta S| = 1$ effective Hamiltonian at the next-to-leading order.
Then we generalize the analytic ansatz of the RG evolution given in the
literature and present a solution, which is finite at all stages of
the calculation. Our solution contains free parameters, which we
show to cancel from the evolution matrix, and compare our
  singularity-free solution with the standard results from the
literature.

\subsection{Singularities in the Evolution Matrix}
\label{sec:singexpl}

The evolution of the Wilson coefficients $v_i$ and $z_i$ from the $W$
boson mass and the charm mass respectively to the hadronic scale 
$\mu$ are given by
\beq
\vec{v} (\mu) & = \hat{U}_3 (\mu, \mu_c)\hat{M}_c (\mu_c) \hat{U}_4 (\mu_c,m_b)\hat{M}_b (m_b) \hat{U}_5 (m_b,M_W) \vec{v}(M_W),
\label{eq:wilsonv}\\
\vec{z} (\mu) & = \hat{U}_3 (\mu, \mu_c) \vec{z}(\mu_c),
\label{eq:wilsonz}
\eeq where $\hat{U}_f (\mu_1,\mu_2)$ is the RG evolution matrix from
$\mu_2$ down to $\mu_1$ and $f $ is the number of the active
flavors between these two energy scales.  The matrices
$\hat{M}_{c,b}$ represent matching matrices between effective theories
with different numbers of flavor and are given in
Ref.~\cite{Buras:1993dy}.  Although the effect of the running of
$\alpha_{EM}$ is numerically negligible for
$\epsilon_{K}'/\epsilon_{K}$ in the Standard Model \cite{Buras:1993dy},  we consider this effect to cover new-physics scenarios with largely
separate scales.

The general form of the evolution matrix is given by 
 \cite{Buras:1979yt, Buchalla:1995vs},
 \beq
  \hat{U}_f (\mu_1,\mu_2) = T_{g_s} \exp \int^{g_s(\mu_1)}_{g_s(\mu_2)} d g_s' \frac{\hat{\gamma}^T \left(g_s' \right)}{\beta \left(g_s'\right)},
  \label{eq:Uint}
 \eeq
 with the  $g_s$-ordering operator $T_{g_s}$ and 
the anomalous dimension matrix $\hat{\gamma}$
and the QCD $\beta$ function. 
The expansions of the latter two quantities and $\alpha_{EM}$ up to NLO read:
\beq
  \label{eq:anomalousNLO}
  \hat{\gamma} \left(g_s (\mu) \right) &= \frac{\alpha_s (\mu)}{4 \pi }
  \hat{\gamma}_s^{(0)} + \frac{\alpha_{EM}(\mu)}{4 \pi}
  \hat{\gamma}_{e}^{(0)} + \frac{\alpha^2_s (\mu)}{\left( 4
      \pi\right)^2} \hat{\gamma}_s^{(1)} + \frac{\alpha_{EM} (\mu)
    \alpha_s(\mu)}{\left( 4 \pi\right)^2} \hat{\gamma}_{se}^{(1)},\\
  \beta \left( g_s (\mu)\right) & = - g_s (\mu) \left(\frac{\alpha_s
      (\mu)}{4 \pi } \beta_0 + \frac{\alpha^2_s
      (\mu)}{\left( 4 \pi\right)^2} \beta_1  + \frac{\alpha_s (\mu) \alpha_{EM}(\mu)}{ ( 4 \pi)^2} \beta^{se}_{1}\right), \\
  \alpha_{EM}(\mu) &=  \alpha_{EM} (M) \left\{ 1 +
      \frac{\alpha_{EM} (M)}{ \alpha_s (\mu)}
      \frac{\beta^{e}_0}{\beta_0} \left( 1 -
        \frac{\alpha_s(\mu)}{\alpha_s (M)} \right)\right\}^{-1}, \eeq
  where $\beta_0 = 11 - 2 f / 3 $, $\beta_1 = 102 - 38f /3 $,
    $\beta^{se}_{1} = -8/9( u + d/4) $, and $\beta^{e}_0 = - 4/3 (
      4u /3 +$ $d/3 +\ell) $ are the leading and next-to-leading
  coefficients of the QCD and QED beta functions,
    and $u,\,d,\,\ell$ are the numbers of the active up-type-quark, 
  down-type-quark, and charged-lepton flavors ($f =
    u+d$).  $\hat{\gamma}^{(0)}_s$ is the LO QCD anomalous dimension
  matrix, and the NLO corrections consist of the three remaining
  matrices, $\hat{\gamma}^{(0)}_e$, $\hat{\gamma}^{(1)}_s$, and
  $\hat{\gamma}^{(1)}_{se}$, which are the leading QED, next-to-leading
  QCD, and combined QCD-QED anomalous dimension matrices, respectively.

The ansatz for the NLO evolution matrix (with
$\mu_1 < \mu_2$) is given by \cite{Ciuchini:1992tj, Ciuchini:1993vr}
\begin{equation}
\label{eq:UNLO}
\hat{U}_f (\mu_1,\mu_2) = \hat{K}(\mu_1) \hat{U}_0 (\mu_1,\mu_2)
\hat{K}^{\prime} (\mu_2),
\end{equation}
where
\begin{align}
\hat{K}(\mu_1) & = \left( \hat{1} + \frac{\alpha_{EM}}{4 \pi}
\hat{J}_{se}\right) \left( \hat{1} + \frac{\alpha_s (\mu_1)}{4 \pi}
\hat{J}_s \right) \left( \hat{1} + \frac{\alpha_{EM}}{\alpha_s (\mu_1)}
\hat{J }_{e}\right), \label{eq:K}\\
\label{eq:Kprime}
\hat{K}^{\prime} (\mu_2) & =  \left( \hat{1} - \frac{\alpha_{EM}}{\alpha_s (\mu_2)} \hat{J }_{e} \right)  \left( \hat{1} - \frac{\alpha_s (\mu_2)}{4 \pi} \hat{J}_{s} \right) \left( \hat{1} - \frac{\alpha_{EM}}{4 \pi} \hat{J}_{se}\right),
\end{align}
and the LO evolution matrix 
\begin{equation}
\hat{U}_0 (\mu_1,\mu_2) = \hat{U}_0 \left(\alpha_s(\mu_1),
\alpha_s(\mu_2)\right) = \exp \left[ \frac{\hat{\gamma}^{(0)T}_s}{2
    \beta_0} \ln \frac{\alpha_s(\mu_2)}{\alpha_s(\mu_1)} 
  \right],
\end{equation} 
where the QED contributions to the beta functions ($
  \beta^{se}_{1},~\beta^{e}_0 $) are discarded in this subsection
    \ref{sec:singexpl}.

The matrices $\hat{K}(\mu_1 )$ and $\hat{K}' (\mu_2 )$ encode the
NLO corrections and depend on the number of active flavors through
the beta function and the anomalous dimension matrices.  The matrices
$\hat{J}_{e}$, $\hat{J}_{s}$ and $\hat{J}_{se}$ govern the leading
electromagnetic, next-to-leading strong, and next-to-leading
combined strong-electromagnetic contributions to the RG evolution.

Differentiating Eqs.~\eqref{eq:UNLO} and \eqref{eq:Uint} with respect to $ g_s (\mu_1)$ yields the following  differential equation for $\hat{K}(g_s (\mu_1))$ \cite{Buras:1979yt,Buras:1991jm},
\beq
\frac{\partial}{\partial g_s (\mu_1)} \hat{K}(g_s(\mu_1)) - \frac{1}{g_s (\mu_1) } \hat{K}(g_s(\mu_1)) \frac{\hat{\gamma}_{s}^{(0)T}}{\beta_0} = \frac{\hat{\gamma}^T (g_s (\mu_1))}{\beta (g_s (\mu_1))}  \hat{K}(g_s(\mu_1)).
\label{eq:diffK}
\eeq The traditional ansatz in the literature is to take $\hat{J}_{e}$,
$\hat{J}_{s}$ and $\hat{J}_{se}$ as constant matrices for any fixed
number of flavors. The differential equation ~\eqref{eq:diffK} then
implies the following equations for the matrices
$\hat{J}_{e}$, $\hat{J}_{s}$ and $\hat{J}_{se}$ \cite{Ciuchini:1993vr},
\beq
\hat{J}_{s} - \left[ \hat{J}_{s} , \frac{\hat{\gamma}^{(0)T}_s}{2 \beta_0} \right ] &= \frac{\beta_1}{\beta_0} \frac{\hat{\gamma}^{(0)T}_s}{2 \beta_0} - \frac{\hat{\gamma}^{(1)T}_s}{2 \beta_0}, 
\label{eq:JsRome}\\
\hat{J }_{e} + \left[ \hat{J }_{e} , \frac{\hat{\gamma}^{(0)T}_s}{2 \beta_0} \right ] &= \frac{\hat{\gamma}^{(0)T}_e}{2 \beta_0}, 
\label{eq:JeRome}\\
\left[ \hat{J}_{se} , \frac{\hat{\gamma}^{(0)T}_s}{2 \beta_0} \right ] & = 
\frac{\hat{\gamma}^{(1)T}_{se}}{2 \beta_0} 
+ \left[  \frac{\hat{\gamma}^{(0)T}_e}{2 \beta_0},  \hat{J}_{s} \right]
- \frac{\beta_1}{\beta_0} \frac{\hat{\gamma}^{(0)T}_e}{2 \beta_0}.
\label{eq:JseRome}
\eeq

It is well known, however, that Eqs.~\eqref{eq:JsRome} and \eqref{eq:JeRome} develop singularities in the case of three flavors.
Furthermore, 
Eq.~\eqref{eq:JseRome} is even singular for any number of flavors.

We now show how these singularities arise. For this purpose, it is
instructional to transform Eqs.~\eqref{eq:JsRome}--\eqref{eq:JseRome}
into the diagonal basis of $\hat{\gamma}^{(0)T}_s $. This is a common
procedure in the literature since it allows to isolate the singularities
and remove them ``by hand''. We stress that this is only for the purpose
of a better understanding of the origin of these singularities. 
A numerical evaluation of 
our solution does not require the diagonalisation of
  $\hat{\gamma}^{(0)T}_s $.
 
Upon transforming Eqs.~\eqref{eq:JsRome}--\eqref{eq:JseRome} into the
basis where $\hat{\gamma}^{(0)T}_{s,D} =\hat{V}^{-1}
\hat{\gamma}^{(0)T}_{s} \hat{V}$ is diagonal, the solutions of
Eqs.~\eqref{eq:JsRome} and \eqref{eq:JeRome} take the form
\begin{equation}
\left( \hat{V}^{-1} \hat{J}_{s,e} \hat{V} \right)_{ij}  = \frac{\cdots}{2
\beta_0 \mp \left( (\hat{\gamma}^{(0)T}_{s,D})_{jj} -
(\hat{\gamma}^{(0)T}_{s,D})_{ii} \right) }.
\label{eq:singularityA}
\end{equation}
We find singular solutions if the difference of two eigenvalues of
$\hat{\gamma}^{(0)T}_{s}$ is equal to $2 \beta_0$, which is the 
case for three flavors: $\hat{\gamma}^{(0)T}_{s,D}$ has the
elements
$2$ and $-16$  and $2 \beta_0^{f=3} = 18$, so that one denominator
in Eq.~\eqref{eq:singularityA} vanishes with a generally non-zero numerator.  
When we transform Eq.~\eqref{eq:JseRome} into the same basis
\begin{equation}
\left( \hat{V}^{-1} \hat{J}_{se} \hat{V} \right)_{ij} = \frac{\cdots}{
  (\hat{\gamma}^{(0)T}_{s,D})_{jj} - (\hat{\gamma}^{(0)T}_{s,D})_{ii}},
\label{eq:singularityB}
\end{equation}
we find singular results for $i=j$ and also for degenerate eigenvalues.

Nonetheless, once all relevant terms have been joined together, all
these singularities cancel and the evolution matrix $\hat{U}_f
(\mu_1,\mu_2)$ becomes finite~\cite{Ciuchini:1993vr}. This procedure,
however, requires taking care of each singularity by hand by adopting
the aforementioned diagonal basis, then regularizing the singularities
and keeping track of them until the end of the calculation.  Indeed,
Buras et al.\ have regulated some of the singularities by a logarithmic
term \cite{Buras:1993dy}.  Subsequently, Adams and Lee have proposed a
systematical solution for all singularities \cite{Adams:2007tk}, which,
however, still requires the adoption of a certain diagonal basis.  The
freedom of choosing the order of the eigenvalues on the diagonal of
$\hat{\gamma}^{(0)T}_{s,D}$ involves an ambiguity. This can pose a
problem in computational implementations, since it is absolutely
necessary to use the same diagonal basis as Adams and Lee do, which is
not the one which orders eigenvalues by their numerical value. The
  solution in Ref.~\cite{Huber:2005ig} follows the same line, after
  diagonalizing $\hat{\gamma}^{(0)T}_{s,D}$ several different cases must
  be considered: whenever two eigenvalues differ by an integer multiple
  of $2\beta_0$ a special implementation is required.  In the next
subsection we propose a solution which does not rely on a specific basis
and permits a much faster, easier and, in particular, more stable
computational algorithm.

\subsection{Removing the Singularities}

In order to eliminate the singularities, we generalize the Roma
group's ansatz \cite{Ciuchini:1992tj, Ciuchini:1993vr} by
adding a logarithmic scale dependence to the $\hat{J}$ matrices
used in Eqs.~\eqref{eq:K},\,\eqref{eq:Kprime} in the following way
\begin{align}
\hat{J}_{s} ~\rightarrow ~& \hat{J}_{s}( \alpha_s(\mu)) = \hat{J}_{s,0}
+ \hat{J}_{s,1}\ln \alpha_s(\mu) , \nonumber\\ 
\hat{J}_{e} ~\rightarrow ~& \hat{J }_{e}( \alpha_s(\mu)) = \hat{J}_{e,0} 
+ \hat{J}_{e,1} \ln \alpha_s(\mu) , \nonumber\\ 
\hat{J}_{se} ~\rightarrow ~& \hat{J}_{se}( \alpha_s(\mu)) =
\hat{J}_{se,0} + \hat{J}_{se,1} \ln \alpha_s(\mu)  +
              \hat{J}_{se,2} \ln^2 \alpha_s(\mu).
\label{eq:newJ}
\end{align}
In addition, we extend Eqs.~\eqref{eq:K},\,\eqref{eq:Kprime} as follows:
\begin{align}
  \hat{K}(\mu_1, \mu_2) =& \left(\hat{1} + \frac{\alpha_{EM}}{4 \pi}
    \hat{J}_{se} ( \alpha_s(\mu_1)) \right) \left(\hat{1} +
    \frac{\alpha_s (\mu_1)}{4 \pi} \hat{J}_{s}( \alpha_s(\mu_1)) \right)
  \non &\times \left(\hat{1} + \frac{\alpha_{EM}}{\alpha_s (\mu_1)}
    \hat{J }_{e}( \alpha_s(\mu_1)) \right.
   \non & ~~~~\left. + \left(\frac{\alpha_{EM}}{\alpha_s
        (\mu_1)} \right)^2  \left( \hat{J}_{ee} ( \alpha_s(\mu_1))   - \frac{\beta^e_0}{\beta_0} \left( 1 - \frac{\alpha_s(\mu_1) }{\alpha_s (\mu_2)} \right) \hat{J}_{e}( \alpha_s(\mu_1))   \right)
  \right), \label{eq:ext} \\ %
  \hat{K}^{\prime} (\mu_2) =& \left( \hat{1} -
    \frac{\alpha_{EM}}{\alpha_s (\mu_2)} \hat{J }_{e} ( \alpha_s(\mu_2))
    - \left(\frac{\alpha_{EM}}{\alpha_s (\mu_2)} \right)^2 \left(
      \hat{J}_{ee} ( \alpha_s(\mu_2)) - \left( \hat{J }_{e} (
        \alpha_s(\mu_2))\right)^2 \right) \right) \non & \times \left(
    \hat{1} - \frac{\alpha_s (\mu_2)}{4 \pi} \hat{J}_{s} (
    \alpha_s(\mu_2)) \right) \left(\hat{1} - \frac{\alpha_{EM}}{4 \pi}
    \hat{J}_{se} ( \alpha_s(\mu_2)) \right), \label{eq:ext2}
\end{align}
which somewhat resembles the NNLO {QCD} result of
Ref.~\cite{Gorbahn:2004my}.  Here 
we use the abbreviation $ \alpha_{EM} \equiv \alpha_{EM} (\mu_2)$ and 
\beq 
\hat{J}_{ee}( \alpha_s(\mu)) =
\hat{J}_{ee,0} + \hat{J}_{ee,1} \ln \alpha_s(\mu).
\label{eq:newJee}
\eeq
We systematically include $\mathcal{O}( \alpha_{EM}^2 / \alpha_s^2
)$ corrections in the RG evolution.  This contribution has not been
considered in the literature. Although appearing as
$\mathcal{O}(\alpha_{EM}^2)$, 
these
terms can become sizable at high energies because of the awkward $ 1/
\alpha_s^2 $ dependence, making them numerically comparable to
$\mathcal{O}(\alpha_s)$.   We note
that this contribution does not receive contributions from higher 
orders of the anomalous dimension matrix in Eq.~\eqref{eq:anomalousNLO},
but only appears at the next-to-leading order.

{With these generalizations we can now solve the} differential
equation in Eq.~\eqref{eq:diffK}.  Inserting our ansatz into
Eq.~\eqref{eq:diffK} we obtain the following nine matrix equations for
the nine constant matrices $\hat{J}$:
\begin{align}
\label{eq:Js1full}
\hat{J}_{s,1} - \left[\hat{J}_{s,1},\frac{\hat{\gamma}^{(0)T}_s}{2 \beta_0}\right] & = 0,  \\
\label{eq:Js0full}
\hat{J}_{s,0} - \left[\hat{J}_{s,0},\frac{\hat{\gamma}^{(0)T}_s}{2 \beta_0}\right] & = \frac{\beta_1}{\beta_0}\frac{\hat{\gamma}^{(0)T}_s}{2 \beta_0} - \frac{\hat{\gamma}^{(1)T}_s}{2 \beta_0} - \hat{J}_{s,1}, \\
\label{eq:Je1}
\hat{J}_{e,1} + \left[\hat{J}_{e,1},\frac{\hat{\gamma}^{(0)T}_s}{2 \beta_0}\right] & = 0, \\
\label{eq:Je0}
\hat{J}_{e,0} + \left[\hat{J}_{e,0},\frac{\hat{\gamma}^{(0)T}_s}{2 \beta_0}\right] & = \frac{\hat{\gamma}^{(0)T}_e}{2 \beta_0} + \hat{J}_{e,1},\\ 
\label{eq:Jse2}
\left[\hat{J}_{se,2}, \frac{\hat{\gamma}^{(0)T}_s}{2 \beta_0}\right] & = 0,\\
\label{eq:Jse1}
\left[\hat{J}_{se,1}, \frac{\hat{\gamma}^{(0)T}_s}{2 \beta_0}\right] & = 
\left[ \frac{\hat{\gamma}^{(0)T}_e}{ 2 \beta_0}, \hat{J}_{s,1} \right]
+ 2  \hat{J}_{se,2}, \\
\label{eq:Jse0}
\left[\hat{J}_{se,0},\frac{\hat{\gamma}^{(0)T}_s}{2 \beta_0}\right] & =
\frac{\hat{\gamma}^{(1)T}_{se}}{2 \beta_0}
 +\left[ \frac{\hat{\gamma}^{(0)T}_e}{ 2 \beta_0}, \hat{J}_{s,0} \right]
  - \frac{\beta_1}{\beta_0}  \frac{\hat{\gamma}^{(0)T}_e}{2 \beta_0} 
  - \frac{\beta^{se}_{1}}{\beta_0} \frac{\hat{\gamma}_s^{(0) T}}{ 2 \beta_0}  + \hat{J}_{se,1},
   \\ 
\label{eq:Jee1}
\hat{J}_{ee,1} + \left[\hat{J}_{ee,1}, \frac{\hat{\gamma}^{(0)T}_s}{4 \beta_0}\right] & = 
\frac{\hat{\gamma}^{(0)T}_e}{4 \beta_0} \hat{J}_{e,1} + \frac{1}{2} \frac{\beta^e_0}{\beta_0} \hat{J}_{e,1 },\\
\label{eq:Jee0}
\hat{J}_{ee,0} + \left[\hat{J}_{ee,0}, \frac{\hat{\gamma}^{(0)T}_s}{4
    \beta_0}\right] & = \frac{\hat{\gamma}^{(0)T}_e}{4 \beta_0}
\hat{J}_{e,0} + \frac{1}{2} \frac{\beta^e_0}{\beta_0} \hat{J}_{e,0 }  + \frac{1}{2} \hat{J}_{ee,1}.
\end{align}
These equations yield finite solutions for $\hat{J}$.  As an effect of
the constant matrices $\hat{J}_{s(,e,se),1}$, the analytic singularities
of Eqs.~\eqref{eq:JsRome}--\eqref{eq:JseRome} do not occur, because for
the problematic matrix elements now both sides of the equations are
zero. We stress that one can solve Eqs.~\eqref{eq:Js1full} to
\eqref{eq:Jee0} without diagonalizing $\hat{\gamma}^{(0)T}_s$; these
equations are mere systems of linear equations for the $100$ elements of
$\hat{J}_{s,e,ee,0,1}$ and $\hat{J}_{se,0,1,2}$ each, which are
{quickly} solved by computer algebra programs
\cite{Lenz:1997aa}. However, there are {multiple solutions} in some
of the inhomogeneous equations, {because} the corresponding
homogeneous equations have a non-trivial null space.  As a consequence,
these solutions for $\hat{J}$ {depend on} arbitrary parameters,
e.g. there are 16 undetermined components in the case of three active
flavors.  These parameters, however, do not produce any ambiguity in
physical results.  In the next subsection, we will show that they
completely drop out after combining terms of the same order and the
evolution matrix in Eq.~\eqref{eq:UNLO} does not depend on these
parameters.  Therefore, one can set them to arbitrary values from the
beginning. In our calculation of $\epsilon_{K}'/\epsilon_{K}$ we kept
the parameters arbitrary as a crosscheck of the consistency of our
calculation.

The procedure to determine the evolution matrix from $\mu_2$ to $\mu_1$
requires algebraically solving the matrix equations
\eqref{eq:Js1full}--\eqref{eq:Jee0} for a given number of active
flavors and inserting the solutions into the full evolution matrix in
Eq.~(\ref{eq:UNLO}).  We use $10\times10$ anomalous dimension matrices
$\hat{\gamma}^{(0)}_s$, $\hat{\gamma}^{(0)}_{e}$, $\hat{\gamma}^{(1)}_s$
and $\hat{\gamma}^{(1)}_{se}$
\cite{Buras:1992tc,Buras:1992zv,Ciuchini:1993vr,Buchalla:1995vs}.  The
solutions for the matrices $\hat{J}$ in the case of three active
flavors (with two active leptons) in naive dimensional regularization (NDR) scheme with
$\overline{\textrm{MS}}$ subtraction, are given as follows: 
\beq 
&\hat{J}_{s,0} = \non & \scalebox{0.69}{$ \left(
\begin{array}{cccccccccc}
 -55/324 & 223/108 & 0 & 0 & 0 & 0 & 0 & 0 & 0 & 0 \\
 223/108 & -55/324 & 0 & 0 & 0 & 0 & 0 & 0 & 0 & 0 \\
-0.7392 & -0.3061 & -2.999 & -0.6652 & 1.457 & 0.2171 & 0 & 0 &
   0.3061 & 0.7392 \\
 0.3814 & -0.1853 & 2.838 & 1.037 & -0.05711 & -0.004122 & 0 & 0 &
   0.1853 & -0.3814 \\
 0.3990 & 0.3264 & 1.850 & 1.444 & -2.514 & 2.750 & 0 & 0 &
   -0.3264 & -0.3990 \\
 -1.181 & -1.776 & -7.095 & -6.691 & 0.6263 & 4.528 & 0 & 0 & 1.776
   & 1.181 \\
 0 & 0 & 0 & 0 & 0 & 0 & -679/648 & 67/24 & 0 & 0 \\
 0 & 0 & 0 & 0 & 0 & 0 & t_s & 3749/648 & 0 & 0 \\
 0 & 0 & 0 & 0 & 0 & 0 & 0 & 0 & -55/324 & 223/108 \\
 0 & 0 & 0 & 0 & 0 & 0 & 0 & 0 & 223/108 & -55/324 \\
\end{array}
\right) \label{Js0result},
$}\\[2mm]
&\hat{J}_{s,1} = 
\scalebox{1.0}{$
\left(
\begin{array}{cccccccccc}
 0 & 0 & 0 & 0 & 0 & 0 & 0 & 0 & 0 & 0 \\
 0 & 0 & 0 & 0 & 0 & 0 & 0 & 0 & 0 & 0 \\
 0 & 0 & 0 & 0 & 0 & 0 & 0 & 0 & 0 & 0 \\
 0 & 0 & 0 & 0 & 0 & 0 & 0 & 0 & 0 & 0 \\
 0 & 0 & 0 & 0 & 0 & 0 & 0 & 0 & 0 & 0 \\
 0 & 0 & 0 & 0 & 0 & 0 & 0 & 0 & 0 & 0 \\
 0 & 0 & 0 & 0 & 0 & 0 & 0 & 0 & 0 & 0 \\
 0 & 0 & 0 & 0 & 0 & 0 & -10/27 & 0 & 0 & 0 \\
 0 & 0 & 0 & 0 & 0 & 0 & 0 & 0 & 0 & 0 \\
 0 & 0 & 0 & 0 & 0 & 0 & 0 & 0 & 0 & 0 \\
\end{array}
\right),
$} \label{Js1result}\\[2mm]
&\hat{J}_{e,0} = \non
&
\scalebox{0.65}{$
\left(
\begin{array}{cccccccccc}
 -4/27 & 0 & 0 & 0 & 0 & 0 & 0 & 0 & 0 & 0 \\
 0 & -4/27 & 0 & 0 & 0 & 0 & 0 & 0 & 0 & 0 \\
 -0.001708 & 0.004962 & 0.002631 & 0.009301 & -0.03258 & -0.08924 & 0.0004431 &
0  & -0.07795 & 0.002792 \\
 -0.004694 & -0.001225 & 0.007331 & 0.01080 & -0.02781 & -0.07666 & -0.006646 &
 0 & -0.01071 & -0.08131 \\
 0.0004270 & 0.003537 & -0.001407 & 0.001703 & -0.008641 & -0.02351 & 0.2102 & 2/5 &
   0.001344 & 0.004454 \\
 -0.001829 & -0.004273 & 0.002924 & 0.0004802 & 0.004780 & 0.01280 & -0.04904 & - 8/135 &
   -0.004205 & -0.006649 \\
 2/15 & -2/135 & 2/135 & -2/15 & 59/270 & 19/90 & t_e
   & 3 t_e - 50/81 & 26/135 & 2/45 \\
 -0.02605 & 0.005587 & -0.01083 & 0.02081 & -0.02530 & 0.06671 & - t_e/3 +  38/729
   & - t_e  + 8/27  & -0.03366 & -0.002023 \\
 0.09942 & 0.02428 & -0.1174 & -0.04438 & -0.1994 & -0.5362 & 2/35 & -8/45 & 0.05967
   & 0.05861 \\
 0.02623 & 0.02072 & 0.04112 & -0.1125 & -0.1951 & -0.5158 & -2/35& - 4/15 & 0.01879
   & -0.06080 \\
\end{array}
\right),
$}\label{Je0result}\\[2mm]
&\hat{J}_{e,1} = 
\scalebox{1.0}{$
\left(
\begin{array}{cccccccccc}
 0 & 0 & 0 & 0 & 0 & 0 & 0 & 0 & 0 & 0 \\
 0 & 0 & 0 & 0 & 0 & 0 & 0 & 0 & 0 & 0 \\
 0 & 0 & 0 & 0 & 0 & 0 & 0 & 0 & 0 & 0 \\
 0 & 0 & 0 & 0 & 0 & 0 & 0 & 0 & 0 & 0 \\
 0 & 0 & 0 & 0 & 0 & 0 & 0 & 0 & 0 & 0 \\
 0 & 0 & 0 & 0 & 0 & 0 & 0 & 0 & 0 & 0 \\
 0 & 0 & 0 & 0 & 0 & 0 & - 4/243 & -4/81 & 0 & 0 \\
 0 & 0 & 0 & 0 & 0 & 0 & 4/729 & 4/243 & 0 & 0 \\
 0 & 0 & 0 & 0 & 0 & 0 & 0 & 0 & 0 & 0 \\
 0 & 0 & 0 & 0 & 0 & 0 & 0 & 0 & 0 & 0 \\
\end{array}
\right),
$}\label{Je1result}\\[2mm]
&\hat{J}_{se,0} = \non
&\scalebox{0.46}{$
\left(
\begin{array}{cccccccccc}
 3/8 & 9/8& 0 &0  &
  0  & 0 &
0 &
 0&0  &
  0 \\
 -9/8& -3/8 &0 & 0    &0  &0  &
 0  &
  0  & 0 &
 0 \\
  -26.08 & 20.94 & -25.20 & 22.07 & 4.847 & 8.717 & 16.02 & 0.00499 & -26.20 & 20.63 \\
21.87 & -25.07 & 31.46 & -15.23 & -5.751 & -8.314 & 7.459 & 0.05014 & 16.21 & -30.05 \\
2.409 & 2.535 & -1.122 & -0.9967 & 0.06192 & -0.1911 &   2 t_s/5 + 142.6 & 0.02577 & 4.175 & 4.300 \\
-1.581 & -1.594 & 0.7172 & 0.7036 & 0.1306 & 0.1116 &  -8  t_s/135 -51.94 & -2.417 & -2.729 & -2.743 \\
 -15.68 & -11.02 & -59.91 & -55.25 & -309.3 & 8.235 & 0.08482 & 0.2545 & 7.761 & 11.53 \\
 - 2 t_s/15 +5.611 &2 t_s/135 +  2.955 &  -2 t_s/135  + 19.78 & 2 t_s/15  + 17.12 & -59 t_s/270 +   102.8 &-19  t_s/90  -3.773 &   -28 t_s/243+  0.4857 & -0.08482 &-26
   t_s/135  -1.473 &-2 t_s/45 -4.129 \\
27.12 & -19.23 & 45.81 & -0.03029 & -8.332 & -7.461 & -8  t_s/45  + 1.621 & -0.3044 & 18.81 & -27.48 \\
-21.04 & 26.43 & -13.67 & 34.30 & 2.682 & 10.09 &  -4  t_s/15 + 3.035 & 0.8012 & -26.07 & 21.45 \\
\end{array}
\right)
$}\non
&
\scalebox{1.0}{$+ \hat{V} 
\left(
\begin{array}{cccccccccc}
 t_{se1}  & 0 & 0 & 0 & 0 & 0 & 0 & 0 & 0 & 0 \\
 0 & t_{se2}  & 0 & 0 & 0 & 0 & 0 & 0 & 0 & 0 \\
 0 & 0 & t_{se3}  & t_{se4} & 0 & 0 & 0 & 0 & 0 & 0 \\
 0 & 0 & t_{se5} & t_{se6} & 0 & 0 & 0 & 0 & 0 & 0 \\
 0 & 0 & 0 & 0 & t_{se7} & 0 & 0 & 0 & 0 & 0 \\
 0 & 0 & 0 & 0 & 0 & t_{se8} & 0 & 0 & 0 & 0 \\
 0 & 0 & 0 & 0 & 0 & 0 & t_{se9} & t_{se10} & 0 & 0 \\
 0 & 0 & 0 & 0 & 0 & 0 & t_{se11} & t_{se12} & 0 & 0 \\
 0 & 0 & 0 & 0 & 0 & 0 & 0 & 0 & t_{se13} & 0 \\
 0 & 0 & 0 & 0 & 0 & 0 & 0 & 0 & 0 & t_{se14} \\
\end{array}
\right)
\hat{V}^{-1}
$},\label{Jse0result}\\[2mm]
&\hat{J}_{se,1} = \non
&
\scalebox{0.65}{$
\left(
\begin{array}{cccccccccc}
   -1.485  & -0.2623  & 0 & 0 & 0 & 0 & 0 & 0 &
0 & 0  \\
-0.2623 & -1.485 & 0 & 0 & 0 & 0 &0 & 0 &
  0 &0  \\
 -0.3914 & 0.9178 & -0.5086 & 0.8458 & 0.1026 & 0.1994 & 0 & 0 & -1.075 & 0.8226 \\
0.9599 & -0.2650 & 1.225 & -0.04511 & -0.1655 & -0.1095 & 0 & 0 & 0.6962 & -1.117 \\
-0.002595 & -0.04387 & -0.09552 & -0.1368 & 0.1728 & -0.03447 & -0.1481 & 0 & 0.04387 & 0.002595 \\
0.05517 & 0.000282 & 0.1661 & 0.1112 & -0.2131 & -0.3630 & 0.02195 & 0 & -0.000282 & -0.05517 \\
 0 & 0 & 0 & 0 & 0 & 0 & - 4 t_s/81  + 1.985 & 0 & 0 & 0 \\
  0.04938 & -0.005487 & 0.005487 & -0.04938 & 0.08093 & 0.07819 &  8 t_s/243 -0.9268 &  4 t_s/81 -0.9234 & 0.07133 & 0.01646 \\
 0.8624 & -0.3145 & 0 & 0 & 0 & 0 & 0.06584 & 0 & -0.1909 & -0.7342 \\
 -0.3145 & 0.8624 & 0 & 0 & 0 & 0 & 0.09877 & 0 & -0.7342 & -0.1909 \\
\end{array}
\right),
$}\label{Jse1result}\\[2mm]
&\hat{J}_{se,2} = 
\scalebox{1.}{$
\left(
\begin{array}{cccccccccc}
 0 & 0 & 0 & 0 & 0 & 0 & 0 & 0 & 0 & 0 \\
 0 & 0 & 0 & 0 & 0 & 0 & 0 & 0 & 0 & 0 \\
 0 & 0 & 0 & 0 & 0 & 0 & 0 & 0 & 0 & 0 \\
 0 & 0 & 0 & 0 & 0 & 0 & 0 & 0 & 0 & 0 \\
 0 & 0 & 0 & 0 & 0 & 0 & 0 & 0 & 0 & 0 \\
 0 & 0 & 0 & 0 & 0 & 0 & 0 & 0 & 0 & 0 \\
 0 & 0 & 0 & 0 & 0 & 0 & 20/2187 & 0 & 0 & 0 \\
 0 & 0 & 0 & 0 & 0 & 0 & - 40/6561 & - 20/2187 & 0 & 0 \\
 0 & 0 & 0 & 0 & 0 & 0 & 0 & 0 & 0 & 0 \\
 0 & 0 & 0 & 0 & 0 & 0 & 0 & 0 & 0 & 0 \\
\end{array}
\right),
$}\label{Jse2result}
\eeq
\beq
&\hat{J}_{ee,0} = \non
&
\scalebox{0.51}{$
\left(
\begin{array}{cccccccccc}
  40/729  &0 & 0 & 0& 0& 0 & 0 & 0 & 0&  0\\
0& 40/729  &0&0 &  0&0 &0 & 0 & 0  & 0 \\
 -0.002519 & -0.003958 & 0.000955 & -0.005971 & 0.03395 & 0.09227 &12 t_e/27083 +  0.001188 &  36 t_e/27083+   0.01576 & 0.02318 & -0.002951 \\
0.000504 & -0.001464 & -0.006771 & -0.003253 & 0.03333 & 0.09097 & - 180 t_e/27083
  +   0.01104 &  - 540 t_e/27083  + 0.03144 & 0.004142 & 0.02686 \\
  0.005995 & -0.003625 & 0.002241 & -0.007379 & 0.01478 & 0.02625 & 8026 t_e/104463 -0.1123 &8026 t_e/34821  -0.2909 & 0.007872 & -0.001747 \\
-0.001130 & 0.002451 & -0.001997 & 0.001584 & -0.003039 & -0.002932 & - 9178 t_e/313389 + 0.03477 &  - 9178 t_e/104463 + 0.08429 & -0.000697 & 0.002884 \\
  -0.02801 & 0.01209 & -0.01239 & 0.02771 & -0.09800 & -0.1928 & - 94 t_e/243 + 0.06658 & - 94 t_e/81 +0.2660 & -0.03582 & 0.004286 \\
 0.008577 & -0.003761 & 0.004577 & -0.007761 & 0.01725 & 0.01575 & 110 t_e/729  -0.03402 & 110 t_e/243 -0.1293 & 0.01058 & -0.001761 \\
 -0.02099 & -0.01189 & 0.02183 & -0.01845 & 0.1185 & 0.2984 &22 t_e/189 - 0.005245 &  22 t_e/63 + 0.02511 & 0.01247 & -0.008604 \\
 -0.009687 & -0.01325 & -0.02604 & 0.01978 & 0.1295 & 0.3422 &  2 t_e/63 + 0.03922 & 2 t_e/21 + 0.1510 & -0.001510 & 0.02511 \\
\end{array}
\right),
$}\label{Jee0result}
\\[2mm]
&\hat{J}_{ee,1} = 
\scalebox{1.0}{$
\left(
\begin{array}{cccccccccc}
 0 & 0 & 0 & 0 & 0 & 0 & 0 & 0 & 0 & 0 \\
 0 & 0 & 0 & 0 & 0 & 0 & 0 & 0 & 0 & 0 \\
 0 & 0 & 0 & 0 & 0 & 0 & - 16/2193723 & - 16/731241 & 0 & 0 \\
 0 & 0 & 0 & 0 & 0 & 0 & 80/731241 & 80/243747 & 0 & 0 \\
 0 & 0 & 0 & 0 & 0 & 0 & - 32104/25384509 & - 32104/8461503 & 0 & 0 \\
 0 & 0 & 0 & 0 & 0 & 0 & 36712/76153527 & 36712/25384509 & 0 & 0 \\
 0 & 0 & 0 & 0 & 0 & 0 &  376/59049 & 376/19683 & 0 & 0 \\
 0 & 0 & 0 & 0 & 0 & 0 & - 440/177147 & - 440/59049 & 0 & 0 \\
 0 & 0 & 0 & 0 & 0 & 0 & - 88/45927 & - 88/15309 & 0 & 0 \\
 0 & 0 & 0 & 0 & 0 & 0 & - 8/15309 & - 8/5103 & 0 & 0 \\
\end{array}
\right), 
$}\label{Jee1result} 
\eeq 
where $t_s$, $t_e$, and $t_{se1,2,\dots,14}$ are the arbitrary
parameters of the matrix equations. Our convention for the matrix
$\hat{V}$ is $(\hat{\gamma}_{s,D}^{(0) T})_{1,1} \leq
(\hat{\gamma}_{s,D}^{(0) T})_{2,2} \leq \dots \leq
(\hat{\gamma}_{s,D}^{(0) T})_{10,10}$.  Although Eq.~\eqref{Jse0result}
{makes explicit reference to the} the diagonal basis, the term
{involving $\hat{V}$} completely drops {out from} the evolution matrix
(see next subsection), and thereby our solution for the latter
does not require any matrix diagonalisation. Our
  Eqs.~(\ref{eq:ext})--(\ref{eq:Jee0}) hold in any operator basis.
  Moreover, if an ordinary four-dimensional basis transformation is
  applied to Eqs.~(\ref{eq:q1})--(\ref{eq:q10}), the corresponding RG
  matrices $\hat J_{\ldots}$ can be simply found by transforming those
  in Eqs.~(\ref{Js0result})--(\ref{Jee1result}) in the same way as
  $\hat{\gamma}_s^{(0)T}$. If the basis transformation is
  $D$-dimensional, meaning that it involves evanescent operators, the
  $\hat J_{\ldots}$ matrices undergo an additional scheme transformation
  \cite{Herrlich:1994kh,Gorbahn:2004my}.  We collect the solutions for
more than three active flavors in Appendix~\ref{app:solutionJ}.

Substituting the generalized ansatz of Eqs.~(\ref{eq:ext}),\,(\ref{eq:ext2}) into Eq.~(\ref{eq:UNLO}),  we find the full next-to-leading order evolution matrix,
\begin{align}
  \hat{U}_f (\alpha_{1},\alpha_{2})=&~ \hat{U}_0 \left( \alpha_1,
    \alpha_2 \right) + \frac{\alpha_{1}}{4 \pi} \hat{U}_{QCD} \left(
    \alpha_1, \alpha_2 \right) + \frac{\alpha_{EM}}{\alpha_{1}}
  \hat{U}_{QED} \left( \alpha_1, \alpha_2 \right) \non &+
  \frac{\alpha_{EM}}{4 \pi} \hat{U}_{QCD\textrm{-}QED} \left( \alpha_1,
    \alpha_2 \right) + \left( \frac{\alpha_{EM}}{\alpha_{1}}\right)^2
  \hat{U}_{QED\textrm{-}QED} \left( \alpha_1, \alpha_2 \right) \non & +
  \mathcal{O}\left(\frac{\alpha_{EM}^2}{\alpha_s}, \alpha_s^2, \alpha_s
    \alpha_{EM}, \alpha_{EM}^2\right),
\label{eq:Ufull}
\end{align}
where we use the abbreviation $\alpha_{1,2} \equiv \alpha_s (\mu_{1,2}
)$ for $\mu_1 < \mu_2$  and $\alpha_{EM} \equiv \alpha_{EM} (\mu_2)$  with
\begin{align}
\label{eq:UQCD}
\hat{U}_{QCD}  \left( \alpha_1, \alpha_2 \right)  & = \hat{J}_{s}(\alpha_{1}) \hat{U}_0  \left( \alpha_1, \alpha_2 \right)  - \frac{\alpha_{2}}{\alpha_{1}} \hat{U}_0   \left( \alpha_1, \alpha_2 \right)  \hat{J}_{s}(\alpha_{2}), \\
\label{eq:UQED}
\hat{U}_{QED}  \left( \alpha_1, \alpha_2 \right)  & = \hat{J }_{e}(\alpha_{1}) \hat{U}_0  \left( \alpha_1, \alpha_2 \right)   - \frac{\alpha_{1}}{\alpha_{2}} \hat{U}_0  \left( \alpha_1, \alpha_2 \right)   \hat{J }_{e}(\alpha_{2}), \\
\label{eq:UQCDQED}
\hat{U}_{QCD\textrm{-}QED} \left( \alpha_1, \alpha_2 \right)  & = \hat{J}_{se}(\alpha_{1}) \hat{U}_0 \left( \alpha_1, \alpha_2 \right)   - \hat{U}_0 \left( \alpha_1, \alpha_2 \right)  \hat{J}_{se}(\alpha_{2}) \nonumber\\ & ~~~ + \hat{J}_{s}(\alpha_{1}) \hat{U}_{QED} \left( \alpha_1, \alpha_2 \right)  - \frac{\alpha_{2}}{\alpha_{1}} \hat{U}_{QED} \left( \alpha_1, \alpha_2 \right)  \hat{J}_{s}(\alpha_{2}),\\
\label{eq:UQEDQED}
\hat{U}_{QED\textrm{-}QED} \left( \alpha_1, \alpha_2 \right)  &=
\hat{J}_{ee} \left( \alpha_1 \right)  \hat{U}_0 \left( \alpha_1, \alpha_2 \right) - \frac{\alpha_1}{\alpha_2}
\hat{U}_{QED} \left( \alpha_1, \alpha_2 \right) \hat{J}_{e} \left( \alpha_2 \right) \non
& ~~~
- \left( \frac{\alpha_1}{\alpha_2} \right)^2  \hat{U}_0 \left( \alpha_1, \alpha_2 \right)  \hat{J}_{ee} \left( \alpha_2 \right)  - \frac{\beta^e_0}{\beta_0} \left( 1 - \frac{\alpha_1} {\alpha_2} \right)    \hat{J }_{e}(\alpha_{1}) \hat{U}_0  \left( \alpha_1, \alpha_2 \right).
\end{align}

\subsection{Cancellation of {Spurious}  Parameters}

We now present some details of the cancellation of the arbitrary parameters. 
First, we take a look at the $\mathcal{O}(\alpha_s)$ part of the evolution matrix in Eq.~\eqref{eq:Ufull},
\begin{align}
\frac{\alpha_1}{4 \pi}  \hat{U}_{QCD} \left( \alpha_1, \alpha_2 \right)  =&~ \frac{\alpha_1}{4 \pi}  \hat{J}_{s,0}  \hat{U}_0 (\alpha_1,\alpha_2) - \frac{\alpha_2}{4 \pi} \hat{U}_0 (\alpha_1,\alpha_2) \hat{J}_{s,0} \nonumber\\
& + \frac{\alpha_1 \ln \alpha_1}{4 \pi}  \hat{J}_{s,1} \hat{U}_0 (\alpha_1,\alpha_2) - \frac{\alpha_2 \ln \alpha_2 }{4 \pi} \hat{U}_0 (\alpha_1,\alpha_2) \hat{J}_{s,1}.
\label{eq:UQCDdetail}
\end{align}

In the three-flavor regime, the matrix $\hat{J}_{s,0}$ in
Eq.~\eqref{Js0result} contains an undetermined component $t_s$.
Since the first and second term of $\hat{U}_{QCD} $ in
Eq.~\eqref{eq:UQCDdetail} depend on different scales, one naively could
argue that the cancellation of any dependence has to take place for
each term independently of the other.  However, we will show that this is
not the case.

We locate the undetermined parameter in $[ \hat{J}_{s,0} ]_{8,7} = t_s$.
The matrix product $\hat{J}_{s,0} \hat{U}_0 (\alpha_1,\alpha_2)$
naturally contains a dependence on $t_s$ in the 8th row.  Actually, this
dependence does cancel for all elements except for $[ \hat{J}_{s,0} U_0
  (\alpha_1,\alpha_2) ]_{8,7} \supset ( \alpha_2 / \alpha_1 )^{1/9}
t_s$.  The matrix product $\hat{U}_0 (\alpha_1,\alpha_2) \hat{J}_{s,0}$
in the second term of $\hat{U}_{QCD} $ naturally obtains the parameter
$t_s$ in the 7th column, and again the product consistently cancels this
dependence for all entries except for $[ \hat{U}_0 (\alpha_1,\alpha_2)
  \hat{J}_{s,0} ]_{8,7} \supset ( \alpha_2 / \alpha_1 )^{-8/9} t_s$.
The full cancellations is thus only achieved by taking both terms of the
first line of Eq.~\eqref{eq:UQCDdetail} into account and takes the form
\begin{align}
\left[\frac{\alpha_1}{4 \pi} \hat{U}_{QCD}  \left( \alpha_1, \alpha_2 \right)  \right]_{8,7} & \supset \left[\frac{\alpha_1}{4 \pi}  \hat{J}_{s,0} \hat{U}_0 (\alpha_1,\alpha_2) - \frac{\alpha_2}{4 \pi} \hat{U}_0 (\alpha_1,\alpha_2) \hat{J}_{s,0}\right]_{8,7} \nonumber\\
& \supset \frac{1}{4 \pi} \left(\alpha_1 \left(\frac{\alpha_2}{\alpha_1}\right)^{\frac{1}{9}} - \alpha_2 \left(\frac{\alpha_2}{\alpha_1}\right)^{- \frac{8}{9}} \right) t_s \non
& = 0.
\label{eq:cancel}
\end{align}

The reason that causes the singularity to arise - eigenvalues of $\hat{\gamma}^{(0)T}_{s}$ differing by $2 \beta_0$ in Eq.~\eqref{eq:singularityA} - is also responsible for the cancellation of the undetermined parameter between the high and low scales. 
 The difference of two eigenvalues of $\hat{\gamma}^{(0)T}_{s}$ by $2 \beta_0$ causes a difference of 1 in the exponents of $( \alpha_2 / \alpha_1 )$ and  indeed
the spectrum of $\hat{\gamma}^{(0)T}_{s} / 2 \beta_0$ contains both $1/9$ and $-8/9$ as eigenvalues. 
Thus,  this difference allows the prefactors $\alpha_1$ and $\alpha_2$ of the first two terms in Eq.~\eqref{eq:UQCDdetail} to exactly cancel these terms between the different scales and entirely independent on the actual size of the scales.

Next, we focus on the arbitrary parameter $t_e$ which appears in
the matrix $\hat{J}_{e,0}$ in Eq.~\eqref{Je0result} in the three flavor
regime and must cancel in the $\hat{U}_{QED}$ part of the evolution
matrix.  Let us denote the $t_e$-dependent piece of
$\hat{J}_{e,0}$ with $\hat{t}_e$, where $[\hat{t}_e]_{7,7}=t_e$,
$[\hat{t}_e]_{7,8}=3 t_e$, $[\hat{t}_e]_{8,7}=- t_e/3$,
$[\hat{t}_e]_{8,8}=- t_e$, and the other components are zero.  Using the
matrix $\hat{V}$ it can be written as $\hat{t}_e = \hat{V} \hat{t}_e^\prime
\hat{V}^{-1}$, where $[\hat{t}_e^\prime]_{10,1} = - t_e $ and the other
components are zero.  Then, in the evolution matrix, the $t_e$
dependence takes the following form: 
\beq 
\frac{\alpha_{EM}}{\alpha_1} \hat{U}_{QED} \left( \alpha_1, \alpha_2
\right) & \supset \alpha_{EM} \left( \frac{1}{\alpha_1} \hat{J}_{e,0}
\hat{U}_0 \left( \alpha_1, \alpha_2 \right) - \frac{1}{\alpha_2}
\hat{U}_0 \left( \alpha_1, \alpha_2 \right) \hat{J}_{e,0} \right)\non &
\supset \alpha_{EM} \hat{V} \left( \frac{1}{\alpha_1} \hat{t}_e^\prime
\hat{U}_{0,D} \left( \alpha_1, \alpha_2 \right) - \frac{1}{\alpha_2}
\hat{U}_{0,D} \left( \alpha_1, \alpha_2 \right) \hat{t}_e^\prime \right)
\hat{V}^{-1},
\label{eq:tecancel}
 \eeq
where $\hat{U}_{0,D}  \left( \alpha_1, \alpha_2 \right) $ is defined as
\beq
\hat{U}_ 0 \left( \alpha_1, \alpha_2 \right) =& \hat{V} \textrm{diag}
\left(\left(\frac{\alpha_2}{\alpha_1}\right)^{\frac{\left(\hat{\gamma}_{s,D}^{(0)T}\right)_{1,1}}{2
    \beta_0}},
\left(\frac{\alpha_2}{\alpha_1}\right)^{\frac{\left(\hat{\gamma}_{s,D}^{(0)T}\right)_{2,2}}{2
    \beta_0}},\dots,
\left(\frac{\alpha_2}{\alpha_1}\right)^{\frac{\left(\hat{\gamma}_{s,D}^{(0)T}\right)_{10,10}}{2
    \beta_0}} \right) \hat{V}^{-1}\\ \equiv& \hat{V} \hat{U}_{0,D}
\left( \alpha_1, \alpha_2 \right) \hat{V}^{-1}.
\eeq
All components except for $(10,1)$ of the parenthesis in
Eq.~\eqref{eq:tecancel} are zero trivially.  The cancellation of the
$(10,1)$ component then proceeds in the same way as in the QCD case:
\beq
\left[ \frac{1}{\alpha_1} \hat{t}_e^\prime \hat{U}_{0,D}  \left( \alpha_1, \alpha_2 \right) - \frac{1}{\alpha_2}  \hat{U}_{0,D}  \left( \alpha_1, \alpha_2 \right) \hat{t}_e^\prime \right]_{10,1}  
 & = \left(  \frac{1}{\alpha_1} \left( \frac{\alpha_2}{\alpha_1} \right)^{-\frac{8}{9}} - \frac{1}{\alpha_2} \left( \frac{\alpha_2}{\alpha_1}\right)^{\frac{1}{9}} \right) \cdot \left( - t_e \right)\non
 & = 0.
\eeq
Therefore, the $t_e$ dependence of $\hat{U}_{QED}$ vanishes.

The cancellation of the parameters
$t_{se1,2,\dots,14}$ in the second matrix
product of Eq.~\eqref{Jse0result} is more trivial.
Let us define the second matrix product as $\hat{V} \hat{t}_{se}
\hat{V}^{-1}$.  In the evolution matrix, the matrix $\hat{t}_{se}$ 
appears only in the $ \hat{U}_{QCD\textrm{-}QED}$ part and the
cancellation can be understood in the following way: 
\beq
\frac{\alpha_{EM}}{ 4 \pi } \hat{U}_{QCD\textrm{-}QED} \left( \alpha_1,
\alpha_2 \right) &\supset \frac{\alpha_{EM}}{4 \pi} \left(
\hat{J}_{se,0} \hat{U}_0 (\alpha_1,\alpha_2) - \hat{U}_0
(\alpha_1,\alpha_2) \hat{J}_{se,0} \right)\\ & \supset
\frac{\alpha_{EM}}{4 \pi} \hat{V}\left[ \hat{t}_{se}, ~\hat{U}_{0,D}
  \left( \alpha_1, \alpha_2 \right) \right] \hat{V}^{-1}\non & = 0,
\eeq
where we use the fact that $(\hat{\gamma}_{s,D}^{(0) T})_{3,3} =
(\hat{\gamma}_{s,D}^{(0) T})_{4,4}$ and $(\hat{\gamma}_{s,D}^{(0)
  T})_{7,7} = (\hat{\gamma}_{s,D}^{(0) T})_{8,8}$ are pairwise
degenerate eigenvalues for any number of active flavors.

On the contrary, 
the cancellation of $t_s$ arising in $ \hat{U}_{QCD\textrm{-}QED}$ and $t_e$ in $ \hat{U}_{QED\textrm{-}QED}$  is highly non-trivial. 
The $t_s$ dependence, for example, resides in $\hat{J}_{s,0}$, $\hat{J}_{se,0}$ and $\hat{J}_{se,1}$ which appear in the matrix $ \hat{U}_{QCD\textrm{-}QED}$. Logarithmic $\alpha_s$ terms are accompanied by $\hat{J}_{se,1}$ and by the matrix products $\hat{J}_{s} \hat{U}_{ QED} $ and $\hat{U}_{ QED} \hat{J}_{s}  $.
Although we do not give an analytic explanation for these cancellations in this paper, 
we have checked that taking the sum of all terms in Eqs.~\eqref{eq:UQCDQED} and \eqref{eq:UQEDQED} eliminates any $t_s$ and $t_e$ dependence of $ \hat{U}_{QCD\textrm{-}QED}$ and $ \hat{U}_{QED\textrm{-}QED}$.

Now we have shown that the evolution matrix in Eq.~\eqref{eq:Ufull}
is independent of the undetermined parameters, so that we can set
them to arbitrary values from the beginning.  These parameters are
directly related to the singular components in
Eqs.~\eqref{eq:singularityA},\,\eqref{eq:singularityB} of the
standard solution in the literature.  Therefore, our method
automatically regularizes all singularities and these parameters
correspond to the choices of the finite pieces of the regulated
  expressions, which can therefore be viewed as scheme parameters.

We have also found that the cancellation of the parameters occurs
between the high and low scales.  This insight is especially important
when considering new physics at a high scale. The Wilson coefficients
for a given model are typically calculated at leading order only.
In the evolution to the scale of 1$\,$\GeV\ appropriate for Kaon
  physics one then usually neglects the corrections to
  $\hat K^\prime$ in Eq.~\eqref{eq:UNLO} justified by the smallness of
  $\alpha_s(\mu_2)$ compared to $\alpha_s(\mu_1)$. In the typical
  applications in flavor physics, which do not involve corrections of
  order $\alpha_{EM}$, this procedure is scheme-independent.  We here
show that such a treatment is inconsistent in view of the cancellation
of the singularity regulating scheme parameters.

This inconsistency does not appear in the QCD and QED parts
 which are nonsingular at $f=4,5,6$.
However the combined QCD-QED part, in which singularities persist
for all numbers of flavors, will yield results depending on unphysical
arbitrary scheme parameters if parts of the evolution matrix are
discarded in the described way. Instead, the pieces of $\hat
  K^\prime$ which depend on the scheme parameters $t_{se}$ must be
  consistently retained.

\subsection{Validation of the Logarithmic Contribution}

Finally, let us comment on the logarithmic contributions $\hat{J}_{s,1}$ and $\hat{J}_{e,1}$.
At the $\mathcal{O} (\alpha_s)$ part, we have the following logarithmic contributions to the evolution matrix,
\beq
\hat{U}_f \left( \alpha_1, \alpha_2 \right) & \supset \frac{\alpha_1}{4 \pi} \hat{U}_{ QCD} \left( \alpha_1, \alpha_2 \right) \non
& \supset \frac{1}{4 \pi} \left(  \alpha_1 \ln \alpha_1 \hat{J}_{s,1} \hat{U}_0  \left( \alpha_1, \alpha_2 \right) - \alpha_2 \ln \alpha_2 \hat{U}_0  \left( \alpha_1, \alpha_2 \right) \hat{J}_{s,1}  \right)
\label{eq:logalphas}\\
& =  \frac{\alpha_1}{4 \pi}  \left(\frac{\alpha_2}{\alpha_1}\right)^{\frac{1}{9} } \ln \frac{\alpha_1}{\alpha_2} \hat{J}_{s,1}.
\label{eq:logalphas2}
\eeq
In the $\hat{J}_{s,1}$ matrix, the only nonzero component is $[
\hat{J}_{s,1}]_{8,7} = - 10/27$.  Using a calculation parallel to the
one in the previous subsection, we find that the only nonzero component
in the matrix product $\hat{J}_{s,1}\hat{U}_0 \left( \alpha_1, \alpha_2
\right) $ is $ [\hat{J}_{s,1}\hat{U}_0 \left( \alpha_1, \alpha_2
\right)]_{8,7} = ( \alpha_2 / \alpha_1 )^{1/9} \cdot (- 10/27)$, and
similarly $ [\hat{U}_0 \left( \alpha_1, \alpha_2 \right)
\hat{J}_{s,1}]_{8,7} = ( \alpha_2 / \alpha_1 )^{- 8/9} \cdot (- 10/27)$.
Then, the $(8,7)$ component in the parenthesis in
Eq.~\eqref{eq:logalphas} becomes $- (10/27) \alpha_1 (\alpha_2 /
\alpha_1)^{1/9 } \ln (\alpha_1 / \alpha_2)$, and we arrive at
Eq.~\eqref{eq:logalphas2}.  We find that this result is consistent with
Eq.\,(40) of Ref.~\cite{Adams:2007tk}, where, in order to regulate the
singularity, a small regulator $\epsilon$ is introduced in the
eigenvalues of $\hat{\gamma}_{s}^{(0)T}$.

{With a similar calculation for} the $\mathcal{O}
(\alpha_{EM}/\alpha_s)$ part we obtain the following term,
\beq 
\hat{U}_f \left( \alpha_1, \alpha_2 \right) & \supset
\frac{\alpha_{EM}}{\alpha_1 } \hat{U}_{ QED} \left( \alpha_1, \alpha_2
\right)\non & \supset \alpha_{ EM} \left( \frac{1}{\alpha_1} \ln
  \alpha_1 \hat{J}_{e,1} \hat{U}_0 \left( \alpha_1, \alpha_2 \right) -
  \frac{1}{\alpha_2} \ln \alpha_2 \hat{U}_0 \left( \alpha_1, \alpha_2
  \right) \hat{J}_{e,1} \right)\non & = \alpha_{EM}
\left(\frac{1}{\alpha_1} \ln \alpha_1 \left( \frac{\alpha_2}{\alpha_1}
  \right)^{- \frac{8}{9}} - \frac{1}{\alpha_2} \ln \alpha_2 \left(
    \frac{\alpha_2}{\alpha_1} \right)^{\frac{1}{9}} \right)
\hat{J}_{e,1} \non & = \frac{\alpha_{EM}}{\alpha_1} \left(
  \frac{\alpha_2}{\alpha_1} \right)^{- \frac{8}{9}} \ln
\frac{\alpha_1}{\alpha_2} \hat{J}_{e,1}.
\eeq
This logarithmic contribution is also consistent with Eq.~(2.28) of
Ref.~\cite{Buras:1993dy}. 

\subsection{Higher orders in  $\alpha_{EM}$ and comparison with
  Ref.~\cite{Huber:2005ig}} 
  
The RG evolution in the pioneering
  papers~\cite{Flynn:1989iu,Buchalla:1989we,Buras:1993dy} discards all
  terms which are quadratic or higher-order in $\alpha_{EM}$. Our solution in
  Eq.~\eqref{eq:Ufull} is correct to order $\alpha_{EM}^2/\alpha_s^2$, but
  neglects terms of order $\alpha_{EM}^2/\alpha_s$ and higher. The extra
  term is numerically unimportant for the SM analysis, but matters in
  studies of new-physics contributions generated at very high scales,
  where $\alpha_s$ is small. We come back to this point in
  Sec.~\ref{sec:co}. The RG evolution derived in
  Ref.~\cite{Huber:2005ig} considers terms quadratic in
  $\alpha_{EM}$, including terms of order
  $\alpha_{EM}^2/\alpha_s$ which we neglect. In particular, the $\mu$
  dependence of $\alpha_{EM}$ affects the RG evolution at order
  $\alpha_{EM}^2/\alpha^2_s$ and is  therefore also included in
  Ref.~\cite{Huber:2005ig}. 
  While Ref.~\cite{Huber:2005ig} addresses
  $B$ decays, the derived formulae equally apply to $\epsilon_K^\prime$
  and were used in Ref.~\cite{Buras:2015yba}.  We argue that the
  inclusion of $\alpha_{EM}^2/\alpha_s$ {terms} in the RGE does not improve the
  prediction of $\epsilon_K^\prime/\epsilon_K$, because other terms of
  the same order are not included in the standard NLO solution: For
  instance, at this order the two-loop pure QED anomalous dimension
  matrix $\hat{\gamma}_{e}^{(1)}$ must be added to $\hat{\gamma}
  \left(g_s (\mu) \right)$ in Eq.~\eqref{eq:anomalousNLO}.

{Another issue are the $\Delta I=1/2$ operators %
\beq %
Q_{11} = \left(
  \bar{s}_{\alpha} d_{\alpha} \right)_{V-A} \, \left(
  \bar{b}_{\beta} b_{\beta} \right)_{V-A}, \qquad \qquad
Q_{12} = \left(
  \bar{s}_{\alpha} d_{\beta} \right)_{V-A} \, \left(
  \bar{b}_{\beta} b_{\alpha} \right)_{V-A}
\eeq %
which are generated by electroweak box diagrams, so that their Wilson 
coefficients are of order $\alpha_{EM}$. In agreement with 
Ref.~\cite{Buchalla:1989we} we find a small impact of these operators,
{contributing ($-0.07 \times 10^{-4}$) to $\epsilon_K^\prime/\epsilon_K$}.
 Furthermore, this contribution dominantly comes from $A_2$ which is entered 
by $Q_{11,12}$ through RG mixing  triggered by $\hat{\gamma}_{e}^{(0)}$
and is thus ${\cal O} (\alpha_{EM}^2/\alpha_s)$ and to be discarded.
While the contribution of $Q_{11,12}$ to $A_0$ is formally part of
the NLO solution for $\epsilon_K^\prime/\epsilon_K$, it is numerically
completely negligible (contributing $-0.01 \times 10^{-4}$).}  

We close this section by comparing our solution of the RG equations
in Eqs.~(\ref{eq:newJ})--(\ref{eq:Jee0}) to the one in 
Ref.~\cite{Huber:2005ig}. Actually, the latter also regulates all the singularities {by} logarithmic terms, and uses the diagonalisation 
of $\hat{\gamma}_{s}^{(0)}$ as described before Eq.~\eqref{eq:singularityA}.
The matrices $\hat J_{\ldots}$ transform into 
$\hat J_{\ldots,D}\equiv\hat{V}^{-1}  \hat J_{\ldots}
\hat{V}$ when passing to the diagonal basis.
    Therefore  Eqs.~(\ref{eq:Js1full})--(\ref{eq:Jee0})
also hold with the replacements $\hat{\gamma}_{s}^{(0)}
\to \hat{\gamma}_{s,D}^{(0)}$ and $\hat J_{\ldots}
\to \hat J_{\ldots,D}$. In this form one can 
most easily compare our result with Eq.~(47) of Ref.~\cite{Huber:2005ig}.
  The $\hat{U}_0$, $\hat{U}_{QCD}$, $\hat{U}_{QED}$,
  $\hat{U}_{QCD-QED}$, and $\hat{U}_{QED-QED}$ correspond to
  $\mathcal{O}(\omega^0 \lambda^0)$, $\mathcal{O}( \omega)$,
  $\mathcal{O}(\lambda)$, $\mathcal{O}(\omega \lambda)$, and
  $\mathcal{O}( \lambda^2)$ terms in Ref.~\cite{Huber:2005ig},
  respectively.  We have checked that our  formulae of
  the RG evolution matrices are numerically equivalent to those in
  Ref.~\cite{Huber:2005ig}. We find that our
    solution is easier to implement and leads to a faster numerical
    evaluation.

\section{$\boldsymbol{\epsilon_{K}^{\prime}/\epsilon_{K}}$ in the
  Standard Model at Next-to-Leading Order}
\label{sec:SM}

In this section, we evaluate $\epsilon_{K}'/\epsilon_{K}$ in the
Standard Model at next-to-leading order, using the evolution matrix
derived in the previous section.

We calculate the Wilson coefficients $v_i$ and $z_i$ in
Eqs.~\eqref{eq:wilsonv} and \eqref{eq:wilsonz} with {the methodology
  of} Ref.\,\cite{Buras:1993dy}.  Throughout this paper, the
$\overline{\textrm{MS}}$--NDR regularization scheme is used.  For the
next-to-leading order RG evolution of the Wilson coefficients, we use
the {singularity-free} evolution matrix in Eq.~\eqref{eq:Ufull} and
systematically discard higher-order contributions.
Table~\ref{tab:wilson} shows our result of the Wilson coefficients at
$\mu = 1.3~\GeV$, where $y_i \equiv v_i - z_i$.  We decompose $y_i$ into
the {LO} contribution $\mathcal{O}(1)$ and the four
$\mathcal{O}(\alpha_{EM}/\alpha_s,\,\alpha_s,\,\alpha_{EM},\alpha^2_{EM}/\alpha^2_s)$
{NLO} {terms}, where $\mathcal{O}(1)$ {refers to} tree-level
$W$-boson exchange {combined with the one-gluon anomalous dimension
  matrix $\hat{\gamma}_{s}^{(0)}$ in the RG evolution.}  Here we take $
\alpha_s (M_Z) = 0.1185$, $\alpha_{EM}(M_W) = 1/128$, $m_t = 163.3~\GeV$,
$m_b = 4.18~\GeV$, and $\mu_c = 1.4 \; \textrm{GeV}$, which is the
threshold {scale} between three and four flavor effective theories
in Eqs.~\eqref{eq:wilsonv} and \eqref{eq:wilsonz}.  Note that we
include $\ln (m_c^2/\mu_c^2) $ contributions {in} the charm quark
threshold correction $z_i (\mu_c)$ in Eq.~\eqref{eq:wilsonz}, where we
use $m_c= 1.275~\GeV$ \cite{Agashe:2014kda}.  {To calculate} $\alpha_s
(\mu)$ we use \texttt{RunDec:v1.0} with two-loop accuracy
\cite{Chetyrkin:2000yt}.
%
\begin{table}[t]
\begin{center}
  \caption{Wilson coefficients at $\mu = 1.3\,\textrm{GeV}$, where the
    7--10th components are divided by $\alpha_{EM}(M_W)$. $y_i$ {is}
    decomposed into the {LO} contribution and the {individual}
    {NLO corrections}.}
\label{tab:wilson}
\small{
\begin{tabular}{l | c |cccccc}
\hline
\hline 
$i $ & $ z_i \left(\mu \right)$ & $y_i \left(\mu \right)$ & $ \mathcal{O}(1)$&$ \mathcal{O}(\alpha_{ EM}/\alpha_s)$&$\mathcal{O}(\alpha_s)$&$\mathcal{O}(\alpha_{EM})$& $\mathcal{O}(\alpha^2_{ EM}/\alpha^2_s)$\\
\hline
$1$ & $-0.3903$ & $ 0$& $ 0$& $ 0$& $ 0$& $ 0$ &$0$\\
$2$ & $1.200$ & $  0$& $ 0$& $ 0$& $ 0$& $ 0$ &$0$\\
$3$ & $0.0044$ & $  0.0275$ &  $0.0254$ & $0.0001$& $0.0007$&$0.0012$ & $0$\\
$4$ & $-0.0131 $ & $ -0.0566$& $-0.0485$ & $ {-0.0002}$ & $-0.0069$ &$-0.0009$&$0$\\
$5$ & $0.0039$ & $ 0.0068$ & $0.0124$ & $ 0.0001$ & $-0.0059$ & $0.0001$&$0$\\
$6$ & $-0.0128 $ & $  -0.0847$ & $-0.0736$& $-0.0003$ & $-0.0099$ &$-0.0008$&$0$\\
$7/\alpha_{EM}$ & $ {0.0040}$ & $   {-0.0321}$ & $ 0$ & $ {-0.1116}$& $ 0$ & $ {0.0760}$&$ {0.0035}$\\
$8/\alpha_{EM}$ & $ {0.0019} $ & $   {0.1148}$& $ 0$ & $ {-0.0227} $& $ 0$& $ {0.1366}$&$ {0.0009}$\\
$9/\alpha_{EM}$ & $ {0.0051}$ & $  { -1.3815}$& $ 0$& $ {-0.1267}$& $ 0$& $ {-1.2581}$&$ {0.0034}$\\
$10/\alpha_{EM}$ & $-0.0013 $ & $   {0.4883}$& $ 0$ & $ {0.0217}$& $ 0$&$ {0.4672}$&$ {-0.0006}$\\
\hline \hline
\end{tabular}
}
\end{center}
\end{table}

Next {we take} the hadronic matrix elements from {a recent lattice
  QCD calculation} \cite{Blum:2011ng,Blum:2012uk, Blum:2015ywa,
  Bai:2015nea}, {using the} real parts ($CP$-conserving parts) of the
isospin amplitudes $A_{I=0,2} = \langle (\pi \pi)_{I = 0,2} \left|
  \mathcal{H}^{|\Delta S| =1}_{\textrm{eff}} \right| K^0 \rangle $
{as additional constraints \cite{Buras:1993dy}. These amplitudes}
have been measured very precisely \cite{Blum:2015ywa},
\beq
\textrm{Re}A_0 &=  \left( 3.3201\pm 0.0018 \right)\times 10^{-7}~\GeV, 
\label{eq:ReA0}\\
\textrm{Re}A_2 &= \left(1.4787 \pm 0.0031\right) \times 10^{-8}~\GeV.
\eeq 
Since the real parts are dominated by Standard-Model tree-level
{coefficients} $z_2$ (see Table~\ref{tab:wilson}), they can be used
to fix one of the hadronic matrix elements $\langle (\pi \pi)_{I} \left|
  Q_i \left( \mu \right) \right| K^0 \rangle$ $\equiv \langle Q_i \left(
  \mu \right) \rangle_I $. $\langle Q_2 \rangle_0 $ {dominates the
  real part of $A_0$, but contributes} to the imaginary part only
through the operator Fierz relations\footnote{The {Fierz} relation
  for $Q_4$ is modified by ${\cal O}(\alpha_s/ 4 \pi)$ corrections
  \cite{Buras:1993dy}, {but} these contributions are numerically
  small \cite{Buras:2015yba}.  } \beq Q_4 = - Q_1 + Q_2 + Q_3,
~~~~~~Q_{10} = Q_2 + \frac{1}{2} \left( Q_1 - Q_3 \right).
\label{eq:Fierz}
\eeq $\langle Q_1 \rangle_0 $ is the second largest contribution and the
remaining matrix elements are almost negligible.  The situation is {more
  handy} in the case of $A_2$, where the real part is parameterized
entirely by $ \langle Q_2 \rangle_2$ due to the fact that $ \langle Q_1
\rangle_2 = \langle Q_2 \rangle_2$ in pure QCD \cite{Buras:1993dy,
  Buchalla:1989we}.  In our analysis we derive values of $\langle Q_2
\rangle_0 $ and $\langle Q_2 \rangle_2$  {at the scale $\mu$} from the
experimental measurements of $\textrm{Re}A_0$ and $\textrm{Re}A_2$,
respectively\footnote{ On the other hand, once one introduces the
    ratio \beq q = \frac{z_{+}(\mu) \left( \langle Q_2 \rangle_0 +
        \langle Q_1 \rangle_0 \right)}{z_{-}(\mu) \left( \langle Q_2
        \rangle_0 - \langle Q_1 \rangle_0 \right)}~~~~~\textrm{with~}
    z_{\pm} (\mu) = z_{2}(\mu) \pm z_{1}(\mu), \eeq one can calculate
    Im$A_{I}/$Re$A_{I}$ without using the fit of $\langle Q_2
    \rangle_{I}$ to the data.  Ref.~\cite{Buras:2015yba} {uses} this
    strategy with  {the} parameter range $0 \leq q \leq 0.1$.
    Basically, the difference with our method  {(corresponding} to the
    $q$-dependent terms in Ref.~\cite{Buras:2015yba})  {only affects}
    {numerically} subleading contributions (the $i=3,\,4,\,9,\,10$
    components of Im$A_{0}/$Re$A_{0}$). In either method
    the hadronic uncertainties are reduced compared to the choice to 
    take all matrix elements from lattice.  }.

The decay amplitude of $K \to (\pi \pi)_{I=0}$ has been computed using a
$2+1$ flavor lattice QCD simulation at the renormalization scale $\mu
=1.531$ GeV \cite{Bai:2015nea}.  In order to combine these matrix
elements with the Wilson coefficients evaluated in {the}
three-flavor regime ---that is, {at} a scale below the charm quark
mass--- we need to evolve the hadronic matrix elements {down to a}
scale below $ \mu_c$.  The isospin amplitude is given as \beq A_I & =
\frac{G_F}{ \sqrt{2}} \lambda_u \langle \vec{Q} ( \mu_1)^T \rangle_I
\vec{C} (\mu_1)\non & = \frac{G_F}{ \sqrt{2}} \lambda_u \langle \vec{Q}
( \mu_1)^T \rangle_I \hat{U}_3 \left( \mu_1,\mu_2 \right)\vec{C}
(\mu_2)\non & = \frac{G_F}{ \sqrt{2}} \lambda_u \langle \vec{Q} (
\mu_2)^T \rangle_I \vec{C} (\mu_2), \eeq where $\mu_1 < \mu_2$ and $C_i
(\mu) \equiv z_i (\mu)+ \tau y_i (\mu) $. In the final line, we use the
fact that the physical amplitude $A_I$ is independent of the
renormalization scale, {so that}
\beq 
\langle \vec{Q} ( \mu_1)^T \rangle_I = \langle \vec{Q} ( \mu_2)^T
\rangle_I \left( \hat{U}_3 \left( \mu_1,\mu_2 \right) \right)^{-1}.
\label{eq:RGHME}
\eeq

In practice, we first evaluate the hadronic matrix elements for the
$I=0$ states at $\mu = 1.3$ GeV from the lattice results
\cite{Bai:2015nea} using a three flavor evolution matrix,
cf. Eq.~\eqref{eq:RGHME}.   {Here we use $\alpha_s^{(3)} (1.531\,\GeV) = 0.353388
$  {as in the lattice calculation of} Ref.~\cite{Bai:2015nea}.}
Then we determine $\langle Q_{2}
(\mu)\rangle_0 $ (and $\langle Q_{4,10} (\mu)\rangle_0 $ through
Eq.~\eqref{eq:Fierz}) from the experimental value of Re$A_0$ using the
Wilson coefficients shown in Table \ref{tab:wilson}. We have taken the
CKM parameters from CKMfitter \cite{Charles:2015gya}.  The results are
shown in Table \ref{tab:bfactors}(a).

\begin{table}
\begin{center}
  \caption{ The hadronic matrix elements (a),\,(b) and $B$ parameters
    (c) extracted from the lattice calculations {for} $I=0$
    \cite{Bai:2015nea} and $I=2$ \cite{Blum:2015ywa}. The experimental
    values of the real parts of the amplitudes have been used
    \cite{Blum:2015ywa}.  The large errors result from the quoted
    lattice errors on the hadronic matrix elements. The experimental
    errors are small in comparison.  We take $\mu = 1.3 \; \textrm{GeV}$.  }
\label{tab:bfactors}
\subtable[]{
\begin{tabular}{lc}
\hline
\hline 
$i $ & $\langle Q_i \left( \mu \right) \rangle_{0}^{{\overline{\textrm{MS}}\textrm{--NDR}}}$$\left(\GEV\right)^3$ \\
\hline
$1$ & $ {-0.144} \pm  0.046$\\
$2$ & $0.105 \pm  0.015$\\
$3$ & $ {-0.040} \pm   {0.068}$\\
$4$ & $ {0.210} \pm   {0.069} $\\
$5$ & $ {-0.179} \pm  0.068$\\
$6$ & $ {-0.338} \pm  {0.121}$\\
$7$ & $ {0.154} \pm  0.065$\\
$8$ & $ {1.540} \pm   {0.372}$\\
$9$ & $ {-0.197} \pm  {0.070}$\\
$10$ & $ {0.053} \pm   {0.038}$\\
\hline \hline
\end{tabular}
}
\hspace{1cm}
\subtable[]{
\begin{tabular}{lc}
\hline
\hline 
$i $ & $\langle Q_i \left( \mu \right) \rangle_{2}^{{\overline{\textrm{MS}}\textrm{--NDR}}}$$\left(\GEV\right)^3$ \\
\hline
$1$ & $0.01006 \pm 0.00002$\\
$2$ & $0.01006 \pm 0.00002$\\
$3$ &  ---\\
$4$ &  ---\\
$5$ &  ---\\
$6$ & ---\\
$7$ & $ {0.127} \pm  {0.012}$\\
$8$ & $ {0.852} \pm  {0.052}$\\
$9$ & $0.01509 \pm 0.00003$\\
$10$ & $0.01509 \pm 0.00003$\\
\hline \hline
\end{tabular}
}
\\
\vspace{0.5cm}
\subtable[]{
\begin{tabular}{lr}
\hline
\hline 
$B_1^{(1/2)} \left(\mu \right)$ & $ {35.5} \pm 11.2$\\
$B_2^{(1/2)} \left(\mu \right)$ & $ {5.17} \pm 0.71$\\
$B_3^{(1/2)} \left(\mu \right)$ & $ {-3.27} \pm  {5.60}$\\
$B_5^{(1/2)} \left(\mu \right)$ & $ {0.88}\pm  {0.33}$\\
$B_6^{(1/2)} \left(\mu \right)$ & $ {0.56} \pm  {0.20}$\\
$B_7^{(1/2)} \left(\mu \right)$ & $ {0.24} \pm  {0.10}$\\
$B_8^{(1/2)} \left(\mu \right)$ & $ {0.98} \pm  {0.24}$\\
\hline
$B_1^{(3/2)} \left(\mu \right)$ & $0.437 \pm 0.001$\\
$B_7^{(3/2)} \left(\mu \right)$ & $ {0.37} \pm  {0.03}$\\
$B_8^{(3/2)} \left(\mu \right)$ & $ {0.77} \pm  {0.05}$\\
\hline \hline
\end{tabular}
}
\end{center}
\end{table}

The decay amplitude of $K \to \left( \pi \pi \right)_{I =2}$ has also
been computed using a $2+1$ flavor lattice QCD simulations, albeit at
the scale $\mu= 3.0$ GeV \cite{Blum:2011ng,Blum:2012uk, Blum:2015ywa}.
According to Ref.~\cite{Blum:2012uk}, one can extract the lattice
results in an operator basis renormalized by the
$\overline{\textrm{MS}}$--NDR regularization scheme.  From
Ref.~\cite{Blum:2015ywa}, which is the latest lattice QCD calculation
for $ I = 2$, we obtain \beq
\mathcal{M}_{(27,1)}^{\overline{\textrm{MS}}\textrm{--NDR} } (3\,\GeV) & = 3 \sqrt{3} \langle Q_1 (3\,\GeV) \rangle_2  =  0.0502  \pm  0.0031~(\GeV)^3\\
\mathcal{M}_{(8,8)}^{\overline{\textrm{MS}}\textrm{--NDR} } (3\,\GeV) &= 2 \sqrt{3} \langle Q_7 (3\,\GeV) \rangle_2  = 0.993  \pm  0.038~(\GeV)^3,\\
\mathcal{M}_{(8,8)_{\textrm{mix}}}^{\overline{\textrm{MS}}\textrm{--NDR}
} (3\,\GeV) &= 2 \sqrt{3} \langle Q_8 (3\,\GeV) \rangle_2 = 4.547 \pm
0.275 ~(\GeV)^3, \eeq where the results of the
$(\Slash{\it{q}},\Slash{\it{q}})$ intermediate scheme are taken as
central value, while the results of the $(\gamma^{\mu}, \gamma^{\mu})$
scheme are taken as uncertainty.  Using the three flavor evolution
matrix in Eq.~\eqref{eq:RGHME}, we obtain the hadronic matrix elements
at $\mu = 1.3$ GeV for the $I = 2$ states. Here, we use the lattice input $\alpha_s$ value: $\alpha_s^{(3)} (3\,\GeV) = 0.24544
$ \cite{Blum:2012uk}.
  Then, from the experimental
value of Re$A_2$ we determine $\langle Q_{2} (\mu)\rangle_2 $ (and
$\langle Q_{1,9,10} (\mu)\rangle_2 $ through Eq.~\eqref{eq:Fierz},
$\langle Q_{1} (\mu)\rangle_2 = \langle Q_{2} (\mu)\rangle_2 $ and $Q_9
= \frac{1}{2} \left( 3 Q_1 - Q_3\right)$ which is a Fierz relation).
The results are shown in Table~\ref{tab:bfactors}(b).  Note that through
the evolution matrices $\hat{U}_{QED} $ and $\hat{U}_{QCD\textrm{-}QED}$
this procedure generates small nonzero values of $\langle
Q_{\textrm{3--6}} (\mu) \rangle_2$, which are regarded as
non-electroweak penguin contributions to Im$A_2$.  Since the lattice
simulations have not calculated them at $3.0$ GeV, one should not use
them at the lower hadronic scale $\mu$.  On the other hand, they have
been calculated {with} chiral perturbation theory
\cite{Cirigliano:2003nn, Cirigliano:2003gt} and are included in the
isospin-violating corrections $\hat{\Omega}_{\textrm{eff}}$ of
Eq.~\eqref{eq:epsilonequation}\footnote {The non-electroweak penguin
  contributions are calculated at $\mu = 1.0 \pm 0.3$ GeV
  \cite{Cirigliano:2003nn, Cirigliano:2003gt}.  }.  Therefore, we
{have} decided to omit these contributions at the hadronic scale $\mu$.

To compare with the literature, we also extract $B$ parameters
from the hadronic matrix elements in Table~\ref{tab:bfactors}(c).  These
$B $ parameters are defined as {in Ref.~}\cite{Buras:2015yba}:
\beq 
\langle Q_6 \left( \mu \right) \rangle_0 & = -4 \sqrt{\frac{3}{2}}
\left( \frac{m_K^2}{m_s (\mu) + m_d ( \mu) }\right)^2 \left( F_K - F_{\pi} \right) B^{(1/2)}_6 \left(\mu\right),\\
\langle Q_8 \left( \mu \right) \rangle_2 & = \sqrt{3} \left(
  \frac{m_K^2}{m_s (\mu) + m_d ( \mu) }\right)^2 F_{ \pi} B^{(3/2)}_8
\left(\mu \right), \eeq 
All other {{$B$ parameters} are defined in Ref.~\cite{Buras:1993dy}.
  For running quark masses, we use  {the} lattice results $m_s
    (2\,\GeV)$ $= 93.8 (2.4)~\MeV$ and $m_d (2\,\GeV) = 4.68 (16) $~MeV
    with the three-flavor RG evolution \cite{Aoki:2013ldr}.  Since the
    uncertainty from the strange quark mass is already included in the
    lattice results of $\langle {Q}_i \rangle_I$ as one of the
    systematic errors, we  {do} not include it in the estimation of
    uncertainties of the $B$ parameters.  The $B$ parameters are
    consistent with Ref.~\cite{Buras:2015yba}, and we also confirmed
    the
    almost $\mu$-independent behavior of $B_6^{(1/2)}(\mu)$ and
    $B_8^{(3/2)}(\mu)$ \cite{Buras:1993dy}.  Note that in the following
  analysis we {will} directly use the hadronic matrix elements $\langle
  Q_i \rangle_I$ {rather than the} $B$ parameters.

  Finally we combine the {short-distance} and long-distance contributions.
  The master equation of $\epsilon_{K}'/\epsilon_{K}$ is given {in}
  Eq.~\eqref{eq:epsilonequation}. Since the isospin-violating correction
  by the electroweak penguins to Im$A_0$ are already subtracted from
  $\hat{\Omega}_{ \textrm{eff}}$ as $\langle Q_{\textrm{7--10}}
  \rangle_0$, {one should evaluate the last term in
    Eq.~\eqref{eq:epsilonequation} as}  
\beq
\left( 1- \hat{\Omega}_{\textrm{eff}}\right)
  \textrm{Im}A_0 = \left( 1- \hat{\Omega}_{\textrm{eff}}\right) \left(
    \textrm{Im}A_0\right)^{\textrm{others}} + \frac{1}{a} \left(
    \textrm{Im}A_0\right)^{\textrm{EWP}}, ~~~~a = 1.017,
\label{eq:Omega}
\eeq {with the two terms representing the contributions from}
$\langle Q_{\textrm{3--6}}\rangle_0$ and $\langle
Q_{\textrm{7--10}}\rangle_0$, {respectively} \cite{Buras:2015yba}.
In addition, the experimental values of Re$A_0$ in Eq.~\eqref{eq:ReA0}
and $| \epsilon_{K} | = 2.228 \times 10^{-3}$ \cite{Agashe:2014kda} are
used.  Our result for $ \epsilon_{K}'/\epsilon_{K}$ in the Standard
Model at the next-to-leading order is
\beq %
\left(\frac{\epsilon_{K}'}{\epsilon_{K}} \right)_{\textrm{SM--NLO}}=
\left( {1.06}\pm  {4.66_{\rm{Lattice}}} \pm {1.91_{\rm{NNLO}}} \pm  {0.59_{\rm{IV}}} \pm  { 0.23_{m_t}} \right) \times 10^{-4}.
\label{eq:smnlo}
\eeq %
The first {error originates} from the lattice-QCD simulations
\cite{Blum:2015ywa, Bai:2015nea} {and is} dominated by the
uncertainty stemming from $ \langle Q_6 \rangle_0 $ {(which is} $\pm
 {4.52}  \times 10^{-4}$) (see Figure~\ref{fig:A0pies}(c)).  The
uncertainties from $ \langle Q_3\rangle_0 $ through Eq.~\eqref{eq:Fierz}
and from $ \langle Q_8\rangle_2 $ are subleading ($\pm  {0.77} \times
10^{-4}$ and $\pm  {0.56} \times 10^{-4}$, respectively).

The second uncertainty comes from {perturbative} higher-order
corrections, {which we estimate in two ways.}  Firstly, we estimate
uncertainties from higher-order corrections to the Wilson coefficients
{by calculating} the RG evolution of the Wilson coefficients with a
different method.  Instead of using the analytic evolution matrices
formulated in Sec.~\ref{section2}, we solve the {corresponding} set
of differential equations numerically.
\beq
\frac{d \vec{v} (\mu)}{d \ln \mu} = \hat{\gamma}^T(g_s (\mu)) \vec{v}
(\mu),~~~~\frac{d \vec{z} (\mu)}{d \ln \mu} = \hat{\gamma}^T(g_s (\mu))
\vec{z} (\mu).
\eeq 
Since this RG evolution contains higher-order {(namely
  $\mathcal{O}(\alpha_s^2, \alpha_s \alpha_{EM})$)} corrections, the
result is interpreted as {a conservative estimate of the}
uncertainty in the short-distance contributions.  As a result, we find
that the Wilson coefficients are shifted by about 10 percent compared
with Table~\ref{tab:wilson}, and we obtain $ \epsilon_{K}' /\epsilon_{K}
=  {-0.32}\times 10^{-4}.  $ Hence, we estimate that the uncertainty from
higher-order corrections is $ \pm  {1.38} \times 10^{-4}$.
%
 \begin{figure}[t]
\begin{center}
\subfigure[$\mu_c$ dependence of $\epsilon'_K/\epsilon_K$ ]
{
 \includegraphics[width=0.45\textwidth, bb = 0 0 360 234]{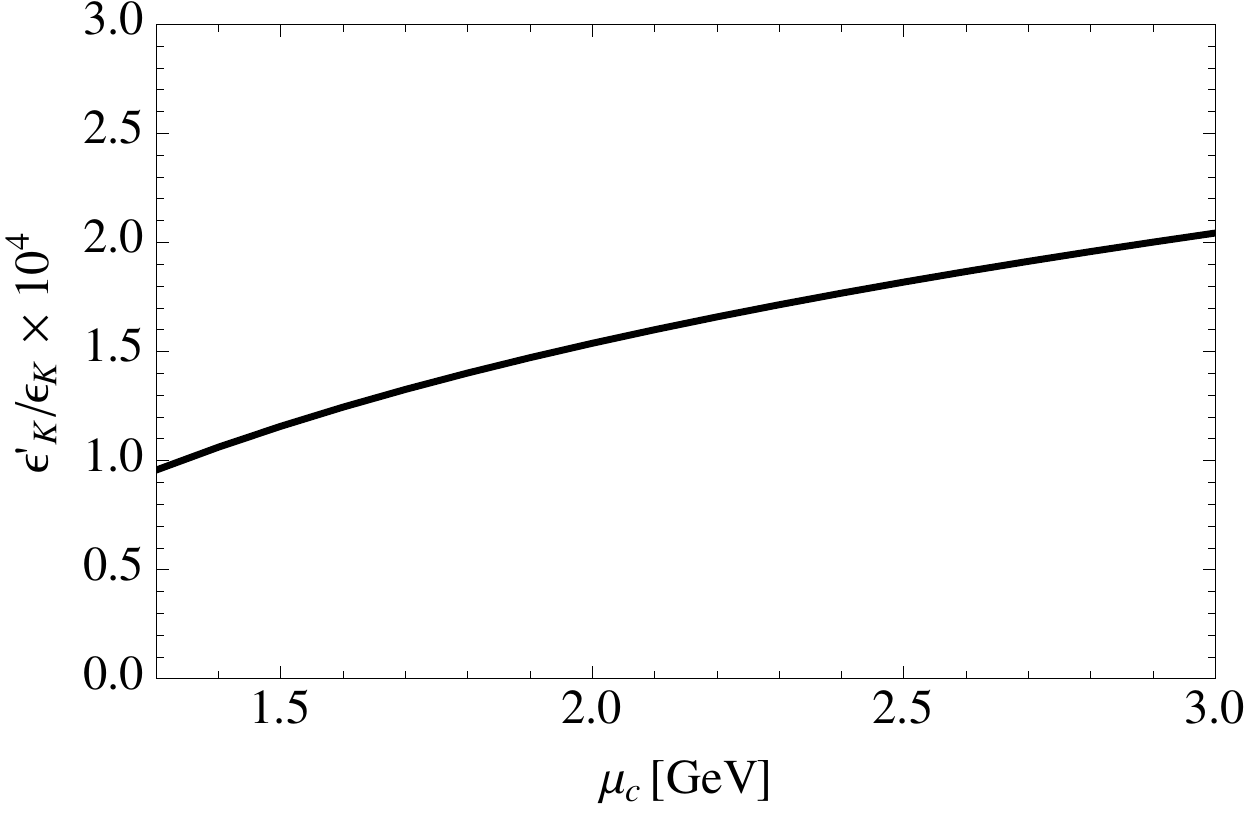}
 }~~~
 \subfigure[$\mu$ dependence of $\epsilon'_K/\epsilon_K$]
 {
  \includegraphics[width=0.45\textwidth, bb = 0 0 360 234]{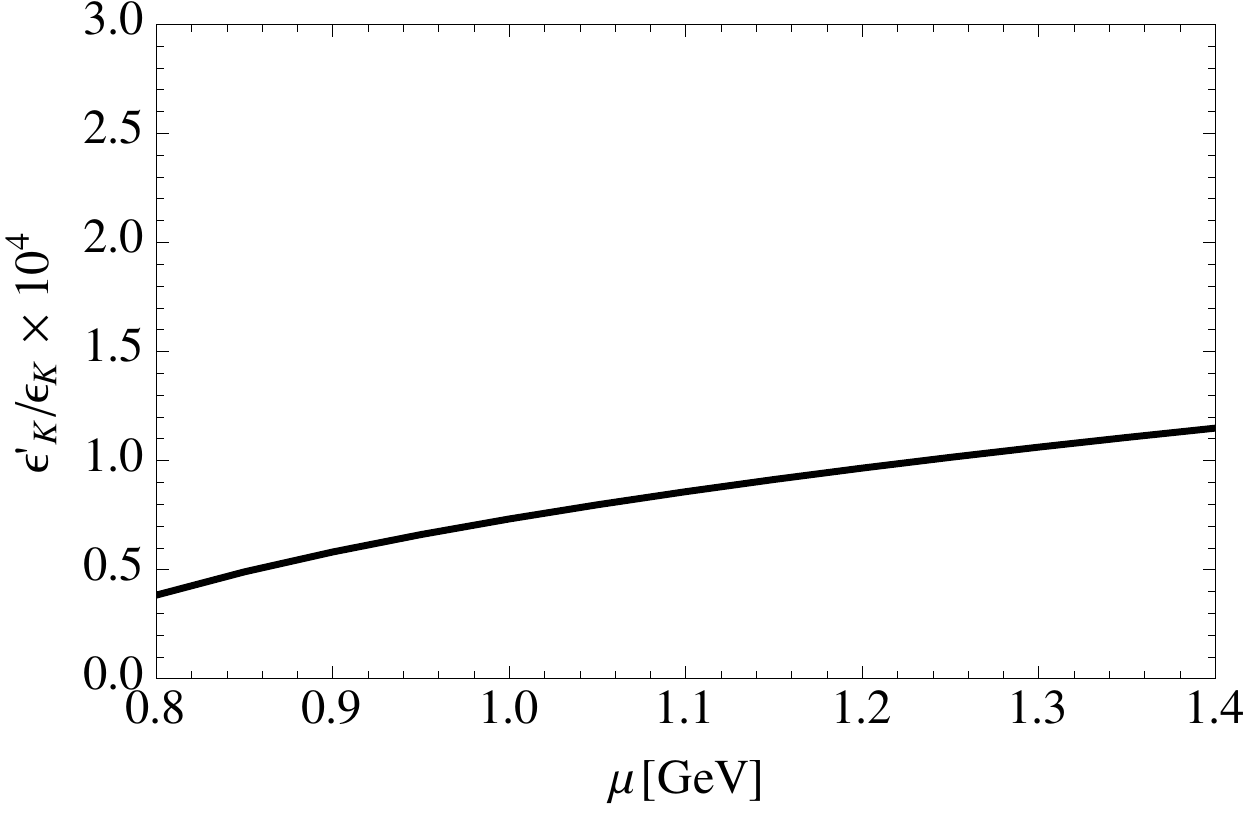}
  }
  \\
 \subfigure[$\mu$ dependence of Im$A_0$]
 {
  \includegraphics[width=0.45\textwidth, bb = 0 0 360 227]{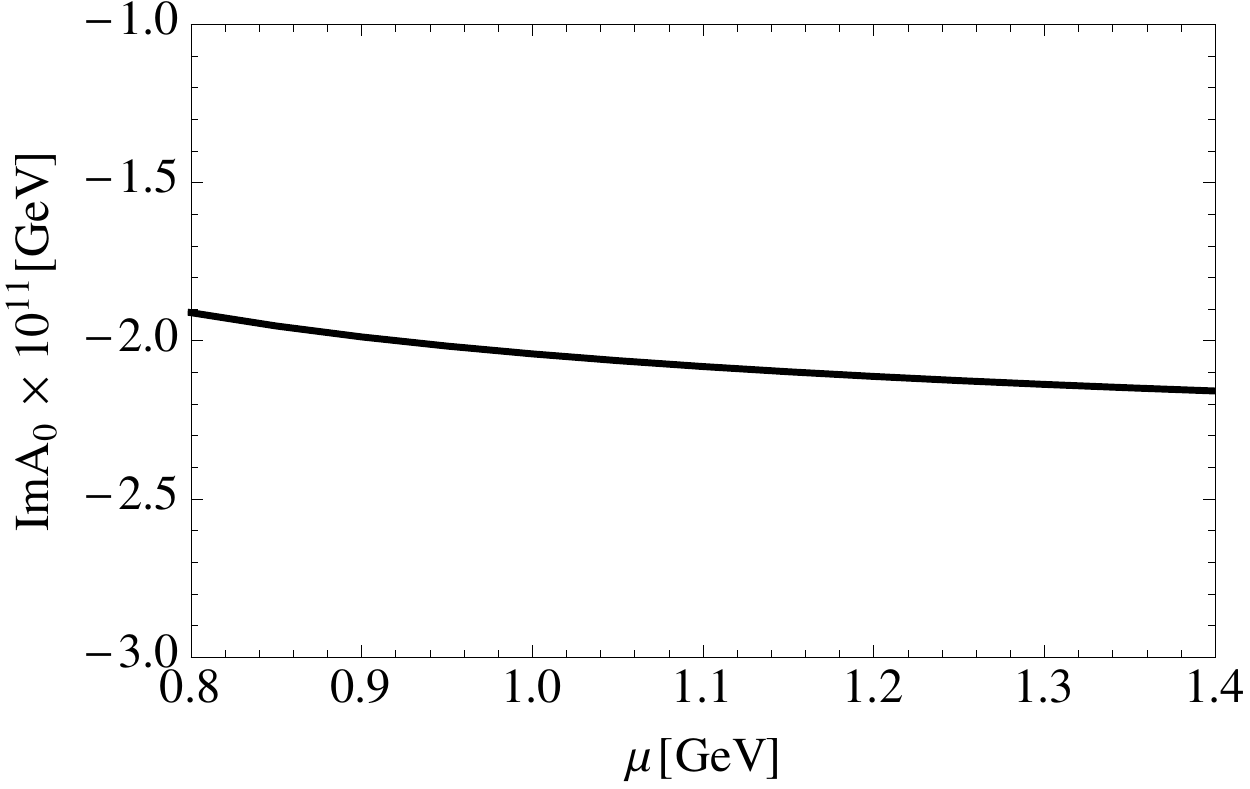}
  }~~~
 \subfigure[$\mu$ dependence of Im$A_2$]
 {
  \includegraphics[width=0.45\textwidth, bb = 0 0 360 227]{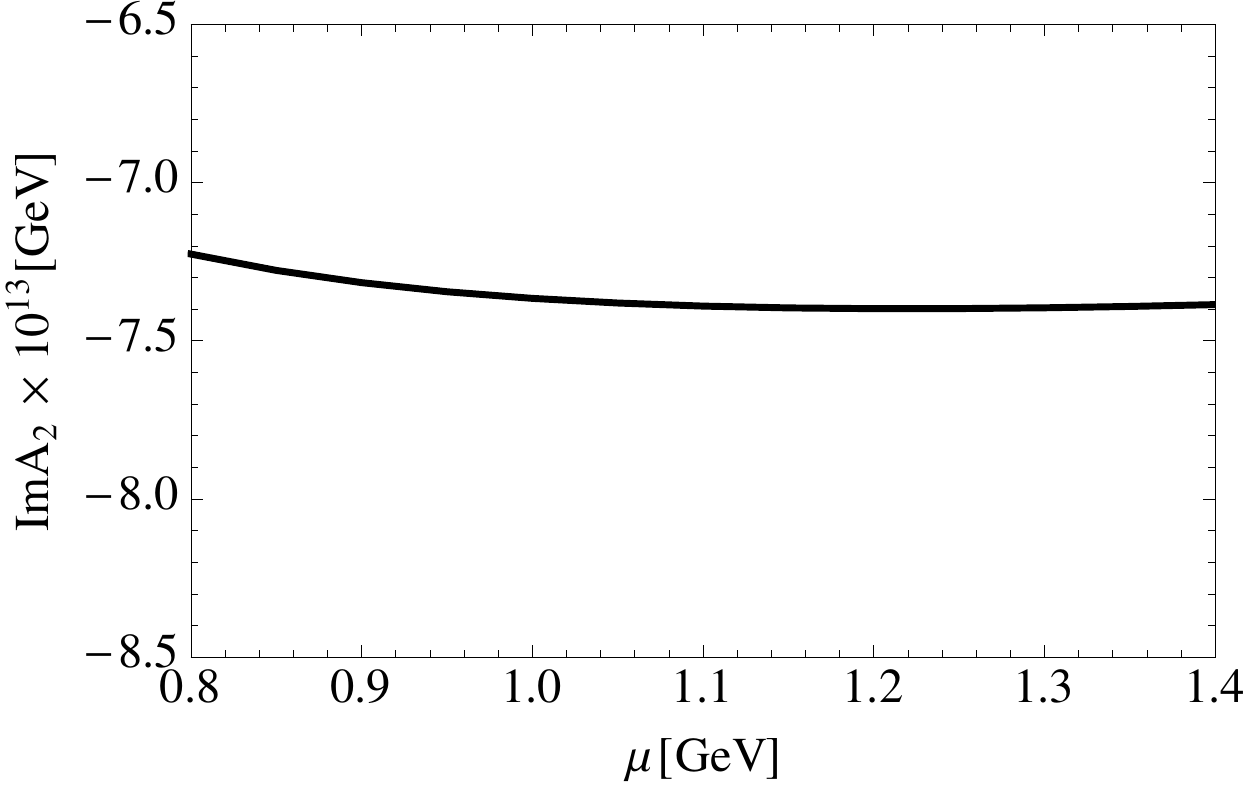}
  }
  \caption{
{(a) The $\mu_c$ dependence of $\epsilon'_{K}/\epsilon_K$ in the range $1.3 < \mu_c <  {3.0}$ GeV with $\mu = 1.3$ GeV.}
The $\mu$ dependence of $\epsilon'_{K}/\epsilon_K$ (b),  Im$A_0$ (c) and Im$A_2$ (d) in the range $ {0.8} < \mu < 1.4 $ GeV with $\mu_c = 1.4$ GeV.
}
 \label{fig:mucdep}
 \end{center}
\end{figure}
%
{Secondly,} we have investigated the $\mu_c$ and $\mu$ dependences of $
\epsilon_{K}'/\epsilon_{K}$.  In Fig.~\ref{fig:mucdep}(a), we show the
$\mu_c$ dependence of $ \epsilon_{K}'/\epsilon_{K}$ in the range $1.3 <
\mu_c <  {3.0}$ GeV with fixed $\mu = 1.3$ GeV.  In
Fig.~\ref{fig:mucdep}(b), we vary $\mu$ with $\mu_c$ fixed at $1.4$ GeV.
We find that the $\mu$ dependence is {small}, $\pm  {0.77} \times
10^{-4}$, while the $\mu_c$ dependence is slightly larger, $\pm  {1.09} 
\times 10^{-4}$.  The {scale $\mu$ enters the prediction in three ways:}
{First, the decomposition of the} isospin-violating corrections in
Eq.~\eqref{eq:Omega} {is imposed at this scale. Second, the omitted}
non-electroweak penguin contributions to Im$A_2$ {depend on $\mu$}, and
third, the experimental values of Re$A_0$ and Re$A_2$ to fix $\langle
Q_{2} (\mu)\rangle_2 $ and $\langle Q_{2} (\mu)\rangle_0 $ {are imposed
  at} the hadronic scale $\mu$.  In this process, we double-count the
uncertainty from the isospin-violating contributions, however, we find
that these uncertainties are very small compared with the uncertainties
stemming from lattice and thus we have not investigated them any
further.  We show the $\mu$ dependences of Im$A_0$ {(and \emph{not} the
  $\mu$ dependence} of $(1-\hat{\Omega}_{\textrm{eff}})$Im$A_0$) and
Im$A_2$ in Figs.~\ref{fig:mucdep}(c) and (d), respectively. We add
  the three uncertainties in quadrature. Strictly speaking, this
  double-counts some pieces of the unknown higher-order corrections.

The third uncertainty {in Eq.~\eqref{eq:smnlo}} stems from
isospin-violating corrections \cite{Cirigliano:2003nn,
  Cirigliano:2003gt}, such as strong isospin violation $(m_u \neq m_d)
$, non-electroweak penguin transitions in the $I=2$ state and $\Delta I
= 5/2$ corrections \cite{Gardner:2000sb,Cirigliano:2000zw}.  The
uncertainty is dominated by the non-electroweak penguin contributions to
Im$A_2$, however, the uncertainty in $\epsilon_{K}'/\epsilon_{K}$ is
small.

The {last} uncertainty {in Eq.~\eqref{eq:smnlo}} comes from the
running mass of the top quark~$m_t ( m_t)$ $= 163.3 \pm 2.7$ GeV
\cite{Alekhin:2012py}.  Since the other uncertainties we have not
elaborated here are negligibly small according to
Ref.~\cite{Buras:2015yba}, we have omitted them in our error
{estimate}.  Therefore, our final result is \beq
\left(\frac{\epsilon_{K}'}{\epsilon_{K}} \right)_{\textrm{SM--NLO}}=
\left(  {1.06} \pm  {5.07} \right) \times 10^{-4}, \eeq which is consistent
with Refs.~\cite{Buras:2015yba} and \cite{Bai:2015nea}.
On the other hand, it is well-known that  the experimental {value} is much larger \cite{Gibbons:1993zq,Barr:1993rx,AlaviHarati:1999xp,Fanti:1999nm,Batley:2002gn, Abouzaid:2010ny}.
The current world average is   \cite{Agashe:2014kda},
\beq
\textrm{Re}\left( \frac{\epsilon_{K}'}{\epsilon_{K}}\right)_{\textrm{exp}} = \left( 16.6 \pm 2.3 \right) \times 10^{-4}.
\eeq
We observe that our prediction of $\epsilon_{K}' /\epsilon_{K}$ in the
Standard Model is $ {2.8}\,\sigma$ below the experimental value.  This
small Standard Model prediction and thus the large tension is supported
by the large-$N_c$ {``dual QCD''} approach \cite{Buras:1987wc,
  Bardeen:1986uz,Bardeen:1986vz,Buras:2014maa,Buras:2015xba,Buras:2016fys},
which is an entirely different approach to low energy QCD than lattice
gauge theory. {There has been a dispute concerning the role of
  final-state interactions (FSI) for the size of $ \langle Q_6 \rangle_0
  $, with the chiral perturbation community favouring an enhancement of
  $ \langle Q_6 \rangle_0 $ by FSI \cite{Pallante:1999qf} and an
  opposing view of the large-$N_c$ community \cite{Buras:2016fys}.
  Modern lattice calculations do include FSI \cite{Lellouch:2000pv} and
  will speak the final word on FSI.}  Since the main uncertainty {of the
  SM prediction for $\epsilon_{K}' /\epsilon_{K}$} comes from
statistical and systematical errors in the lattice calculation of the
hadronic matrix elements for $A_0$, {the expected progress in this field
  will sharpen the Standard Model prediction in the near future
  \cite{Bai:2015nea}.}

We note that in absence of a lattice result for the hadronic matrix
element and the smallness of the corresponding Wilson coefficient, we
omit the contribution from the chromomagnetic penguin operators $Q_{8g}
= m_s g_s/(16 \pi^2) \overline{s} T^a \sigma_{\mu \nu} (1 - \gamma_5) d
G^{\mu \nu\,a}$ (and {the opposite-chirality analogue}
$\tilde{Q}_{8g}$).  According to Ref.~\cite{Buras:2015yba},
{chromomagnetic penguins} contribute
$|0.2$--$0.7|\times10^{-4}$\footnote{The sign depends on the sign of the
  hadronic matrix element.  The preliminary lattice calculation of
  $\langle \pi | Q_{8g} | K \rangle$ \cite{Lubicz:2014qfa} and
  calculations in the chiral quark model \cite{Bertolini:1993rc,
    Deshpande:1994vp, Bertolini:1994qk} imply that a contribution to
  $\epsilon'_K / \epsilon_K$ is positive at the leading order.  However,
  next-to-leading order contributions to $\langle (\pi \pi)_{I=0} |
  Q_{8g} | K^{0} \rangle$ are expected to mess up the leading order
  estimate because of a parametric enhancement $\propto 1/N_c \cdot
  m_K^2 / m_{\pi}^2$ \cite{Barbieri:1999ax, Buras:1999da}.} to
$\epsilon'_K/\epsilon_K$, which rather small compared with {the}
{QCD-penguin} and QED-penguin contributions (see
Figure~\ref{fig:A0pies}(c)).  Even if we {add} this contribution as $
+ 0.7\times10^{-4}$ to the central value (to the higher-order
uncertainty) of $\epsilon'_K/\epsilon_K$, the discrepancy still persists
at $2.7\,(2.8)\,\sigma$.

\begin{figure*}[]
\subfigure[$ \textrm{Im}A_0 $ ]{
 \includegraphics[width=0.32\textwidth, bb = 0 0 267 267]{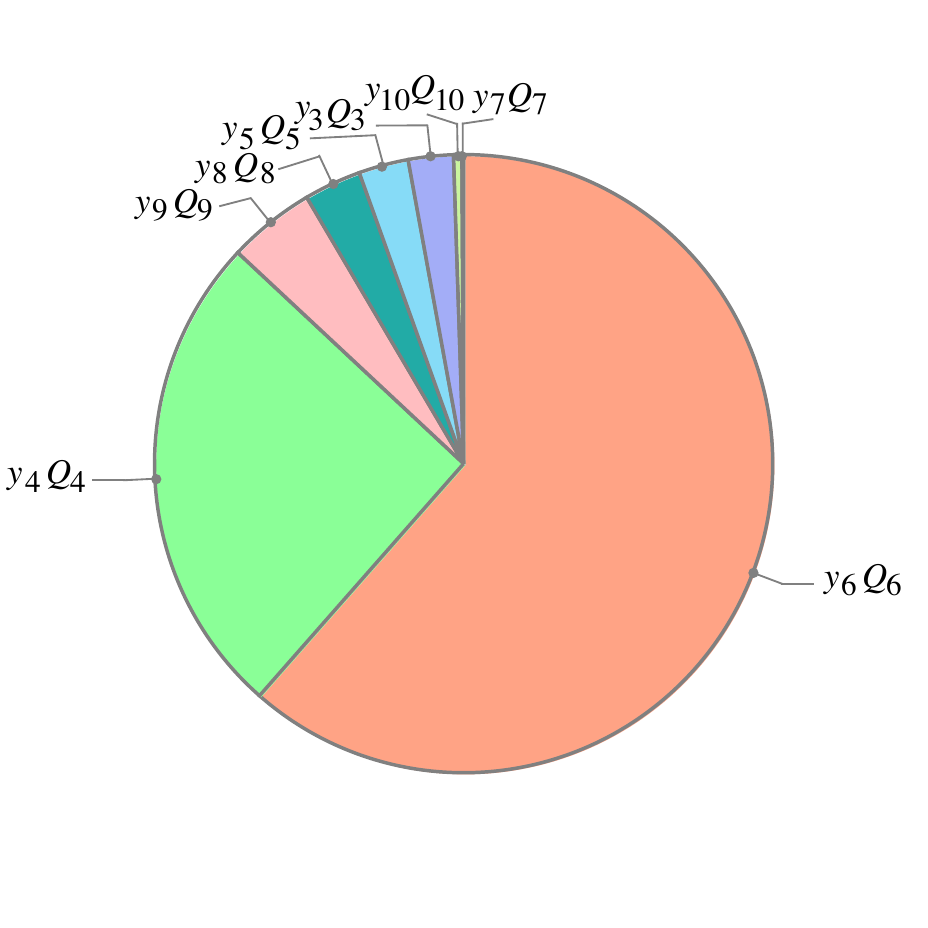}
}
  \hspace{-1.1cm}
\subfigure[$\textrm{Im}A_2 $]{
  \includegraphics[width=0.32\textwidth, bb = 0 0 265 265]{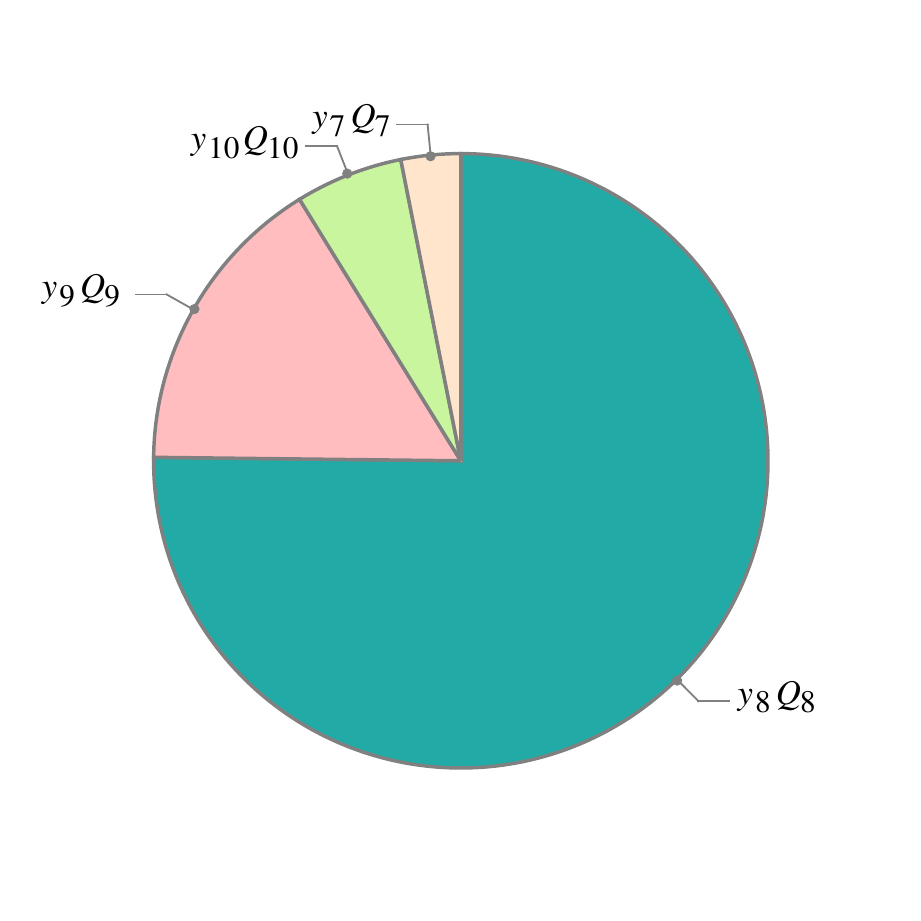}
  }
  \hspace{-1.5cm}
  \subfigure[$ \epsilon_{K}'/\epsilon_{K}$]{
    \includegraphics[width=0.45\textwidth, bb = 0 0 372 262]{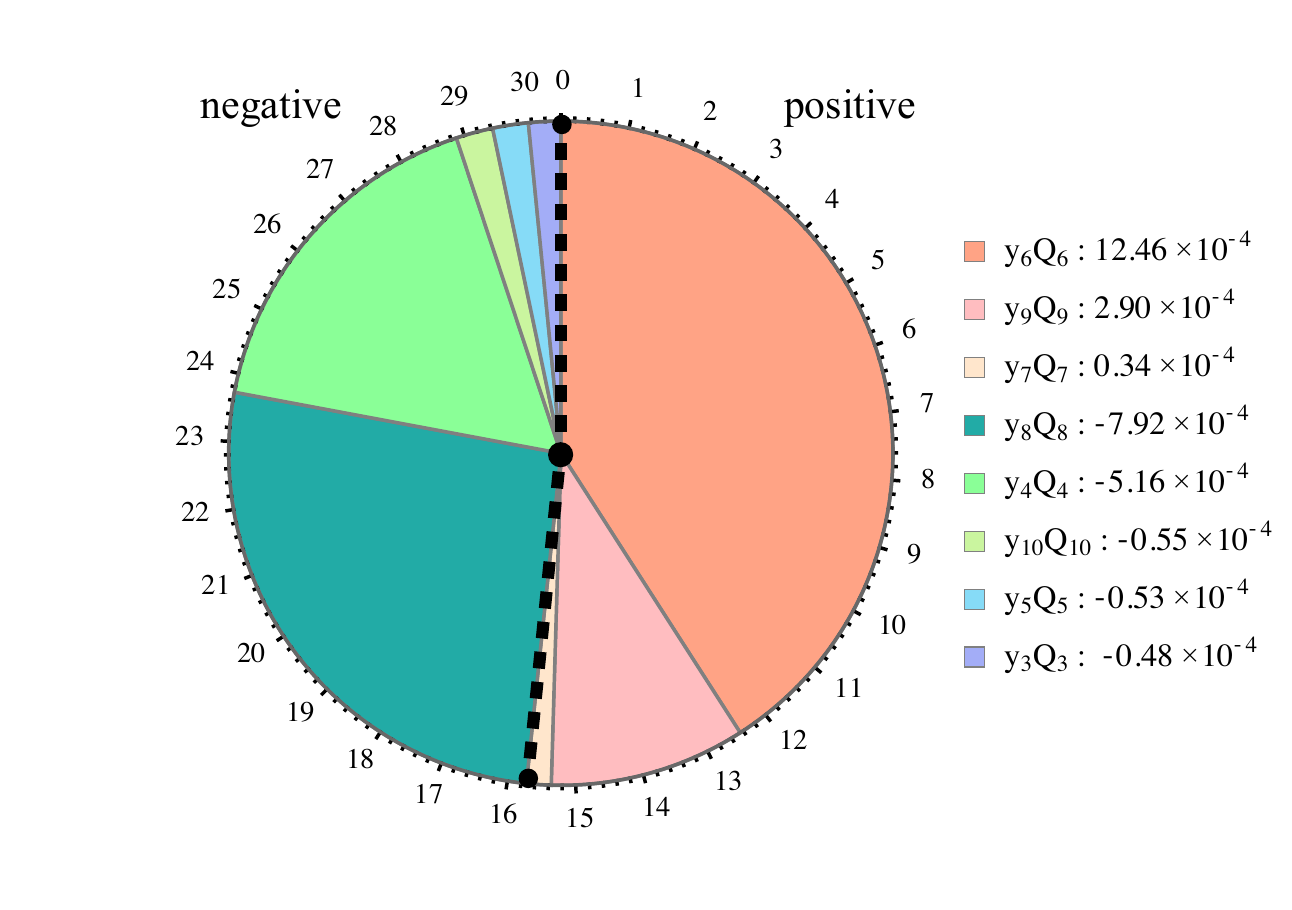}
}
\caption{Composition of Im$A_0$, Im$A_2$ and $
  \epsilon_{K}'/\epsilon_{K}$ with respect to the operator basis. We
  take $\mu = 1.3$ GeV.  In subfigure (c), the right (left) side of the
  dashed line represents positive (negative) contributions.  }
 \label{fig:A0pies}
\end{figure*}
%
In Fig.~\ref{fig:A0pies} we show the composition of Im$A_0$, Im$A_2$
and $ \epsilon_{K}'/\epsilon_{K}$ with respect to the operator basis.
We observe that the positive dominant contribution to $
\epsilon_{K}'/\epsilon_{K}$ comes from $Q_6$ while $Q_9$ is
subdominant. The dominant negative contribution comes from $Q_8$ while
$Q_4$ is subdominant.  Remarkably, their sum almost cancels at
next-to-leading order. This leads to an extremely small {central
  value of the} Standard Model prediction for $
\epsilon_{K}'/\epsilon_{K}$.

Although the results of the Wilson coefficients 
by themselves are slightly different when compared to the result of
Ref.~\cite{Buras:2015yba}, the products  {with the hadronic matrix elements} are well consistent\footnote{
  Indeed, the values of $ {y_6 \langle Q_6 \rangle_0 }$ and $ {y_8 \langle Q_8 \rangle_2}$ are in
  good agreement with Ref.~\cite{Buras:2015yba}.}.  The main difference
{between this reference and our analysis} is in the subleading
contributions.  In Ref.~\cite{Buras:2015yba}, the hadronic matrix
elements $\langle Q_3 (\mu) \rangle_0$, $\langle Q_5 (\mu) \rangle_0$
and $\langle Q_7 (\mu) \rangle_0$ are set to be $0$ as central values,
while we have evaluated them from the lattice data.  The numerical
difference in $\epsilon'_K/\epsilon_K$ is $\sim -1\times 10^{-4} $.  We
also find that the contribution of
$\mathcal{O}(\alpha_{EM}^2/\alpha^2_s)$ terms, which has not been
considered in the literature so far, only contributes to $\epsilon'_K /
\epsilon_K $ as {little} as $ {-0.10} \times 10^{-4}$. {This term,
  however,  {can} be relevant in new-physics models with TeV-scale
  isospin violation.}

\section{Beyond the Standard Model}
\label{sec:NP}

\subsection{Preliminaries}

Upon integrating out heavy degrees of freedom in models of new physics,
{new} contributions to Wilson coefficients of the Standard Model
operators $Q_i$ (and their opposite-chirality {analogues}
$\tilde{Q}_i$) arise.

As we have shown in the previous section, the Standard Model prediction
of $\epsilon'_K/\epsilon_K$ is significantly below the experimental
data.  Although the discrepancy is only $ {2.8}\,\sigma$ at present, its
confirmation {with higher significance} by future lattice {results}
{may establish a}  footprint of new physics.
Indeed, several new physics models can alleviate the
$\epsilon'_K/\epsilon_K$ tension, like generic flavor-violating $Z$ and
$Z^{\prime}$ models \cite{Buras:2014sba,Buras:2015yca,Buras:2015jaq},
331 models \cite{Buras:2014yna,Buras:2015kwd,Buras:2016dxz}, the
Littlest Higgs model with $T$-parity \cite{Blanke:2015wba},
flavor-violating additional pseudo-scalar models \cite{Goertz:2015nkp},
and the Minimal Supersymmetric Standard Model
\cite{Tanimoto:2016yfy,Kitahara:2016otd}. 

Since $\epsilon_{K}^{\prime} / \epsilon_{K}$ is linear in the Wilson
coefficients, the {SM and new-physics contributions are simply
additive:}
\begin{align}
\frac{\epsilon_{K}^{\prime}}{\epsilon_{K}} & = \left(
  \frac{\epsilon_{K}^{\prime}}{\epsilon_{K}} \right)_{\textrm{SM}} +
\left( \frac{\epsilon_{K}^{\prime}}{\epsilon_{K}} \right)_{\textrm{NP}} .
\end{align}
Using the following effective Hamiltonian for the new physics
contributions, %
\beq%
\mathcal{H}^{|\Delta S| = 1}_{\textrm{eff,~NP}} = \frac{G_F}{\sqrt{2}}
\sum_{i=1}^{10} \left( Q_i (\mu) s_i (\mu) + \tilde{Q}_i (\mu)
  \tilde{s}_i (\mu) \right) + \textrm{H.c.},
\label{eq:effHamiNP}
\eeq %
where the opposite-chirality operators $\tilde{Q}_i $ are {found
  from} $Q_i$ by interchanging $V-A \leftrightarrow V+A $, the new
physics contribution is given by 
\beq
\left( \frac{\epsilon_{K}^{\prime}}{\epsilon_{K}} \right)_{\textrm{NP}}
&= \frac{G_F \omega_{+}}{2 \left| \epsilon_{K}^{\textrm{exp}}\right|
  \textrm{Re} A_0^{\textrm{exp}}} \non & \times \left[
  \frac{1}{\omega_{+}} \langle \vec{Q} ( \mu)^T \rangle_2 \textrm{Im}
  \left[ \vec{s}(\mu) - \vec{\tilde{s}} (\mu) \right] - \langle \vec{Q}
  ( \mu)^T \rangle_0 (1-\hat{\Omega}_{\textrm{eff}}) \textrm{Im} \left[
    \vec{s}(\mu) - \vec{\tilde{s}} (\mu) \right] \right]\non & =
\frac{G_F \omega_{+}}{2 \left| \epsilon_{K}^{\textrm{exp}}\right|
  \textrm{Re} A_0^{\textrm{exp}}} \left[ \frac{1}{\omega_{+}} \langle
  \vec{Q} ( \mu)^T \rangle_2 - \langle \vec{Q} ( \mu)^T \rangle_0
  (1-\hat{\Omega}_{\textrm{eff}}) \right] \textrm{Im} \left[
  \vec{s}(\mu) - \vec{\tilde{s}} (\mu) \right],\non & = \frac{G_F
  \omega_{+}}{2 \left| \epsilon_{K}^{\textrm{exp}}\right| \textrm{Re}
  A_0^{\textrm{exp}}} \langle \vec{Q}_{\epsilon'_K} ( \mu)^T \rangle
\textrm{Im} \left[ \vec{s}(\mu) - \vec{\tilde{s}} (\mu) \right] \non & =
\frac{G_F \omega_{+}}{2 \left| \epsilon_{K}^{\textrm{exp}}\right|
  \textrm{Re} A_0^{\textrm{exp}}} \langle \vec{Q}_{\epsilon'_K} ( \mu)^T
\rangle \hat{U} \left(\mu, \mu_{\textrm{NP}} \right) \textrm{Im} \left[
  \vec{s}(\mu_{\textrm{NP}}) - \vec{\tilde{s}} (\mu_{\textrm{NP}})
\right],
\label{eq:NPeps}
\eeq 
where the isospin-violating correction in Eq.~\eqref{eq:Omega} is
\begin{eqnarray}
\left(1-\hat{\Omega}_{\textrm{eff}} \right)_{ij} = 
 \left\{ \begin{array}{ll}
    0.852 & (i=j=\textrm{1--6}) \\
    0.983 & (i=j=\textrm{7--10})\\
    0 &  (i \neq j),
  \end{array} \right.
\end{eqnarray}
and we employed $\langle \tilde{Q}_i ( \mu) \rangle_I  = - \langle Q_i ( \mu) \rangle_I$ and defined  $ \langle \vec{Q}_{\epsilon'_K} \rangle$ as
\beq
 \langle \vec{Q}_{\epsilon'_K} ( \mu)^{T} \rangle  \equiv 
 \frac{1}{\omega_{+}} \langle \vec{Q} ( \mu)^T \rangle_2   -    \langle \vec{Q} ( \mu)^T \rangle_0 (1-\hat{\Omega}_{\textrm{eff}}).    
\eeq
The evolution matrix  in Eq.~\eqref{eq:NPeps} is given by
\beq
\hat{U} \left(\mu, \mu_{\textrm{NP}} \right) \equiv \hat{U}_3 \left(\mu, \mu_{c} \right) \hat{M}_c \left(\mu_c\right) \hat{U}_4 \left(\mu_c,m_b \right) \hat{M}_b (m_b)
\hat{U}_5 \left(m_b, m_t \right) \hat{M}_t (m_t) \hat{U}_6 \left(m_t, \mu_{\textrm{NP}} \right),
\label{eq:U3456}
\eeq Since the matching matrices depend only on the difference of the
number of active up- and down-type quark flavors, we take $\hat{M}_t
(m) = \hat{M}_c (m)$.  Note that the RG evolution {of the}
opposite-chirality operators is {the} same as for the Standard Model
operators and that these {two sets of operators do not mix with each
  other}.  We also note that the chromomagnetic operators are omitted in
our analysis.

In this section, we give a useful formula for the new physics
contributions to $\epsilon'_K / \epsilon_K$ considering the analytic
solutions of the next-to-leading order evolutions matrices and the
hadronic matrix elements we derived.  We note that we omit the weak
boson exchanges in the RG evolutions from $\mu_{\textrm{NP}}$ to $M_W$,
where $\mu_{\textrm{NP}}$ represents the matching scale between the new
physics and the effective Hamiltonian in Eq.~\eqref{eq:effHamiNP}.
{Like the} photon exchanges {one should treat} weak boson exchanges as
next-to-leading contributions.  {Note that} {large isospin violation in
  new-physics models enters $\epsilon'_K / \epsilon_K$ through 
  the initial conditions of the Wilson coefficients  and not through
  the RG evolution.} 
  
We also should comment on the running of $\alpha_{EM}$.  Above $M_W$
  scale, we use $e(\mu_{\textrm{NP}}) = g (\mu_{\textrm{NP}})
  g'(\mu_{\textrm{NP}}) / \sqrt{g^2(\mu_{\textrm{NP}}) +
    g^{'2}(\mu_{\textrm{NP}})}$, and $\beta^{e}_0 = \beta_0^{g'} / \cos
  \theta_W^2(M_Z) $, where $\beta_0^{g'} = -53/9 ~(\mu < m_t)$ or $ -
  41/6 ~(\mu > m_t) $.  Strictly speaking, we have to consider the
  running of $\theta_W$ for consistency.  {However, we have checked that the
    numerical effect for an $\mathcal{O}(10$ TeV) scale of new physics is
    small. Therefore we use a fixed value: $\sin^2 \theta_W = 0.231$.}

\subsection{Counting of Orders\label{sec:co}}

In a full next-to-leading order estimation, we have to consider the
leading order term $\mathcal{O}(1)$ arising from the one-loop QCD RG
evolution as well as the terms defined as next-to-leading order, which
are: the one-loop QED correction $\mathcal{O}(\alpha_{EM}/ \alpha_s)$,
the QCD two-loop correction $\mathcal{O}(\alpha_s)$, and the two-loop
term including a photon and a gluon  {at}
$\mathcal{O}(\alpha_{EM})$. The next-to-leading order RG evolution
matrix has an additional $\mathcal{O}(\alpha_{EM}^2/\alpha^2_s)$
correction, which appears only at this order.  Hereafter, we will always
refer to these orders when labelling perturbative quantities of the
Wilson coefficients and the evolution matrices as
$\vec{s}_0,~\vec{s}_e,~\vec{s}_s,~\vec{s}_{se},$ {$\vec{s}_{ee}$} and
$\hat{U}_{0},~\hat{U}_{e},~\hat{U}_{s},~\hat{U}_{se},~\hat{U}_{ee}$,
respectively.

When we multiply two quantities which are given by a perturbation
series, we have to carefully keep track of and consistently discard
higher orders {of the perturbative series}.  This is a subtle {and
  cumbersome} feature which complicates mathematical expressions.  In
this context, equations of the RG evolution should be more of a symbolic
character which are exact in the limit of expanding the corresponding
quantities to all orders.  Since we necessarily truncate the
perturbation expansion of the Wilson coefficients as well as the
evolution matrices at some point, a product of them at next-to-leading
order is represented as follows:
\begin{align}
\label{eq:NPRGexpand}
&\langle \vec{Q}_{\epsilon'_K} ( \mu)^{T} \rangle  \left( \vec{s} (\mu) - \vec{\tilde{s}} (\mu)\right) \non
&~~=    \langle \vec{Q}_{\epsilon'_K} ( \mu)^{T} \rangle   \hat{U} (\mu,\mu_{\textrm{NP}})\left(  \vec{s} (\mu_{\textrm{NP}})- \vec{\tilde{s}} (\mu_{\textrm{NP}})\right) \non
& \stackrel{(\textrm{NLO})}{=}  \langle \vec{Q}_{\epsilon'_K} ( \mu)^{T} \rangle  \big( \hat{U}_{0} + \hat{U}_{e}  + \hat{U}_{s}  + \hat{U}_{se}  + \hat{U}_{ee} \big) \big( \vec{s}_{0}  + \vec{s}_{e}  + \vec{s}_{s}  + \vec{s}_{se} {+\vec{s}_{ee}} \big) \nonumber\\
&~~=  \langle \vec{Q}_{\epsilon'_K} ( \mu)^{T} \rangle   \left( \underbrace{\hat{U}_{0} \vec{s}_{0}}_{=: \vec{s}_{0} (\mu)} + \underbrace{\hat{U}_{0}\vec{s}_{e} + \hat{U}_{e}  \vec{s}_{0}}_{=: \vec{s}_{e} (\mu)} + \underbrace{ \hat{U}_{0}\vec{s}_{s} +  \hat{U}_{s} \vec{s}_{0}}_{=: \vec{s}_{s} (\mu)} \right. \non
& ~~~~~~ \left.+ \underbrace{\hat{U}_{0}\vec{s}_{se}  + \hat{U}_{e} \vec{s}_{s}  +  \hat{U}_{s}\vec{s}_{e} +  \hat{U}_{se}\vec{s}_{0}}_{=: \vec{s}_{se} (\mu)}
 +  \underbrace{\hat{U}_{e} \vec{s}_{e} +  \hat{U}_{ee} \vec{s}_{0} {+  \hat{U}_{0} \vec{s}_{ee}  } }_{=: \vec{s}_{ee} (\mu)}  \right) 
+ \mathcal{O}  \left(\frac{\alpha_{EM}^2}{\alpha_s }, \alpha_s^2, \alpha_s \alpha_{EM},  \alpha_{EM}^2\right)\nonumber\\
&~~= \langle \vec{Q}_{\epsilon'_K} ( \mu)^{T} \rangle  \left( \underbrace{\vec{s}_{0} (\mu) + \vec{s}_{e} (\mu) + \vec{s}_{s} (\mu) + \vec{s}_{se} (\mu) + \vec{s}_{ee} (\mu)}_{=:\vec{s}_{\textrm{NLO}} (\mu)} \right) + \mathcal{O} \left(\frac{\alpha_{EM}^2}{\alpha_s }, \alpha_s^2, \alpha_s \alpha_{EM},  \alpha_{EM}^2\right)\nonumber\\
&~~= \langle \vec{Q}_{\epsilon'_K} ( \mu)^{T} \rangle   \vec{s}_{\textrm{NLO}} (\mu)  + \mathcal{O}\left(\frac{\alpha_{EM}^2}{\alpha_s }, \alpha_s^2, \alpha_s \alpha_{EM},  \alpha_{EM}^2\right).
\end{align}
Here we {have} suppressed the opposite-chirality coefficients
$\vec{\tilde{s}}$ and the arguments of $\hat{U} (\mu,\mu_{\textrm{NP}})$
and $\vec{s} (\mu_{\textrm{NP}})$ for {better} readability.  This procedure
defines $\vec{s}_{\textrm{NLO}} (\mu)$ as a next-to-leading order
quantity, where higher orders have been discarded consistently.

In view of undetermined Wilson coefficients, it is beneficial to arrange
{the} terms {above} according to the Wilson coefficients
{evaluated} at the new physics scale as 
\beq 
\langle \vec{Q}_{\epsilon'_K} ( \mu)^{T} \rangle \vec{s} (\mu) &
\stackrel{(\textrm{NLO})}{=} \langle \vec{Q}_{\epsilon'_K} ( \mu)^{T}
\rangle\left[ \left( \hat{U}_{0} + \hat{U}_{e} + \hat{U}_{s} +
    \hat{U}_{se} + \hat{U}_{ee} \right) \vec{s}_{0} \right.\non &
~~~~~~~~~~~ \left.+ \left( \hat{U}_{0} + \hat{U}_{e} + \hat{U}_{s}
  \right) \vec{s}_{e} + \left( \hat{U}_{0} + \hat{U}_{e} \right)
  \vec{s}_{s} + \hat{U}_{0} \vec{s}_{se} {+ \hat{U}_{0} \vec{s}_{ee}
  } \right],
 \label{eq:QUs}
\eeq
where we {have} again suppressed $\vec{\tilde{s}}$ and the arguments of
$\hat{U} (\mu,\mu_{\textrm{NP}})$ and $\vec{s} (\mu_{\textrm{NP}})$.
{For} given numerical values for the hadronic matrix elements at a
low scale and {with our} evolution matrices {connecting}
$\mu_{\textrm{NP}}$ with the low scale $\mu$, we can determine the
\emph{weights}\ {which multiply the} 
Wilson coefficients Im$[ \vec{s} (\mu_{\textrm{NP}})
-\vec{\tilde{s}} (\mu_{\textrm{NP}})]$ {in Eq.~\eqref{eq:QUs}}
for {any chosen} scale of new physics. \\

\subsection{Evolution Matrices at the TeV scale}

Above the electroweak scale we observe an approximately logarithmic
behavior of the evolution matrix $\hat{U}(\mu, \mu_{ \textrm{NP}})$ in
Eq.~\eqref{eq:U3456} with increasing energy scale.  {This
  observation}  allows us to {derive} an approximation for the evolution
matrix in the high energy region, {which has an error of only a few 
percent}.
We give approximate functions for all {components} of the evolution
matrix {linking} the new physics scale to the hadronic scale. Cast in
the form
\begin{equation}
\hat{U}_{0,e,s,se,ee}(\mu, \mu_{ \textrm{NP}}) = \hat{U}_{1,fit} + \hat{U}_{2,fit} \ln\frac{\mu_{\textrm{NP}}}{1\,\TeV},
\end{equation}
we combine them in terms of Eq.~\eqref{eq:QUs}.

Using the analytic evolution matrices evaluated in Sec.~\ref{section2} and the next-to-leading order matching matrices $\hat{M}_{c,b,t}$,
we obtain
\beq
\hat{U}_{0} \left( \mu, \mu_{\textrm{NP}} \right)+ \hat{U}_{e} \left( \mu, \mu_{\textrm{NP}} \right)+ \hat{U}_{s} \left( \mu,  \mu_{\textrm{NP}} \right)+ \hat{U}_{se} \left( \mu, \mu_{\textrm{NP}} \right) + \hat{U}_{ee}\left( \mu, \mu_{\textrm{NP}} \right) \non
\simeq \hat{U}_{0,1,fit} + \hat{U}_{ 0,2,fit} \ln \frac{\mu_{\textrm{NP}}}{1\,\TeV},
\label{eq:app0}
\eeq
for the $\mathcal{O}(1)$ Wilson coefficients at the $\mu_{\textrm{NP}}$ scale, and
\beq
\hat{U}_{0} \left( \mu, \mu_{\textrm{NP}} \right)+ \hat{U}_{e} \left( \mu, \mu_{\textrm{NP}} \right)+ \hat{U}_{s} \left( \mu,  \mu_{\textrm{NP}} \right)   &
\simeq \hat{U}_{e,1,fit} + \hat{U}_{e,2,fit} \ln \frac{\mu_{\textrm{NP}}}{1\,\TeV},\\
\hat{U}_{0} \left( \mu, \mu_{\textrm{NP}} \right)+ \hat{U}_{e} \left( \mu, \mu_{\textrm{NP}} \right) &
\simeq \hat{U}_{s,1,fit} + \hat{U}_{s,2,fit} \ln \frac{\mu_{\textrm{NP}}}{1\,\TeV},\\
\hat{U}_{0} \left( \mu, \mu_{\textrm{NP}} \right) &
\simeq \hat{U}_{se,1,fit} + \hat{U}_{se,2,fit} \ln \frac{\mu_{\textrm{NP}}}{1\,\TeV},
\label{eq:appse}
\eeq
for the
$\mathcal{O}(\alpha_{EM}/\alpha_s),~\mathcal{O}(\alpha_s),~\mathcal{O}(\alpha_{EM})$
{(or $\mathcal{O}(\alpha_{EM}^2/\alpha^2_s)$)} Wilson coefficients at
the $\mu_{\textrm{NP}}$ scale, respectively.  Here $\mu = 1.3$ GeV and
$\mu_c = 1.4$ GeV are taken, and the fitting matrices $\hat{U}_{fit}$
are given in Appendix~\ref{app:approximation}.
 We find that these approximate evolution matrices are highly accurate
 in the range of $500\,\textrm{GeV}$--$10\,\textrm{TeV}$.

In order to estimate which Wilson coefficients are expected to gain large enhancements through the RG evolution, we calculate \emph{weights}\ for the Wilson coefficients at the $\mu_{ \textrm{NP}}$ scale.
We regard the coefficients of $\langle \vec{Q}_{\epsilon'_K} ( \mu)^{T} \rangle \sum_i$ $\hat{U}_{i} \left(\mu, \mu_{ \textrm{NP}} \right) $ $( \vec{s}(\mu_{\textrm{NP}}) - \vec{\tilde{s}} (\mu_{\textrm{NP}}) )$  in Eq.~\eqref{eq:QUs} as weights of the Wilson coefficients.

%
%
\begin{table}[tbph]
\begin{center}
\caption{The coefficient $ \langle \vec{Q}_{\epsilon^{\prime}_K}(\mu)^{T} \rangle ( \hat{U}_0 + \hat{U}_{e} + \hat{U}_{s} + \hat{U}_{se} + \hat{U}_{ee} )  $ for the  $\mathcal{O}(1)$ Wilson coefficients  at the scale $\mu_{\textrm{NP}} $ in units of $(\GeV)^{3}$, where $\mu = 1.3 $ GeV.
}
\label{table:weight0}
  \begin{tabular}{crrrr}
Coefficients & \multicolumn{4}{c}{ $ \langle \vec{Q}_{\epsilon^{\prime}_K}(\mu)^{T} \rangle \left( \hat{U}_0 + \hat{U}_{e} + \hat{U}_{s} + \hat{U}_{se} + \hat{U}_{ee} \right)$} \\
\cmidrule{2-5}
$\mu_{\textrm{NP}}~[\textrm{TeV}]$ & $ 1$  & $3$&  $ 5 $ & $10$  \\
\midrule
$s_{0,1}-\tilde{s}_{0,1}$ &  0.265 &0.236 &0.221  & 0.199 \\
$s_{0,2}-\tilde{s}_{0,2}$ &   $-0.062$ &  $-0.085$&  $-0.095$& $-0.108$ \\
$s_{0,3}-\tilde{s}_{0,3}$ &  0.045& $0.006$&$-0.014$& $-0.044$ \\
$s_{0,4}-\tilde{s}_{0,4}$ & $ -0.193$ & $-0.178$ & $-0.168$& $-0.153$\\
$s_{0,5}-\tilde{s}_{0,5}$ & 0.081 & $-0.016$&$-0.067$ &$-0.145$\\
$s_{0,6}-\tilde{s}_{0,6}$ & 0.305 & 0.147& $0.058$ & $-0.076$\\
$s_{0,7}-\tilde{s}_{0,7}$ &  26.16 &  29.97 &  31.76& 34.19\\
$s_{0,8}-\tilde{s}_{0,8}$ &  88.61 &  100.46 & 106.02 &113.60\\
$s_{0,9}-\tilde{s}_{0,9}$ &  0.117 &$-0.024$ &$-0.097$ & $-0.201$\\
$s_{0,10}-\tilde{s}_{0,10}$ &  $-0.084$ &  $-0.147$&  $-0.177$ &$-0.219$\\
  \end{tabular}
 \end{center}
\vspace{0.5cm}
\begin{center}
\caption{The coefficient $ \langle \vec{Q}_{\epsilon^{\prime}_K}(\mu)^{T} \rangle ( \hat{U}_0 + \hat{U}_{e} + \hat{U}_{s} )  $ for the $\mathcal{O}(\alpha_{EM}/\alpha_s)$ Wilson coefficients  at the scale $\mu_{\textrm{NP}} $ in units of $(\GeV)^{3}$, where $\mu = 1.3 $ GeV.
}
\label{table:weighte}
  \begin{tabular}{crrrr}
Coefficients & \multicolumn{4}{c}{ $ \langle \vec{Q}_{\epsilon^{\prime}_K}(\mu)^{T} \rangle \left( \hat{U}_0 + \hat{U}_{e} + \hat{U}_{s} \right)$} \\
\cmidrule{2-5}
$\mu_{\textrm{NP}}~[\textrm{TeV}]$ & $ 1$  & $3$&  $ 5 $ & $10$  \\
\midrule
$s_{e,1}-\tilde{s}_{e,1}$ & 0.290 & 0.267 &0.255 &0.237 \\
$s_{e,2}-\tilde{s}_{e,2}$ &   $-0.076$ &  $-0.101$&  $-0.112$& $-0.127$ \\
$s_{e,3}-\tilde{s}_{e,3}$ & 0.090& $0.065$&$0.051 $& $0.030$ \\
$s_{e,4}-\tilde{s}_{e,4}$ & $ -0.234$ & $-0.228$ & $ -0.222 $& $-0.213$\\
$s_{e,5}-\tilde{s}_{e,5}$ & 0.144& $0.066$&$0.023$ &$ -0.042$\\
$s_{e,6}-\tilde{s}_{e,6}$ & 0.423 &0.301 &0.230 & $0.120$\\
$s_{e,7}-\tilde{s}_{e,7}$ &  26.29&  30.14&  31.93 & 34.38\\
$s_{e,8}-\tilde{s}_{e,8}$ & 88.77  &  100.67  & 106.24& 113.85\\
$s_{e,9}-\tilde{s}_{e,9}$ &  0.216 &$0.101$ &$0.041$ & $-0.045$\\
$s_{e,10}-\tilde{s}_{e,10}$ &  $-0.096$ &  $-0.162$&  $-0.193$ &$-0.236$\\
  \end{tabular}
 \end{center}
\end{table}
\begin{table}[tbph]
\begin{center}
\caption{The coefficient $ \langle \vec{Q}_{\epsilon^{\prime}_K}(\mu)^{T} \rangle ( \hat{U}_0 + \hat{U}_{e} )  $ for the  $\mathcal{O}(\alpha_s)$ Wilson coefficients  at the scale $\mu_{\textrm{NP}} $ in units of $(\GeV)^{3}$, where $\mu = 1.3 $ GeV.
}
\label{table:weights}
  \begin{tabular}{crrrr}
Coefficients & \multicolumn{4}{c}{ $ \langle \vec{Q}_{\epsilon^{\prime}_K}(\mu)^{T} \rangle \left( \hat{U}_0 + \hat{U}_{e}  \right)$} \\
\cmidrule{2-5}
$\mu_{\textrm{NP}}~[\textrm{TeV}]$ & $ 1$  & $3$&  $ 5 $ & $10$  \\
\midrule
$s_{s,1}-\tilde{s}_{s,1}$ & 0.288 & 0.266 &0.254 &0.236 \\
$s_{s,2}-\tilde{s}_{s,2}$ &   $-0.086$ &  $-0.111$&  $-0.122$& $-0.136$ \\
$s_{s,3}-\tilde{s}_{s,3}$ &  0.096& $0.071 $&$0.058 $& $0.037$ \\
$s_{s,4}-\tilde{s}_{s,4}$ & $-0.219$ & $-0.208$ & $-0.200$& $-0.188$\\
$s_{s,5}-\tilde{s}_{s,5}$ & 0.091& $ 0.004 $&$-0.043$ &$-0.113$\\
$s_{s,6}-\tilde{s}_{s,6}$ & 0.264 &0.119 &0.038 & $-0.086$\\
$s_{s,7}-\tilde{s}_{s,7}$ &  22.30 & 25.42&  26.88 & 28.86\\
$s_{s,8}-\tilde{s}_{s,8}$ &  75.45 & 85.00 & 89.47 & 95.57\\
$s_{s,9}-\tilde{s}_{s,9}$ & 0.208 &$0.092 $ &$0.032$ & $-0.055$\\
$s_{s,10}-\tilde{s}_{s,10}$ &  $-0.108$ &  $-0.173$&  $-0.204$ &$-0.246$\\
  \end{tabular}
 \end{center}
\vspace{0.5cm}
\begin{center}
\caption{The coefficient $ \langle \vec{Q}_{\epsilon^{\prime}_K}(\mu)^{T} \rangle  \hat{U}_0 $ for the  $\mathcal{O}(\alpha_{EM})$ and $\mathcal{O}(\alpha^2_{EM}/\alpha^2_s )$ Wilson coefficients  at the scale  $\mu_{\textrm{NP}} $ in units of $(\GeV)^{3}$, where $\mu = 1.3 $ GeV.
}
\label{table:weightse}
  \begin{tabular}{crrrr}
Coefficients & \multicolumn{4}{c}{ $ \langle \vec{Q}_{\epsilon^{\prime}_K}(\mu)^{T} \rangle  \hat{U}_0 $} \\
\cmidrule{2-5}
$\mu_{\textrm{NP}}~[\textrm{TeV}]$ & $ 1$  & $3$&  $ 5 $ & $10$  \\
\midrule
$s_{se,1}-\tilde{s}_{se,1}$ & 0.391 &0.401 &0.406 & 0.412 \\
$s_{se,2}-\tilde{s}_{se,2}$ &   $-0.075$ &  $-0.098 $&  $-0.108 $& $-0.121$ \\
$s_{se,3}-\tilde{s}_{se,3}$ &0.154 & $0.167 $&$0.173 $& $0.181$ \\
$s_{se,4}-\tilde{s}_{se,4}$ & $-0.356$ & $-0.387$ & $-0.402 $& $-0.421$\\
$s_{se,5}-\tilde{s}_{se,5}$ & 0.448 & $0.495$&$0.517$ &$ 0.546$\\
$s_{se,6}-\tilde{s}_{se,6}$ & 1.126&1.251 & 1.309 & $1.388$\\
$s_{se,7}-\tilde{s}_{se,7}$ &  22.77 &26.06 &  27.60 & 29.70 \\
$s_{se,8}-\tilde{s}_{se,8}$ &  76.05 &85.80 & 90.38 & 96.62\\
$s_{se,9}-\tilde{s}_{se,9}$ & 0.556 &$0.568 $ &$0.574$ & $0.582$\\
$s_{se,10}-\tilde{s}_{se,10}$ &  $0.004$ &  $-0.027 $&  $-0.040$ &$-0.058$\\
  \end{tabular}
 \end{center}
\end{table}
%
 In Table~\ref{table:weight0}, we list the coefficient $
\langle \vec{Q}_{\epsilon^{\prime}_K}(\mu)^{T} \rangle ( \hat{U}_0 +
\hat{U}_{e} + \hat{U}_{s} + \hat{U}_{se} + \hat{U}_{ee} ) $ for the
$\mathcal{O}(1)$ Wilson coefficients at the scale $\mu_{\textrm{NP}} =
1,\,3,\,5$ and $10\,\TeV$ in units of $(\GeV)^{3}$, where the hadronic
matrix elements of Table~\ref{tab:bfactors} are taken.  Similarly, the
weights of the
$\mathcal{O}(\alpha_{EM}/\alpha_s),~\mathcal{O}(\alpha_s)$, and
$\mathcal{O}(\alpha_{EM})$ {(or
  $\mathcal{O}(\alpha_{EM}^2/\alpha^2_s)$)} Wilson coefficients are
given in Tables~\ref{table:weighte},\,\ref{table:weights}, and
\ref{table:weightse}, respectively.  Note that these values are not
obtained by fitting but using the exact analytic evolution matrices.  We
observe that these values are of course dominated by $\hat{U}_0$,
{with the sub-dominant contribution stemming from} $\hat{U}_{e}$ because of
the $1/\omega_{+}$ enhancement {and $\hat{U}_{s}$}.  We also find, that the largest weights
come in the $7$ and $8$ components, and they are further enhanced
through the RG evolution in the high energy regime.  Compared with the
coefficients at the weak scale,
\beq &\langle \vec{Q}_{\epsilon'_K} ( \mu)^{T} \rangle \hat{U}_0 (\mu,
M_W) \non &~~= \left(
 {0.37,\, -0.02,\, 0.12,\, -0.29,\, 0.34,\, 0.83,\, 15.33,\, 54.09,\, 0.53,\, 0.08}
\right),\\ &\langle
\vec{Q}_{\epsilon'_K} ( \mu)^{T} \rangle \left( \hat{U}_0 + \hat{U}_{e}
+ \hat{U}_{s} + \hat{U}_{se} + \hat{U}_{ee} \right) (\mu, 1\,\TeV) \non
&~~={ \left(
 {0.27,\, -0.06,\, 0.05,\,-0.19, \,0.08, \,0.31,\, 26.16,\, 88.61,\, 0.12,\,-0.08}
  \right),} \label{eq:weight1}\\ &\langle \vec{Q}_{\epsilon'_K} (
\mu)^{T} \rangle \left( \hat{U}_0 + \hat{U}_{e} + \hat{U}_{s} +
\hat{U}_{se} + \hat{U}_{ee} \right) (\mu, 10\,\TeV) \non &~~={ \left(
 {0.20,\, -0.11, \,-0.04, \,-0.15,\, -0.15,\, -0.08,\, 34.19,\, 113.60, -0.20,\, -0.22}
  \right),}
\eeq 
{the weights of the $7$ and $8$ components increase}
by $50$--$100$\,\% through the RG evolution {at the scale of
  $1$--$10$\,TeV.} If one omits the NLO correction $
\hat{U}_{e}+\ldots\hat{U}_{ee} $ in Eq.~\eqref{eq:weight1}, one finds
$ {22.77} $ and $ {76.05}$ for the 7th and 8th element (see
Tab.~\ref{table:weightse}), which shows the impact of the NLO
corrections on these elements.  Although the enhancement factor from the
RG evolution has been pointed out before in
Ref.~\cite{Buras:2015kwd,Buras:2015jaq} within a leading-order analysis,
it has not been considered in {most} of the literature.  We emphasize
that this factor should be included when one {studies} TeV-scale
new-physics contributions to the QED-penguin operators in order to
alleviate the $\epsilon'_K/\epsilon_K$ discrepancy.

\section{Conclusions and Discussion}
\label{sec:summary}

Based on the first complete lattice calculation of the hadronic matrix
elements for the $K \to \pi \pi $ decay, we have evaluated the
Standard-Model prediction of $\epsilon^{\prime}_K/\epsilon_K$ at the
next-to-leading order.  It is well known that the analytic RG evolution
matrices for the $\Delta S = 1$ nonleptonic effective Hamiltonian at the
next-to-leading order contains singularities in intermediate steps of
the calculation. These singularities make practical calculation
laborious even though {appropriate regulators} disappear {from} the
final (physical) result.  {In this paper, we have generalized the
  analytic ansatz {of the Roma group \cite{Ciuchini:1992tj,
      Ciuchini:1993vr} to solve the RG equations and derive a
    singularity-free solution by adding} logarithmic terms {to the
    ansatz}. As a novel feature of our solution compared to
    Refs.~\cite{Adams:2007tk,Huber:2005ig} we do neither require the
    diagonalization of the LO anomalous dimension matrix nor
    case-by-case implementations for different 
    eigenvalues of this matrix. Instead, the
    different cases are encoded in the $\hat J$ matrices given in
    Eqs.~\eqref{Js0result}--\eqref{Jee1result} and
    Appendix~\ref{app:solutionJ}.  The singular nature of the RG
  equations leads to the presence of spurious parameters which cancel
  between the high-scale and low-scale NLO terms in the RG evolution
  matrix and thereby do not produce any ambiguity and play the role of
  scheme parameters with respect to the regularization of the
  singularities.  {Thus} we have explicitly proven that all
  singularities are automatically {treated in the proper way} without
  the need for a manual regularization of the evolution matrix.  This
  feature also leads to a subtlety whenever the NLO evolution matrix is
  combined with LO initial conditions for the Wilson coefficients, as
  one usually does in studies of new-physics contributions to
  $\epsilon_K^\prime$.}

Using the improved RG evolution matrices and applying the recent lattice
results, we have calculated $\epsilon_{K}'/\epsilon_{K}$ in the Standard
Model at the next-to-leading order.  Our final results is
$\epsilon_{K}'/\epsilon_{K} = \left(  {1.06} \pm  {5.07} \right) \times
10^{-4} $, which is $ {2.8}\,\sigma$ below the measured value.
Our {result is} consistent with the {recent} literature and
highlights a tension between the Standard-Model prediction and
experiment.  The uncertainty is dominated by the lattice result of
$\langle (\pi \pi)_0 | Q_6 | K^0 \rangle$.  Therefore, {upcoming}
improvements of lattice calculations will reveal whether this tension
really {calls} for new physics or not.

We have also evaluated the evolution matrices in the high energy region
for calculations of new physics contributions to
$\epsilon^{\prime}_K/\epsilon_K$.  {To this end} we have {further}
obtained an {easy-to-use} approximate formula for the RG evolution
matrices in the TeV region at the next-to-leading order and {have}
also calculated the weights for each of the Wilson coefficients at the
{scale of new physics}.  We observe that the largest weights come in the
$7$ and $8$ components of the Wilson coefficients and that they are
further enhanced through the RG evolution {between electroweak and
  TeV scales}. Here we confirm the feature noticed at LO in 
Ref.~\cite{Buras:2015kwd,Buras:2015jaq} and find a further enhancement
by the NLO corrections to the evolution matrices. 
Especially the Wilson coefficients of the QED-penguin
operators at the {scale of $1$--$10$\,TeV} increase by $50$--$100$\,\%
compared with the Wilson coefficients at the weak scale.

\section*{Acknowledgements}

The authors thank Andrzej Buras and Chris Sachrajda for illuminating
discussions and Christoph Bobeth for pointing our attention to
Ref.~\cite{Huber:2005ig}. {We are grateful to Andrzej Buras, Martin
  Gorbahn, and Sebastian J\"ager for alerting us to a mistake in an
  earlier version of this paper. We thank the referee {for guiding} us to a
  simpler version of Eq.~(\ref{eq:JseRome}) and
  Eqs.~(\ref{eq:Jse2})--(\ref{eq:Jse0})} {and for pointing out the issue of the running of $\alpha_{EM}$.}
The work of UN is supported by BMBF under
grant no.~05H15VKKB1.  PT acknowledges support from the DFG-funded
doctoral school {\it KSETA}.

\appendix
\section{Solutions for the matrices $\boldsymbol{\hat{J}}$}
\label{app:solutionJ}
In this appendix, we summarize the solutions for the matrices $\hat{J}$
of Eqs.~\eqref{eq:Js1full}--\eqref{eq:Jee0}.  Here we set all arbitrary
parameters to be zero, which does not affect the evolution matrix in
Eq.~\eqref{eq:Ufull}.  We find that the matrices $\hat{J}_{s,1}$,
$\hat{J}_{e,1}$, $\hat{J}_{se,2}$ and $\hat{J}_{ee,1}$ are zero matrices
in the case where the active number of flavours is four, five or six.

In the case of four active quark 
and three active lepton flavors, the matrices $\hat{J}$ are given as
follows: \beq &\hat{J}_{s,0} = \non & \scalebox{0.7}{$ \left(
\begin{array}{cccccccccc}
 -0.05587 & 1.848 & 0 & 0 & 0 & 0 & 0 & 0 & 0 & 0 \\
 1.848 & -0.05587 & 0 & 0 & 0 & 0 & 0 & 0 & 0 & 0 \\
 -0.9365 & -0.4668 & -4.736 & -2.337 & 0.003212 & 0.3418 &
   0.0008031 & 0.08546 & -0.4697 & 0.3586 \\
 0.5649 & -0.07649 & 3.954 & 2.101 & 2.963 & -0.1944 & 0.7408 &
   -0.04860 & 0.6414 & -0.3081 \\
 0.4272 & 0.3745 & 2.458 & 1.908 & -3.758 & 2.824 & -0.6655 &
   0.01002 & 0.05269 & -0.1638 \\
 -1.279 & -1.705 & -8.527 & -8.045 & -11.11 & 5.288 & -5.422 &
   -0.3542 & 0.4257 & -0.09234 \\
 0 & 0 & 0 & 0 & 0 & 0 & -1.096 & 2.784 & 0 & 0 \\
 0 & 0 & 0 & 0 & 0 & 0 & 10.58 & 6.705 & 0 & 0 \\
 0 & 0 & 0 & 0 & 0 & 0 & 0 & 0 & -0.05587 & 1.848 \\
 0 & 0 & 0 & 0 & 0 & 0 & 0 & 0 & 1.848 & -0.05587 \\
\end{array}
\right),
$}\\
&\hat{J}_{e,0} = \non
&
\scalebox{0.65}{$
\left(
\begin{array}{cccccccccc}
 -0.16 & 0 & 0 & 0 & 0 & 0 & 0 & 0 & 0 & 0 \\
 0 & -0.16 & 0 & 0 & 0 & 0 & 0 & 0 & 0 & 0 \\
 0.003439 & -0.005106 & 0.001425 & -0.01567 & 0.03486 & 0.09318 &
   -0.006889 & -0.02519 & -0.07039 & -0.007484 \\
 -0.0005069 & -0.01002 & 0.006128 & -0.01290 & 0.01379 & 0.05192 &
   -0.01036 & 0.006656 & -0.004585 & -0.1036 \\
 0.005392 & -0.006428 & -0.002926 & -0.02657 & 0.07003& 0.1814 &
   0.1252 & 0.1139 & 0.01764 & -0.006001 \\
 -0.003848 & -0.0005921 & 0.003105 & 0.009616 & -0.03212 & -0.07626 &
   -0.01869 & 0.05272 & -0.01310 & -0.006584 \\
 0.1939 & -0.1130 & 0.1686 & -0.4453 & 1.036 & 2.138 & 0.1791 &
   -0.4654 & 0.4974 & -0.1164 \\
 -0.04539 & 0.03811 & -0.04301 & 0.1240 & -0.2759 & -0.5575 &
   0.01411 & 0.2606 & -0.1147 & 0.05233 \\
 0.1096 & 0.02356 & -0.01178 & -0.02391 & -0.1192 & -0.5136 &
   0.1171 & -0.2515 & 0.1747 & 0.08262 \\
 0.03175 & 0.02141 & 0.08054 & -0.1001 & -0.1933 & -0.5465 &
   -0.08608 & -0.4136 & 0.05499 & -0.04569 \\
\end{array}
\right),
$}\\
&\hat{J}_{se,0} = \non
&
\scalebox{0.7}{$
\left(
\begin{array}{cccccccccc}
 0.375 & 1.125 & 0 & 0 & 0 & 0 & 0 & 0 & 0 & 0 \\
 -1.125 & -0.375 & 0 & 0 & 0 & 0 & 0 & 0 & 0 & 0 \\
 -6.983 & 4.245 & -14.00 & 8.408 & 3.891 & 5.925 & 15.05 & 0.5512
   & -13.32 & 8.650 \\
 4.789 & -6.528 & 14.16 & -7.925 & -4.057 & -6.102 & 5.913 &
   -0.09229 & 6.279 & -15.36 \\
 5.844 & 4.699 & 22.01 & 19.72 & 86.73 & -2.892 & 121.9 &
   -0.9073 & 6.526 & 4.236 \\
 -1.550 & -2.109 & -6.160 & -7.277 & -27.45 & 1.337 & -42.88 &
   -8.161 & -1.571 & -2.688 \\
 -16.91 & -11.78 & -84.30 & -74.62 & -347.6 & 11.87 & -86.80 &
   3.237 & -5.914 & 2.873 \\
 4.544 & 3.345 & 26.74 & 24.34 & 110.6 & -6.837 & 26.53 & -1.799 &
   0.2636 & -2.134 \\
 7.741 & -3.157 & 27.85 & 6.262 & -0.2248 & -5.260 & -0.06204 &
   -1.963 & 11.01 & -11.03 \\
 -3.806 & 7.758 & -3.113 & 20.22 & 8.580 & 7.661 & 2.683 & 3.178 &
   -11.43 & 11.46 \\
\end{array}
\right),
$}\\
&\hat{J}_{se,1} = \non
&
\scalebox{0.672}{$
\left(
\begin{array}{cccccccccc}
 -1.437 & -0.3260 & 0 & 0 & 0 & 0 & 0 & 0 & 0 & 0 \\
 -0.3260 & -1.437 & 0 & 0 & 0 & 0 & 0 & 0 & 0 & 0 \\
 -0.1961 & 0.4893 & -0.6832 & 0.6795 & 0.2602 & 0.1364 & 0.06504 & 0.03410 & -0.9651 & 0.9651 \\
 0.5214 & -0.09977 & 1.000 & -0.2338 & -0.05872 & -0.01202 & -0.01468 & -0.003004 & 0.9009 & -0.9009 \\
 0.05525 & 0.004740 & 0.2305 & 0.1295 & 1.006 & -0.09577 & 0.06746 & -0.02394 & 0.05052 & -0.05052 \\
 0.01468 & 0.006783 & 0.07229 & 0.05649 & -0.7178 & -0.8597 & -0.08158 & -0.1054 & 0.007898 & -0.007898 \\
 0 & 0 & 0 & 0 & 0 & 0 & 0.7365 & 0 & 0 & 0 \\
 0 & 0 & 0 & 0 & 0 & 0 & -0.3915 & -0.4379 & 0 & 0 \\
 0.4798 & -0.07969 & 0 & 0 & 0 & 0 & 0 & 0 & 0.002400 & -0.5651 \\
 -0.07969 & 0.4798 & 0 & 0 & 0 & 0 & 0 & 0 & -0.5651 & 0.002400 \\
\end{array}
\right),
$}\\
&\hat{J}_{ee,0} = \non
&
\scalebox{0.65}{$
\left(
\begin{array}{cccccccccc}
 0.09387 & 0 & 0 & 0 & 0 & 0 & 0 & 0 & 0 & 0 \\
 0 & 0.09387 & 0 & 0 & 0 & 0 & 0 & 0 & 0 & 0 \\
 -0.007568 & 0.004171 & -0.003074 & 0.01400 & -0.02073 & -0.04983 & 0.009393 & 0.05187 & 0.02577 & 0.005510 \\
 -0.003805 & 0.008022 & -0.01301 & 0.01704 & -0.002762 & -0.01491 & 0.02298 & 0.05596 & -0.004908 & 0.06248 \\
 0.006342 & -0.003123 & 0.01278 & -0.006150 & 0.002694 & -0.01347 & -0.08904 & -0.1721 & 0.01264 & -0.006294 \\
 -0.001112 & 0.003255 & -0.006809 & 0.001925 & 0.007354 & 0.02220 & 0.02320 & 0.02432 & 0.000069 & 0.008803 \\
 -0.08221 & 0.08779 & -0.06815 & 0.2718 & -0.6678 & -1.605 & 0.02205 & 0.3006 & -0.2125 & 0.1274 \\
 0.02919 & -0.03213 & 0.02574 & -0.09691 & 0.2203 & 0.5159 & -0.02866 & -0.1783 & 0.07471 & -0.04794 \\
 -0.02286 & -0.03780 & 0.01157 & -0.1058 & 0.3251 & 0.8688 & 0.02406 & 0.1580 & 0.01950 & -0.06052 \\
 -0.01653 & -0.02727 & -0.06727 & -0.001290 & 0.2567 & 0.6911 & 0.1135 & 0.4163 & -0.01594 & 0.01271 \\
\end{array}
\right).
$}
\eeq

In the case of five active flavours, the matrices $\hat{J}$ are given as follows:
\beq
&\hat{J}_{s,0} = \non
&
\scalebox{0.7}{$
\left(
\begin{array}{cccccccccc}
 0.09940 & 1.528 & 0 & 0 & 0 & 0 & 0 & 0 & 0 & 0 \\
 1.528 & 0.09940 & 0 & 0 & 0 & 0 & 0 & 0 & 0 & 0 \\
 -0.8769 & -0.5324 & -5.350 & -3.443 & 6.908 & 0.01534 & 0.6908 &
   0.001534 & 0.09398 & 0.5551 \\
 0.3241 & -0.2016 & 2.745 & 1.406 & -5.349 & 0.05042 & -0.5349 &
   0.005042 & 0.3637 & -0.2583 \\
 0.5565 & 0.5109 & 3.804 & 3.112 & -3.433 & 2.928 & -0.2259 &
   0.01534 & -0.2326 & -0.3566 \\
 0.1455 & -0.6772 & -0.6268 & -1.428 & 13.75 & 4.877 & 0.5228 &
   -0.3080 & 0.7500 & -0.3175 \\
 0 & 0 & 0 & 0 & 0 & 0 & -1.174 & 2.775 & 0 & 0 \\
 0 & 0 & 0 & 0 & 0 & 0 & 8.519 & 7.957 & 0 & 0 \\
 0 & 0 & 0 & 0 & 0 & 0 & 0 & 0 & 0.09940 & 1.528 \\
 0 & 0 & 0 & 0 & 0 & 0 & 0 & 0 & 1.528 & 0.09940 \\
\end{array}
\right),
$}\\
&\hat{J}_{e,0} = \non
&
\scalebox{0.65}{$
\left(
\begin{array}{cccccccccc}
 -0.1739 & 0 & 0 & 0 & 0 & 0 & 0 & 0 & 0 & 0 \\
 0 & -0.1739 & 0 & 0 & 0 & 0 & 0 & 0 & 0 & 0 \\
 -0.00008430 & 0.001850 & 0.0008760 & 0.006336 & -0.01477 & -0.03845
   & -0.01774 & -0.05406 & -0.08765 & 0.002382 \\
 -0.004461 & -0.006971 & 0.008984 & 0.0005819 & -0.009264 & -0.01468 &
   -0.01741 & 0.004557 & -0.01788 & -0.1082 \\
 0.005095 & -0.003528 & -0.0004455 & -0.02594 & 0.05269 & 0.1345 &
   0.1129 & 0.05212 & 0.01551 & 0.002383 \\
 -0.004149 & -0.003101 & 0.003889 & 0.006674 & -0.01610 & -0.03635 &
   -0.01215 & 0.08836 & -0.01439 & -0.01264 \\
 0.1078 & 0.09271& 0.04484 & 0.3330 & -0.6222& -2.183 & 0.08561 &
   -0.8183 & 0.3009 & 0.1117 \\
 -0.01577 & -0.03068 & -0.01027 & -0.1407 & 0.2713 & 0.8851 &
   0.05531 & 0.3929 & -0.04218 & -0.02168 \\
 0.1440 & -0.01065 & -0.09022 & -0.05549 & -0.06738 & -0.2646 &
   0.2167 & -0.08880 & 0.3031 & -0.004212 \\
 0.06094 & -0.01177 & 0.04954 & -0.1338 & -0.1006 & -0.2569 &
   -0.02778 & -0.3062 & 0.1580 & -0.1423 \\
\end{array}
\right),
$}\\
&\hat{J}_{se,0} = \non
&
\scalebox{0.7}{$
\left(
\begin{array}{cccccccccc}
 0.375 & 1.125 & 0  &
0 &
  0  &
 0 &
0 &
 0  &
 0  &
0 \\
 -1.125 & -0.375 & 0  &
  0  &
 0  &
0  &
 0  &
  0 &
  0 &
  0  \\
 -2.500 & 1.4315 & -3.851 & 2.898 & 0.9962 & 2.635 & 10.87 &
   -0.1166 & -5.228 & 3.122 \\
 1.642 & -2.134& 4.912 & -1.460 & -1.726 & -2.608 & 6.969 &
   0.4229 & 1.620 & -5.702 \\
 2.180 & 1.968 & 3.317 & 2.955 & 13.61 & -0.6743 & 69.30 &
   -0.1624 & 4.882 & 4.428 \\
 -0.6997 & -1.109 & -0.9273 & -0.8342 & -4.444 & 0.3460 & -27.91 &
   -5.246 & -1.635 & -2.910 \\
 -6.153 & -3.916 & -35.13 & -28.90 & -136.4 & 5.962 & -13.59 &
   0.7483 & 1.775 & 3.592 \\
 1.549 & 0.8194 & 12.28 & 9.893 & 45.61 & -3.670 & 3.698 &
   -0.4178 & -1.493 & -2.488 \\
 3.019 & -0.8036 & 16.92& 5.807 & -9.717 & -3.182 & -1.440 &
   -1.383 & 2.308 & -3.745 \\
 -1.305 & 3.483 & -3.346 & 11.38 & -1.864 & 6.249 & 0.2816 &
   2.227 & -3.812 & 3.052 \\
\end{array}
\right),
$}\\
&\hat{J}_{se,1} = \non
&
\scalebox{0.656}{$
\left(
\begin{array}{cccccccccc}
 -1.361 & -0.3748 & 0 & 0 & 0 & 0 & 0 & 0 & 0 & 0 \\
 -0.3748 & -1.361 & 0 & 0 & 0 & 0 & 0 & 0 & 0 & 0 \\
 0.1224 & 0.1600 & -0.3109 & 0.4207 & 0.1276 & 0.1172 & 0.01276 & 0.01172 & -0.1577 & 0.08237 \\
 0.1835 & 0.1929 & 0.6790 & 0.08838 & -0.09612 & 0.000429 & -0.009612 & 0.000043 & 0.02365 & -0.1460 \\
 0.01271 & -0.01502 & 0.03352 & -0.04966 & 0.3448 & -0.06405 & -0.07195 & -0.006405 & 0.02137 & -0.02022 \\
 0.01922 & 0.01290 & 0.1219 & 0.1030 & -0.3008 & -0.3655 & 0.03423 & 0.04996 & -0.003290 & -0.01277 \\
 0 & 0 & 0 & 0 & 0 & 0 & 1.064 & 0 & 0 & 0 \\
 0 & 0 & 0 & 0 & 0 & 0 & -0.6431 & -0.8651 & 0 & 0 \\
 0.2542 & 0.009287 & 0 & 0 & 0 & 0 & 0 & 0 & -0.5983 & -0.3469 \\
 0.009287 & 0.2542 & 0 & 0 & 0 & 0 & 0 & 0 & -0.3469 & -0.5983 \\
\end{array}
\right),
$}\\
&\hat{J}_{ee,0} = \non
&
\scalebox{0.65}{$
\left(
\begin{array}{cccccccccc}
 0.1159 & 0 & 0 & 0 & 0 & 0 & 0 & 0 & 0 & 0 \\
 0 & 0.1159 & 0 & 0 & 0 & 0 & 0 & 0 & 0 & 0 \\
 -0.006758 & -0.000220 & 0.001054 & -0.007266 & 0.02571 & 0.06903 & 0.01810 & 0.08137 & 0.03717 & 0.002972 \\
 -0.001817 & 0.005815 & -0.01259 & -0.005803 & 0.04023 & 0.09858 & 0.03282 & 0.07166 & 0.000845 & 0.07832 \\
 -0.001142 & 0.01273 & 0.001550 & 0.06754 & -0.1345 & -0.3735 & -0.09641 & -0.1631 & -0.004201 & 0.004407 \\
 0.002002 & -0.001365 & -0.003840 & -0.02566 & 0.05563 & 0.1504 & 0.02425 & 0.009857 & 0.007927 & 0.008735 \\
 -0.007347 & -0.07684 & -0.01964 & -0.3217 & 0.5632 & 1.675 & 0.1525 & 0.6839 & -0.01222 & -0.06966 \\
 0.002651 & 0.02899 & 0.004425 & 0.1344 & -0.2542 & -0.7479 & -0.07896 & -0.3310 & 0.005741 & 0.01977 \\
 -0.07157 & 0.03534 & 0.03101 & 0.08707 & -0.08769 & -0.2702 & -0.06562 & -0.03941 & -0.1143 & 0.06248 \\
 -0.05563 & 0.02067 & -0.05241 & 0.08527 & 0.06749 & 0.1530 & 0.06268 & 0.3315 & -0.1407 & 0.1353 \\
\end{array}
\right).
$}
\eeq
Above the scale $M_W$ in the $f=5$ case
only $\hat{J}_{ee,0}$ is replaced by
\beq
&\hat{J}_{ee,0} = \non
&
\scalebox{0.65}{$
\left(
\begin{array}{cccccccccc}
 0.1020 & 0 & 0 & 0 & 0 & 0 & 0 & 0 & 0 & 0 \\
 0 & 0.1020 & 0 & 0 & 0 & 0 & 0 & 0 & 0 & 0 \\
 -0.006807 & 0.000021 & 0.001232 & -0.006242 & 0.02332 & 0.06286 & 0.01511 & 0.07266 & 0.02996 & 0.003183 \\
 -0.002382 & 0.005057 & -0.01143 & -0.005667 & 0.03882 & 0.09606 & 0.03033 & 0.07109 & -0.001430 & 0.06900 \\
 -0.000380 & 0.01202 & 0.001590 & 0.06336 & -0.1261 & -0.3519 & -0.08537 & -0.1542 & -0.001936 & 0.004376 \\
 0.001464 & -0.001586 & -0.003427 & -0.02457 & 0.05312 & 0.1445 & 0.02260 & 0.01625 & 0.006104 & 0.007528 \\
 -0.000419 & -0.06346 & -0.01667 & -0.2598 & 0.4436 & 1.319 & 0.1296 & 0.5317 & 0.007078 & -0.06051 \\
 0.001486 & 0.02427 & 0.003764 & 0.1115 & -0.2096 & -0.6173 & -0.06592 & -0.2708 & 0.002577 & 0.01708 \\
 -0.05788 & 0.03230 & 0.02696 & 0.08653 & -0.1037 & -0.3173 & -0.05763 & -0.07181 & -0.08512 & 0.05362 \\
 -0.04767 & 0.01862 & -0.04578 & 0.07839 & 0.04976 & 0.1078 & 0.05242 & 0.2757 & -0.1201 & 0.1187 \\
\end{array}
\right).
$}
\eeq

In the case of six active flavours, the matrices $\hat{J}$ are given as follows:
\beq
&\hat{J}_{s,0} = \non
&
\scalebox{0.7}{$
\left(
\begin{array}{cccccccccc}
 0.3146 & 1.056 & 0 & 0 & 0 & 0 & 0 & 0 & 0 & 0 \\
 1.056 & 0.3146 & 0 & 0 & 0 & 0 & 0 & 0 & 0 & 0 \\
 -1.600 & -1.535 & -12.36 & -12.02 & 8.793 & -1.131 & 2.198 &
   -0.2828 & -0.8657 & -0.8686 \\
 0.8862 & 0.5280 & 7.429 & 7.255 & -5.576 & 0.6539 & -1.394 &
   0.1635 & 0.8013 & 0.4058 \\
 0.6905 & 0.7180 & 5.579 & 5.023 & -4.997 & 3.346 & -0.9249 &
   0.1452 & 0.3177 & 0.2199 \\
 0.1455 & -0.6299 & -0.3870 & -1.489 & 7.913 & 5.885 & -0.09758 &
   -0.9513 & 0.8482 & -0.5904 \\
 0 & 0 & 0 & 0 & 0 & 0 & -1.298 & 2.766 & 0 & 0 \\
 0 & 0 & 0 & 0 & 0 & 0 & 8.303 & 9.690 & 0 & 0 \\
 0 & 0 & 0 & 0 & 0 & 0 & 0 & 0 & 0.3146 & 1.056 \\
 0 & 0 & 0 & 0 & 0 & 0 & 0 & 0 & 1.056 & 0.3146 \\
\end{array}
\right),
$}\\
&\hat{J}_{e,0} = \non
&
\scalebox{0.65}{$
\left(
\begin{array}{cccccccccc}
 -0.1905 & 0 & 0 & 0 & 0 & 0 & 0 & 0 & 0 & 0 \\
 0 & -0.1905 & 0 & 0 & 0 & 0 & 0 & 0 & 0 & 0 \\
 -0.004325 & 0.007185 & 0.009385 & 0.009390 & -0.02752 & -0.06628 &
   -0.02694 & -0.07676 & -0.1194 & 0.02764 \\
 -0.009919 & -0.002219 & 0.03288 & -0.01917 & -0.02621 & -0.02129 &
   -0.02834 & 0.01474 & -0.06108 & -0.09564 \\
 0.005509 & -0.002168 & -0.01074 & -0.009225 & 0.03980 & 0.08293 &
   0.1274 & 0.05083 & 0.03016 & -0.005145 \\
 -0.004879 & -0.003778 & 0.01343 & -0.008532 & -0.01554 & -0.01330 &
   -0.01789 & 0.1048 & -0.02867 & -0.01273 \\
 0.1518 & 0.04324 & 0.2197 & 0.2298 & -0.1970 & -1.348 & 0.2127 &
   -0.8371 & 0.5734 & 0.07970 \\
 -0.02942 & -0.01451 & -0.04693 & -0.1085 & 0.1550& 0.6298 &
   0.03875 & 0.4193 & -0.1089 & -0.01106 \\
 0.1723 & -0.02276 & 0.05556 & -0.02633 & -0.008854 & -0.3349 &
   0.3311 & -0.08373 & 0.5572 & -0.08924 \\
 0.08303 & -0.02141 & 0.1183 & -0.1089 & -0.1361 & -0.3994 &
   -0.03402 & -0.4332 & 0.3145 & -0.2324 \\
\end{array}
\right),
$}\\
&\hat{J}_{se,0} = \non
&
\scalebox{0.7}{$
\left(
\begin{array}{cccccccccc}
 0.375 & 1.125 & 0 &
   0  &
 0  &
0  &
  0 &
 0  &
   0  &
0  \\
 -1.125 & -0.375 & 0  &
  0 &
 0 &
  0  &
  0 &
  0  &
  0 &
  0 \\
 -1.717 & 0.4502 & -4.641 & 1.340 & 1.346 & 3.309 & 6.823 &
   -0.04493 & -4.718 & 1.585 \\
 1.485 & -0.6824 & 4.572 & -1.354 & -0.9936 & -3.700 & 6.745 &
   0.4676 & 3.500 & -2.081 \\
 2.474 & 2.178 & 8.318 & 7.311 & 25.97 & -1.158 & 48.94 &
   -0.4794 & 6.973 & 6.144 \\
 -0.6938 & -0.9269 & -2.447 & -2.263 & -8.583 & 0.05831 & -20.43 &
   -7.770 & -1.899 & -3.039 \\
 -4.430 & -2.583 & -32.52 & -26.58 & -103.8 & 4.885 & -26.00 &
   1.075 & 0.3241 & 3.002 \\
 1.087 & 0.2394 & 11.23 & 7.972 & 34.52 & -1.077 & 7.739 &
   -0.2204 & -0.7237 & -2.909 \\
 1.392 & -0.5260 & 11.12 & 3.336 & -9.094 & -3.990 & -4.493 &
   -2.786 & 3.083 & -2.243 \\
 -0.9953 & 1.721 & -6.237 & 4.523 & -0.8947 & 6.267 & -1.189 &
   3.698 & -3.152 & 3.108 \\
\end{array}
\right),
$}\\
&\hat{J}_{se,1} = \non
&
\scalebox{0.646}{$
\left(
\begin{array}{cccccccccc}
 -1.215 & -0.4779 & 0 & 0 & 0 & 0 & 0 & 0 & 0 & 0 \\
 -0.4779 & -1.215 & 0 & 0 & 0 & 0 & 0 & 0 & 0 & 0 \\
 0.03954 & 0.1511 & -0.5005 & 0.3909 & 0.2004 & 0.09779 & 0.05009 & 0.02445 & -0.1793 & 0.2458 \\
 0.1726 & 0.1039 & 0.6485 & -0.07127 & -0.05979 & 0.04647 & -0.01495 & 0.01162 & 0.2136 & -0.1041 \\
 0.02675 & -0.003629 & 0.1532 & 0.03172 & 0.4150 & -0.09556 & 0.04074 & -0.02389 & 0.04375 & -0.03219 \\
 0.009964 & 0.01439 & 0.08856 & 0.1063 & -0.4370 & -0.6094 & 0.002421 & 0.1197 & 0.000558 & 0.01162 \\
 0 & 0 & 0 & 0 & 0 & 0 & 0.2520 & 0 & 0 & 0 \\
 0 & 0 & 0 & 0 & 0 & 0 & -0.4467 & -1.088 & 0 & 0 \\
 0.1161 & -0.000709 & 0 & 0 & 0 & 0 & 0 & 0 & -0.6925 & -0.4811 \\
 -0.000709 & 0.1161 & 0 & 0 & 0 & 0 & 0 & 0 & -0.4811 & -0.6925 \\
\end{array}
\right),
$}\\
&\hat{J}_{ee,0} = \non
&
\scalebox{0.65}{$
\left(
\begin{array}{cccccccccc}
 0.1391 & 0 & 0 & 0 & 0 & 0 & 0 & 0 & 0 & 0 \\
 0 & 0.1391 & 0 & 0 & 0 & 0 & 0 & 0 & 0 & 0 \\
 -0.003752 & -0.006197 & -0.01666 & -0.01541 & 0.04809 & 0.1273 & 0.03381 & 0.1356 & 0.06098 & -0.02018 \\
 0.000326 & 0.001314 & -0.05390 & 0.004611 & 0.07978 & 0.1672 & 0.06923 & 0.1516 & 0.02842 & 0.07314 \\
 0.002820 & 0.007589 & 0.02592 & 0.04766 & -0.1033 & -0.2861 & -0.1243 & -0.2192 & -0.000268 & 0.01032 \\
 -0.000053 & 0.001118 & -0.02262 & -0.01220 & 0.05444 & 0.1313 & 0.04123 & 0.03780 & 0.01107 & 0.01113 \\
 -0.004754 & -0.04852 & -0.04424 & -0.2157 & 0.2605 & 0.8907 & 0.1751 & 0.6774 & 0.000729 & -0.1105 \\
 0.009495 & 0.01675 & 0.02216 & 0.1040 & -0.1569 & -0.5103 & -0.08580 & -0.3685 & 0.03165 & 0.02339 \\
 -0.06667 & 0.03419 & 0.000002 & 0.06611 & -0.03570 & -0.1442 & -0.07226 & -0.1137 & -0.1609 &
   0.1208 \\
 -0.07613 & 0.02859 & -0.1156 & 0.05588 & 0.1459 & 0.3689 & 0.1048 & 0.5101 & -0.2848 & 0.2398 \\
\end{array}
\right).
$}
\eeq

\section{Approximation of Evolution Matrices}
\label{app:approximation}
In this appendix we list the approximate evolution matrices
$\hat{U}_{fit}$ of Eqs.~\eqref{eq:app0}--\eqref{eq:appse}.

The evolution matrices for the $\mathcal{O}(1)$ Wilson coefficients are
\beq
&\hat{U}_{0,1,fit} = \non
&
\scalebox{0.6}{$
\left(
\begin{array}{cccccccccc}
 1.381 & -0.6586 & 0 & 0 & 0 & 0 & 0 & 0 & 0 & 0 \\
 -0.6579 & 1.383 & 0 & 0 & 0 & 0 & 0 & 0 & 0 & 0 \\
 -0.02657 & 0.04460 & 1.347 & -0.5061 & 0.1084 & 0.3509 & 0.01854 & 0.05795 & -0.05577 & 0.05648 \\
 0.03219 & -0.08950 & -0.7163 & 1.076 & -0.1140 & -0.5954 & -0.02081 & -0.1003 & 0.1029 & -0.0863 \\
 0.006732 & 0.009857 & 0.06040 & 0.03421 & 0.8730 & 0.3916 & -0.009192 & 0.006706 & -0.002316 & -0.002685 \\
 0.04266 & -0.1445 & -0.1220 & -0.5808 & 1.196 & 3.844 & -0.04416 & -0.2361 & 0.1676 & -0.1556 \\
 -0.006910 & -0.000813 & -0.005331 & 0.005611 & -0.01547 & -0.01260 & 0.8834 & 0.3349 & -0.01980 & -0.006946 \\
 -0.006196 & 0.000386 & -0.005099 & 0.009086 & -0.02131 & -0.05363 & 1.377 & 5.070 & -0.02191 & -0.004320 \\
 -0.008814 & -0.000827 & 0.005846 & 0.000204 & -0.007655 & -0.005259 & -0.02659 & -0.01518 & 1.348 & -0.6634 \\
 0.002569 & -0.000270 & -0.004176 & 0.01113 & 0.002071 & -0.000822 & 0.008502 & 0.004499 & -0.6453 & 1.377 \\
\end{array}
\right),
$}\\
&\hat{U}_{0,2,fit} = \non
&
\scalebox{0.6}{$
\left(
\begin{array}{cccccccccc}
 0.04902 & -0.06578 & 0 & 0 & 0 & 0 & 0 & 0 & 0 & 0 \\
 -0.06571 & 0.04921 & 0 & 0 & 0 & 0 & 0 & 0 & 0 & 0 \\
 -0.002858 & 0.004911 & 0.04273 & -0.04448 & 0.01743 & 0.06239 & 0.003106 & 0.01145 & -0.005919 & 0.007786 \\
 0.004467 & -0.006409 & -0.05801 & 0.02537 & -0.02718 & -0.09404 & -0.004846 & -0.01681 & 0.008209 & -0.01020 \\
 0.000051 & -0.000773 & 0.000153 & -0.005655 & 0.004400 & 0.04479 & -0.000598 & -0.000873 & 0.001332 & -0.001328 \\
 0.007679 & -0.01904 & -0.001577 & -0.08913 & 0.1533 & 0.3766 & -0.01304 & -0.05873 & 0.02464 & -0.03331 \\
 -0.000971 & 0.000008 & -0.001421 & 0.000965 & -0.003372 & -0.003838 & -0.000828 & 0.04034 &
   -0.003464 & -0.000859 \\
 -0.002034 & 0.000038 & -0.002690 & 0.002769 & -0.008055 & -0.01680 & 0.2019 & 0.6221 & -0.008119 & -0.001580 \\
 -0.001513 & -0.000057 & 0.000438 & -0.000632 & -0.002303 & -0.001513 & -0.005736 & -0.003896 & 0.04208 & -0.06626 \\
 0.000823 & -0.000001 & -0.000722 & 0.001962 & 0.000930 & -0.000258 & 0.002827 & 0.001194 & -0.06154
   & 0.04840 \\
\end{array}
\right).
$}
\eeq

The evolution matrices for the $\mathcal{O}(\alpha_{EM}/\alpha_s)$ Wilson coefficients are
\beq
&\hat{U}_{e,1,fit} = \non
&
\scalebox{0.6}{$
\left(
\begin{array}{cccccccccc}
 1.384 & -0.6596 & 0 & 0 & 0 & 0 & 0 & 0 & 0 & 0 \\
 -0.6596 & 1.384 & 0 & 0 & 0 & 0 & 0 & 0 & 0 & 0 \\
 -0.02644 & 0.04449 & 1.347 & -0.5063 & 0.1084 & 0.3508 & 0.01869 & 0.05786 & -0.05398 & 0.05588 \\
 0.03221 & -0.08918 & -0.7165 & 1.076 & -0.1138 & -0.5947 & -0.02013 & -0.09927 & 0.1027 & -0.08501 \\
 0.006614 & 0.009876 & 0.06024 & 0.03434 & 0.8727 & 0.3913 & -0.01040 & 0.007747 & -0.002679 & -0.002541 \\
 0.04268 & -0.1441 & -0.1221 & -0.5803 & 1.196 & 3.845 & -0.04218 & -0.2321 & 0.1682 & -0.1550 \\
 -0.007192 & -0.001077 & -0.005373 & 0.005309 & -0.01602 & -0.005198 & 0.8816 & 0.3370 & -0.0213 & -0.006792 \\
 -0.004750 & -0.000363 & -0.002599 & 0.006771 & -0.01759 & -0.04805 & 1.385 & 5.079 & -0.01599 & -0.005018 \\
 -0.009682 & -0.001318 & 0.009455 & 0.000347 & -0.008310 & -0.004251 & -0.02899 & -0.01618 & 1.347 & -0.6649 \\
 0.003240 & 0.000337 & -0.006239 & 0.01228 & 0.002650 & -0.000994 & 0.009981 & 0.004416 & -0.6448 & 1.380 \\
\end{array}
\right),
$}\\
&\hat{U}_{e,2,fit} = \non
&
\scalebox{0.6}{$
\left(
\begin{array}{cccccccccc}
 0.04959 & -0.06613 & 0 & 0 & 0 & 0 & 0 & 0 & 0 & 0 \\
 -0.06613 & 0.04959 & 0 & 0 & 0 & 0 & 0 & 0 & 0 & 0 \\
 -0.002814 & 0.004885 & 0.04276 & -0.04452 & 0.01748 & 0.06243 & 0.003194 & 0.01153 & -0.005538 & 0.007572 \\
 0.004455 & -0.006361 & -0.05804 & 0.02537 & -0.02712 & -0.09391 & -0.004734 & -0.01663 & 0.008081 & -0.009949 \\
 0.000025 & -0.000747 & 0.000109 & -0.005599 & 0.004337 & 0.04466 & -0.000661 & -0.000532 & 0.001256 & -0.001236 \\
 0.007654 & -0.01896 & -0.001614 & -0.08899 & 0.1534 & 0.3766 & -0.01270 & -0.05792 & 0.02470 & -0.03318 \\
 -0.001020 & -0.000042 & -0.001512 & 0.000625 & -0.002880 & -0.001394 & -0.000771 & 0.04163 & -0.003738 & -0.000825 \\
 -0.001700 & -0.000089 & -0.001936 & 0.002320 & -0.007113 & -0.01521 & 0.2036 & 0.6241 & -0.006620 & -0.001725 \\
 -0.001794 & -0.000100 & 0.000773 & -0.000779 & -0.002622 & -0.001291 & -0.006709 & -0.004276 & 0.04126 & -0.06661 \\
 0.000955 & 0.000077 & -0.001109 & 0.002185 & 0.001279 & -0.000036 & 0.003500 & 0.001712 & -0.06132 & 0.04896 \\
\end{array}
\right).
$}
\eeq

The evolution matrices for the $\mathcal{O}(\alpha_s)$ Wilson coefficients are
\beq
&\hat{U}_{s,1,fit} = \non
&
\scalebox{0.6}{$
\left(
\begin{array}{cccccccccc}
 1.411 & -0.7127 & 0 & 0 & 0 & 0 & 0 & 0 & 0 & 0 \\
 -0.7127 & 1.411 & 0 & 0 & 0 & 0 & 0 & 0 & 0 & 0 \\
 -0.009379 & 0.03849 & 1.428 & -0.5372 & 0.05796 & 0.3276 & 0.009884 & 0.05309 & -0.03652 & 0.03661 \\
 0.01918 & -0.06796 & -0.7546 & 1.116 & -0.1185 & -0.6070 & -0.02120 & -0.1018 & 0.07697 & -0.06536 \\
 -0.005195 & 0.01615 & 0.009571 & 0.06192 & 0.8626 & 0.1512 & 0.001064 & 0.02609 & -0.01979 & 0.01784 \\
 0.02719 & -0.1165 & -0.1074 & -0.5139 & 1.020 & 3.416 & -0.03532 & -0.1846 & 0.1343 & -0.1238 \\
 -0.007192 & -0.001077 & -0.005373 & 0.005309 & -0.01602 & -0.005198 & 0.8044 & -0.01284 & -0.02130 & -0.006792 \\
 -0.004750 & -0.000363 & -0.002599 & 0.006771 & -0.01759 & -0.04805 & 1.166 & 4.362 & -0.01599 & -0.005018 \\
 -0.009682 & -0.001318 & 0.009455 & 0.000347 & -0.008310 & -0.004251 & -0.02899 & -0.01618 & 1.373 & -0.7181 \\
 0.003240 & 0.000337 & -0.006239 & 0.01228 & 0.002650 & -0.000994 & 0.009981 & 0.004416 & -0.6980 & 1.406 \\
\end{array}
\right),
$}\\
&\hat{U}_{s,2,fit} = \non
&
\scalebox{0.6}{$
\left(
\begin{array}{cccccccccc}
 0.05237 & -0.06874 & 0 & 0 & 0 & 0 & 0 & 0 & 0 & 0 \\
 -0.06874 & 0.05237 & 0 & 0 & 0 & 0 & 0 & 0 & 0 & 0 \\
 -0.001807 & 0.004179 & 0.04830 & -0.04670 & 0.01449 & 0.05948 & 0.002493 & 0.01111 & -0.003888 & 0.006388 \\
 0.003138 & -0.005261 & -0.06167 & 0.02726 & -0.02594 & -0.09351 & -0.004500 & -0.01681 & 0.006327 & -0.008337 \\
 -0.000731 & 0.000706 & -0.001981 & 0.001873 & -0.004333 & 0.01923 & 0.000445 & 0.003261 & -0.001199 & 0.001503 \\
 0.005578 & -0.01553 & -0.002278 & -0.07695 & 0.1317 & 0.3191 & -0.01019 & -0.04632 & 0.01977 & -0.02734 \\
 -0.001020 & -0.000042 & -0.001512 & 0.000625 & -0.002880 & -0.001394 & -0.01510 & -0.002932 & -0.003738 & -0.000825 \\
 -0.001700 & -0.000089 & -0.001936 & 0.002320 & -0.007113 & -0.01521 & 0.1673 & 0.5064 & -0.006620 & -0.001725 \\
 -0.001794 & -0.000100 & 0.000773 & -0.000779 & -0.002622 & -0.001291 & -0.006709 & -0.004276 & 0.04404 & -0.06921 \\
 0.000955 & 0.000077 & -0.001109 & 0.002185 & 0.001279 & -0.000036 & 0.003500 & 0.001712 & -0.06393 & 0.05174 \\
\end{array}
\right).
$}
\eeq

The evolution matrices for the $\mathcal{O}(\alpha_{EM})$ and  $\mathcal{O}(\alpha^2_{EM}/\alpha^2_s)$ Wilson coefficients are
\beq
&\hat{U}_{se,1,fit} = \non
&
\scalebox{0.6}{$
\left(
\begin{array}{cccccccccc}
 1.394 & -0.7045 & 0 & 0 & 0 & 0 & 0 & 0 & 0 & 0 \\
 -0.7045 & 1.394 & 0 & 0 & 0 & 0 & 0 & 0 & 0 & 0 \\
 -0.009453 & 0.03820 & 1.428 & -0.5374 & 0.05790 & 0.3279 & 0.009686 & 0.05425 & -0.04508 & 0.03987 \\
 0.01931 & -0.06738 & -0.7551 & 1.116 & -0.1184 & -0.6078 & -0.02089 & -0.1042 & 0.08184 & -0.07190 \\
 -0.005225 & 0.01599 & 0.009690 & 0.06182 & 0.8626 & 0.1514 & 0.005876 & 0.02676 & -0.01998 & 0.01742 \\
 0.02740 & -0.1156 & -0.1079 & -0.5135 & 1.020 & 3.415 & -0.02789 & -0.1626 & 0.1356 & -0.1214 \\
 0 & 0 & 0 & 0 & 0 & 0 & 0.8305 & 0 & 0 & 0 \\
 0 & 0 & 0 & 0 & 0 & 0 & 1.188 & 4.394 & 0 & 0 \\
 0 & 0 & 0 & 0 & 0 & 0 & 0 & 0 & 1.394 & -0.7045 \\
 0 & 0 & 0 & 0 & 0 & 0 & 0 & 0 & -0.7045 & 1.394 \\
\end{array}
\right),
$}\\
&\hat{U}_{se,2,fit} = \non
&
\scalebox{0.6}{$
\left(
\begin{array}{cccccccccc}
 0.04914 & -0.06644 & 0 & 0 & 0 & 0 & 0 & 0 & 0 & 0 \\
 -0.06644 & 0.04914 & 0 & 0 & 0 & 0 & 0 & 0 & 0 & 0 \\
 -0.001818 & 0.004086 & 0.04834 & -0.04679 & 0.01446 & 0.05962 & 0.002471 & 0.01164 & -0.005626 & 0.007251 \\
 0.003145 & -0.005101 & -0.06177 & 0.02741 & -0.02592 & -0.09379 & -0.004515 & -0.01783 & 0.007658 & -0.009488 \\
 -0.000730 & 0.000669 & -0.001957 & 0.001837 & -0.004334 & 0.01930 & 0.001106 & 0.003520 & -0.001234 & 0.001403 \\
 0.005619 & -0.01524 & -0.002424 & -0.07670 & 0.1318 & 0.3187 & -0.007642 & -0.03995 & 0.02017 & -0.02644 \\
 0 & 0 & 0 & 0 & 0 & 0 & -0.01058 & 0 & 0 & 0 \\
 0 & 0 & 0 & 0 & 0 & 0 & 0.1759 & 0.5171 & 0 & 0 \\
 0 & 0 & 0 & 0 & 0 & 0 & 0 & 0 & 0.04914 & -0.06644 \\
 0 & 0 & 0 & 0 & 0 & 0 & 0 & 0 & -0.06644 & 0.04914 \\
\end{array}
\right).
$}
\eeq




\end{document}